\documentclass[fleqn,usenatbib]{mnras}

\usepackage[T1]{fontenc}

\DeclareRobustCommand{\VAN}[3]{#2}
\let\VANthebibliography\thebibliography
\def\thebibliography{\DeclareRobustCommand{\VAN}[3]{##3}\VANthebibliography}

\usepackage{graphicx}	% Including figure files
\usepackage{amsmath}	% Advanced maths commands
\usepackage{amssymb}	% Extra maths symbols
\usepackage[version=4]{mhchem}
\usepackage{newtxtext,newtxmath}
\usepackage[dvipsnames]{xcolor}

\usepackage{pdflscape}
\usepackage{multirow}
\usepackage{lscape}
\usepackage{tablefootnote}

\title[On dust evolution in binary discs -- comparison with ALMA observations]{On dust evolution in planet-forming discs in binary systems. II -- Comparison with Taurus and $\rho$~Ophiuchus (sub-)millimetre observations: discs in binaries have small dust sizes}

\author[F. Zagaria et al.]{
Francesco Zagaria$^{1,2,3,4}$\thanks{E-mail: fz258@cam.ac.uk},
Giovanni P. Rosotti$^{4,5}$,
Giuseppe Lodato$^{6}$
\\
$^{1}$Dipartimento di Fisica, Università degli Studi di Pavia, Via Agostino Bassi 6, I-27100 Pavia, Italy\\
$^{2}$Scuola Universitaria Superiore IUSS Pavia, Piazza della Vittoria 15, I-27100 Pavia, Italy\\
$^{3}$Institute of Astronomy, University of Cambridge, Madingley Road, Cambridge CB3 0HA, UK\\
$^{4}$Leiden Observatory, Leiden University, P.O.~Box 9513, NL-2300~RA Leiden, the Netherlands\\
$^{5}$School of Physics and Astronomy, University of Leicester, Leicester LE1 7RH, UK\\
$^{6}$Dipartimento di Fisica, Università degli Studi di Milano, Via Giovanni Celoria 16, I-20133 Milano, Italy
}

\date{Accepted XXX. Received YYY; in original form ZZZ}

\pubyear{2020}

\begin{document}
\label{firstpage}
\pagerange{\pageref{firstpage}--\pageref{lastpage}}
\maketitle

\begin{abstract}
    The recently discovered exoplanets in binary or higher-order multiple stellar systems sparked a new interest in the study of proto-planetary discs in stellar aggregations. Here we focus on disc solids, as they make up the reservoir out of which exoplanets are assembled and dominate (sub-)millimetre disc observations. These observations suggest that discs in binary systems are fainter and smaller than in isolated systems. In addition, disc dust sizes are consistent with tidal truncation only if they orbit very eccentric binaries. In a previous study we showed that the presence of a stellar companion hastens the radial migration of solids, shortening disc lifetime and challenging planet formation. In this paper we confront our theoretical and numerical results with observations: disc dust fluxes and sizes from our models are computed at ALMA wavelengths and compared with Taurus and $\rho$~Ophiuchus data. A general agreement between theory and observations is found. In particular, we show that the dust disc sizes are generally smaller than the binary truncation radius due to the combined effect of grain growth and radial drift: therefore, small disc sizes do not require implausibly high eccentricities to be explained. Furthermore, the observed binary discs are compatible within $1\sigma$ with a quadratic flux-radius correlation similar to that found for single-star discs and show a close match with the models. However, the observational sample of resolved binary discs is still small and additional data are required to draw more robust conclusions on the flux-radius correlation and how it depends on the binary properties.
\end{abstract}


%%%%%%%%%%%%%%%%%%%%%%%%%%%%%%%%%%%%%%%%%%%%%%%%%%

%%%%%%%%%%%%%%%%% BODY OF PAPER %%%%%%%%%%%%%%%%%%

\section{Introduction}
About a half of the main-sequence stars are part of binary or higher-order multiple stellar systems \citep[e.g.,][]{Raghavan+10_2010ApJS..190....1R,Moe&DiStefano17_2017ApJS..230...15M} and this fraction is expected to increase significantly in the case of pre-main-sequence stars \citep[e.g.,][]{Duchene&Kraus13_2013ARA&A..51..269D,Chen+13_2013ApJ...768..110C}. Given the mounting evidence that extra-solar planets are almost ubiquitous in our Galaxy \citep[e.g.,][]{Winn&Fabrycky15_2015ARA&A..53..409W}, multiple stellar aggregations should be regarded as the most natural environment in which exoplanets are assembled. 

How planet formation takes place is a long-standing problem and stellar multiplicity is expected to influence significantly this processes \citep[e.g.,][]{Thebault&Haghighipour15_2015pes..book..309T, Marzari&Thebault20_2019Galax...7...84M}. Although it is generally thought that the presence of a stellar companion challenges planet formation \citep[e.g.,][]{Kraus+12_2012ApJ...745...19K, Chen+13_2013ApJ...768..110C, Kraus+16_2016AJ....152....8K}, the increasing (and puzzling!) evidence of binary stars hosting exoplanets \citep[e.g.,][]{Hatzes16_2016SSRv..205..267H, Martin18_2018haex.bookE.156M}, either orbiting one component of the system or both, proves that planets can be assembled in such a rough environment too. 

It is widely acknowledged that planets are born in so-called proto-planetary discs, disc-like objects orbiting young stars and mainly composed of gas and dust. Being these systems non-static, studying the dynamics of their constituents is fundamental to understand %how planets are assembled. In particular, focusing on the evolution of the planet-formation potential in discs in the presence of a stellar companion appears to be a compelling way both to figure out 
how the currently observed population of exoplanets (in multiple stellar systems) may have originated and to explain their properties. 

The effects of stellar multiplicity on disc evolution have been extensively studied in the case of gas. In particular, it has been shown that the angular momentum exchange between a disc and an embedded satellite promotes the truncation of the former at a fraction of the binary separation \citep[e.g.,][]{Goldreich&Tremaine79_1979ApJ...233..857G,Goldreich&Tremaine80_1980ApJ...241..425G,Lin&Papaloizou86_1986ApJ...309..846L}. The final location of the truncation depends both on the mass ratio and the eccentricity of the system \citep[e.g.,][]{Papaloizou&Pringle77_1977MNRAS.181..441P,Artymowicz&Lubow94_1994ApJ...421..651A,Pichardo+05_2005MNRAS.359..521P}, as well as on the binary orbit to disc plane misalignment (e.g., \citealt{Lubow+15_2015ApJ...800...96L}). As a consequence, proto-planetary discs in binaries are expected to be fainter, smaller and therefore \citep[e.g.,][]{Pringle81_1981ARA&A..19..137P} shorter-lived than the single-star ones. 

On the contrary, the evolution of the dust in discs in multiple stellar systems is still poorly constrained from the theoretical point of view. Given that our current knowledge of the properties of binary discs (fluxes and sizes) relies almost entirely on dust observations, it is surprising that only a few studies have focused on this topic so far \citep[e.g.,][]{Zsom+11_2011A&A...527A..10Z}. Indeed, even though dust grains make up only a tiny fraction of the total disc mass, they play a fundamental role in disc evolution. Solids are the building blocks of planets and minor bodies, such as comets, meteors and asteroids. Moreover, dust dominates the (sub-)millimetre thermal continuum observations and sets the conditions for molecular line emission. To fill this gap, in a previous paper (\citealt{Zagaria+21_2021MNRAS.504.2235Z}, Paper~I in the following) we addressed the issue of the secular evolution of dust grains in planet-forming discs in binary systems. In particular, we showed that the radial migration of solids is hastened in those systems, suggesting that dust disperses faster in binary rather than in single-star discs. In this paper we confront the numerical outcomes of Paper~I with the observations.

In the young ($1-3\text{ Myr}$ old) Taurus-Auriga, $\rho$~Ophiuchus and Lupus star-forming regions, \citet{Harris+12_2012ApJ...751..115H}, using SMA data, as well as \citet{Cox+17_2017ApJ...851...83C,Akeson+19_2019ApJ...872..158A,Zurlo+20_2020MNRAS.496.5089Z} and \citet{Zurlo+21_2021MNRAS.501.2305Z}, using ALMA data, showed that binary discs are significantly fainter in the continuum than the single-star ones. Moreover, \citet{Harris+12_2012ApJ...751..115H,Akeson+19_2019ApJ...872..158A,Zurlo+20_2020MNRAS.496.5089Z} and \citet{Zurlo+21_2021MNRAS.501.2305Z} showed that binary disc fluxes increase with the stellar separation and that discs in wide binaries are almost as bright as the single-star ones. On the contrary, in the older ($5-11\text{ Myr}$ old) Upper Scorpius OB-association, \citet{Barenfeld+19_2019ApJ...878...45B} found no statistically relevant differences between binary and single-star disc brightness distribution. As for the disc sizes, almost all the previously cited studies lack the necessary angular resolution to perform a detailed analysis. Recently, \citet{Manara+19_2019A&A...628A..95M} provided the first homogeneous survey of discs in multiple stellar systems in the unbiased sample of Taurus sources in \citet{Long+18_2018ApJ...869...17L,Long+19_2019ApJ...882...49L} whose resolution ($\sim0.12$~arcsec) was high enough to spatially resolve both the circumstellar binary discs. \citet{Manara+19_2019A&A...628A..95M} found that binary discs tend to be smaller than those around single stars. %, in agreement with our theoretical expectations. However, no correlation between dust fluxes and the binary separation was witnessed, probably due to the smallness of their sample.

%Before comparing the observational results with the models, 
First of all, 
%relying on the dust continuum emission data from the previous samples, we will study the dependence of the disc continuum emission in each binary pair on their separation for the targets in the youngest star-forming regions consistently. Furthermore, both in the case of $\rho$~Ophiuchus, for the common sources in \citet{Cox+17_2017ApJ...851...83C,Cieza+19_2019MNRAS.482..698C,Williams+19_2019ApJ...875L...9W} and \citet{Zurlo+20_2020MNRAS.496.5089Z}, as well as Lupus, using \citet{Ansdell+16_2016ApJ...828...46A,Ansdell+18_2018ApJ...859...21A} and \citet{Zurlo+20b_2020MNRAS.tmp.3477Z} data, we can compute disc-integrated spectral indices between $0.9\text{ mm}$ and $1.3\text{ mm}$ and discuss their dependence on the stellar multiplicity as \citet{Akeson&Jensen14_2014ApJ...784...62A} did in Taurus. 
we will gather the available samples from the literature and discuss the trends found in the data. Then, we will compute disc dust fluxes and sizes from our models in Paper~I and compare them with those in Taurus \citep{Manara+19_2019A&A...628A..95M} and $\rho$~Ophiuchus \citep{Cox+17_2017ApJ...851...83C}. The angular resolution of the latter survey ($\sim0.20$~arcsec) is high-enough to resolve the primary component of all binaries and, at least marginally, several secondaries. For this reason, to be consistent with the analysis in \citet{Manara+19_2019A&A...628A..95M}, we fit the dust continuum emission in \citet{Cox+17_2017ApJ...851...83C} discs in multiple stellar systems in the visibility plane employing the state-of-the-art techniques used in Taurus \citep{Tazzari+18_2018MNRAS.476.4527T}. This is needed for a homogeneous treatment of the data in the two samples. We will use the resulting fluxes and sizes in our analysis.

In comparing our models with the observations, there are two issues in particular we would like to focus on: the relationship between disc dust sizes and the tidal truncation radius, as well as the flux-radius correlation. As for the disc sizes, under the assumption that they trace the position of the truncation radius, \citet{Cox+17_2017ApJ...851...83C} and \citet{Manara+19_2019A&A...628A..95M} showed that discs in binaries are much smaller than what is expected from tidal truncation theory \citep[e.g.,][]{Artymowicz&Lubow94_1994ApJ...421..651A}. The only possible way to explain this inconsistency is by invoking very high eccentricities (typically $e>0.5$). Clearly, this is in contrast with the known eccentricity distribution in the field (with a median of $e\sim0.3$, e.g., \citealt{Raghavan+10_2010ApJS..190....1R,Duchene&Kraus13_2013ARA&A..51..269D}). This problem can be circumvented by assuming a reasonable ($\gtrsim2$, e.g., \citealt{Ansdell+18_2018ApJ...859...21A,Sanchis+21_2021arXiv210111307S} in Lupus singles and wide binaries, as well as \citealt{Rodriguez+18_2018ApJ...859..150R} in the binary RW~Aur) conversion factor between gas and dust disc sizes \citep{Manara+19_2019A&A...628A..95M}. Here we can directly compare the disc dust sizes inferred from our zero-eccentricity models with the results of \citet{Cox+17_2017ApJ...851...83C} and \citet{Manara+19_2019A&A...628A..95M}.

In single-star discs \citet{Tripathi+17_2017ApJ...845...44T}, using results from SMA, reported a correlation between (sub-)millimetre disc dust sizes (the radius enclosing 68 per cent of the total dust emission) and dust fluxes, $R_\mathrm{68,obs}\propto L_\mathrm{mm}^{0.5}$, known as the flux-radius correlation. This correlation was later confirmed by \citet{Tazzari+17_2017A&A...606A..88T,Andrews+18a_2018ApJ...865..157A} and \citet{Tazzari20b_2020arXiv201002249T} in Lupus, by \citet{Long+19_2019ApJ...882...49L} in Taurus and by \citet{Barenfeld+17_2017ApJ...851...85B} in Upper Scorpius, using ALMA data (see \citealt{Hendler20_2020ApJ...895..126H} for a summary view). Originally the flux-radius correlation has been justified hypothesising that discs are optically thick with a filling factor of $\sim0.3$ (due to e.g., the presence of sub-structures in the disc, as proposed by \citealt{Tripathi+17_2017ApJ...845...44T} and \citealt{Andrews+18a_2018ApJ...865..157A}). More recently, \citet{Zhu+19_2019ApJ...877L..18Z} suggested that, in the presence of dust self-scattering, a high albedo can also account for the observed relation. Finally, \citet{Rosotti+19b_2019MNRAS.486L..63R} showed that the flux-radius correlation can be explained if radial drift is the main process limiting dust growth. We want to assess whether a similar relation holds in binary discs too, and how it is influenced by the binary separation. To this aim we will both examine any correlations in the models in Paper~I as a function of the tidal truncation radius and assess if similar results occur in the data.

This Paper is organised as follows. In Section~\ref{sec.2} the main observational results of the binary disc surveys in Taurus, $\rho$~Ophiuchus and Lupus are summarised. In particular, we analyse the dependence of the (sub-)millimetre fluxes on the binary separation% and compute the disc-integrated spectral indices in $\rho$~Ophiuchus and Lupus discs
. Section~\ref{sec.3} deals with the determination of disc fluxes and sizes from our models as exemplified in Appendix~\ref{app:1}. We defer to Appendix~\ref{app:2} a detailed discussion of the fits of \citet{Cox+17_2017ApJ...851...83C} targets in the visibility plane. In Section~\ref{sec.5} disc sizes from models and observations are confronted, while in Section~\ref{sec.4} we discuss how the flux-radius correlation in binaries depends on the truncation radius, firstly in the observations and then in our models, further dealing with their relationship.  Finally, in Section~\ref{sec.6} we draw our conclusions.

\section{A logbook of binary disc observations}\label{sec.2}

\begin{figure*}
    \centering 
	\includegraphics[width=0.8\textwidth]{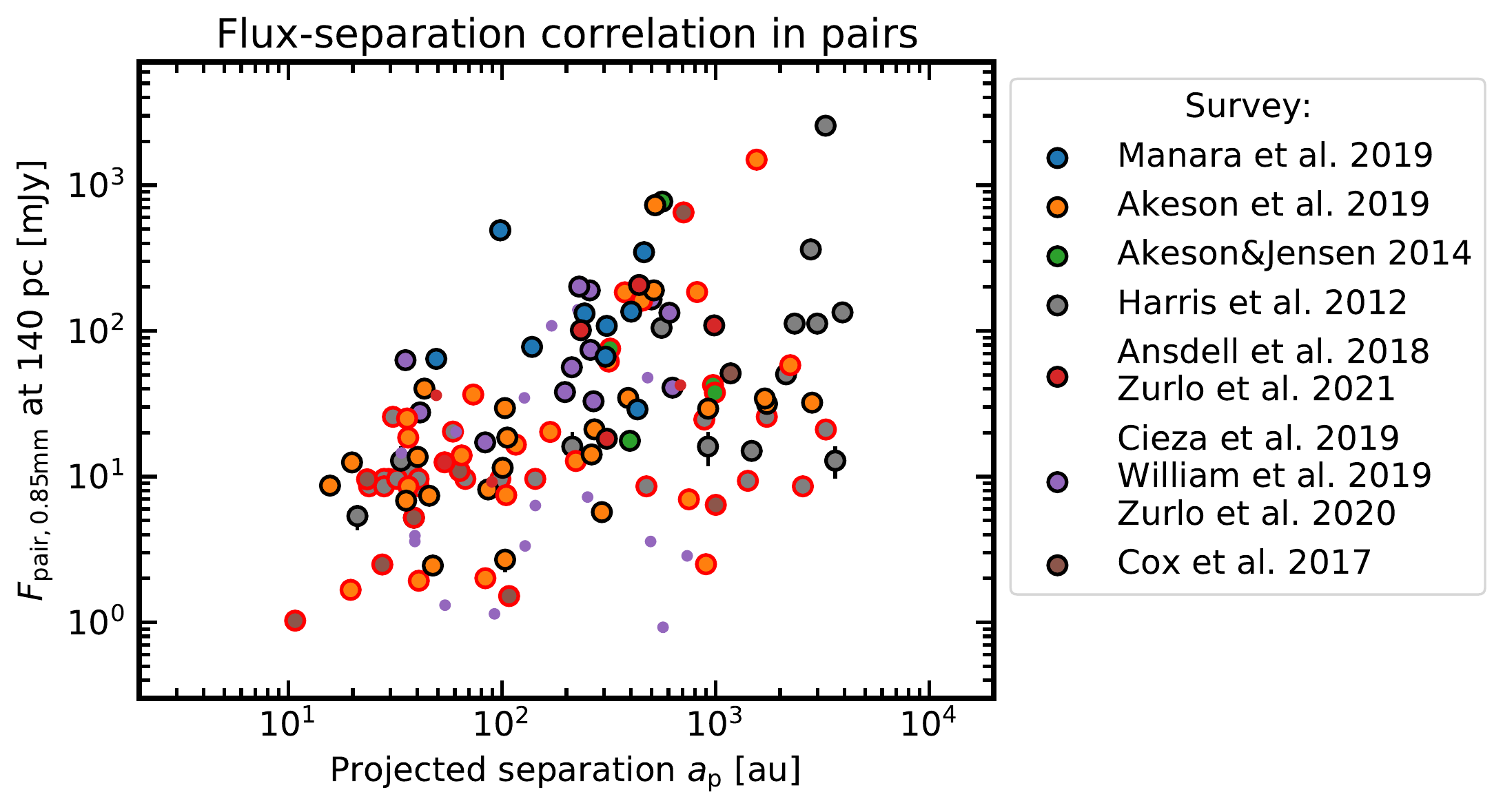}
    \caption{The $0.85\text{ mm}$ flux in binary pairs, $F_\text{pair}$, as a function of their projected separation, $a_\text{p}$, for multiple stellar discs in Taurus, $\rho$~Ophiuchus and Lupus. Detected pairs are identified by the black edges, while upper limits to the non-detections are indicated by the red ones. Smaller dots with no edges refer to those pairs where only one disc was observed and detected. Known circumbinary discs were excluded from the sample. The plot is colour-coded according to the observation survey.}
    \label{fig.2.1a}
\end{figure*}

To put our work in context we look back to the proto-planetary disc surveys in multiple stellar systems in the literature. We focus on the dependence of the disc (sub-)millimetre dust emission in binaries on their projected separation. This
%\subsection{Binary disc fluxes: a summary view}
%The dependence of the disc (sub-)millimetre dust emission in binaries on their projected separation 
was studied for the first time by \citet{Harris+12_2012ApJ...751..115H} in Taurus using SMA data. They found that the flux of each stellar pair\footnote{A pair is defined as \textquotedblleft any subset of the system that could potentially interact dynamically\textquotedblright~\citep{Harris+12_2012ApJ...751..115H}.}, the sum of the fluxes of the two pair components, increases with its separation in discrete jumps (roughly by a factor of five for binaries wider than $a_\text{p}=30\text{ au}$ and $300\text{ au}$). We collect archival data from multiple stellar disc surveys in Taurus \citep{Harris+12_2012ApJ...751..115H,Akeson&Jensen14_2014ApJ...784...62A,Akeson+19_2019ApJ...872..158A,Manara+19_2019A&A...628A..95M}, $\rho$~Ophiuchus \citep{Cox+17_2017ApJ...851...83C,Cieza+19_2019MNRAS.482..698C,Williams+19_2019ApJ...875L...9W,Zurlo+20_2020MNRAS.496.5089Z} and Lupus \citep{Ansdell+18_2018ApJ...859...21A,Zurlo+21_2021MNRAS.501.2305Z} with the aim of discussing if \citet{Harris+12_2012ApJ...751..115H} trends are still valid in a larger sample, with discs from different star-forming regions of the same age. 

In Fig.~\ref{fig.2.1a} the $0.85\text{ mm}$ flux of each binary pair, $F_\text{pair}$, re-scaled to a distance $d=140\text{ pc}$, is plotted as a function of the binary projected separation, $a_\text{p}$. To convert the fluxes to $0.85\text{ mm}$, in this paper we assume that $F_\nu\propto\nu^2$. This scaling relation is valid in the Taurus-Auriga region between $0.85\text{ mm}$ and $1.33\text{ mm}$\citep{Akeson&Jensen14_2014ApJ...784...62A}. However, we consider it to be valid also in $\rho$~Ophiuchus and Lupus. In Appendix~\ref{app:0} a motivation for this choice will be provided. Following \citet{Akeson+19_2019ApJ...872..158A}, we consider a binary pair to be detected only if both the binary disc components were detected. In Fig.~\ref{fig.2.1a} the detections are plotted using black edges, while the upper limits on the non-detections\footnote{For \citet{Akeson&Jensen14_2014ApJ...784...62A} binaries, three times the continuum image rms of $0.40$~mJy~beam$^{-1}$ was used as upper limit if the fluxes are not reported.} are identified by the red ones. If only one component of the pair was observed and detected a smaller dot without edges is shown. In the same context, if only one component of the pair was observed but not detected, it is not plotted in Fig.~\ref{fig.2.1a}. When the same system was observed in different surveys, in our analysis we always considered the most recent one\footnote{For \citet{Akeson+19_2019ApJ...872..158A} sources in common with \citet{Harris+12_2012ApJ...751..115H} without reported projected separation, we consider those in \citet{Harris+12_2012ApJ...751..115H}. Instead, for FU~Tau, that is not in the \citet{Harris+12_2012ApJ...751..115H} sample, we use the \citet{Monin+13_2013A&A...551L...1M} estimate.} (generally with higher angular resolution and sensitivity). Known circumbinary discs were excluded from the sample with the exception of hierarchical higher-order multiple stellar systems. In this case pairs composed by circumstellar and circumbinary discs were also taken into account \citep{Harris+12_2012ApJ...751..115H}. The discs around SSTc2d~J162413.5-241822, SSTc2d~J162435.2-242620 and SSTc2d~J162755.2-242839 in \citet{Williams+19_2019ApJ...875L...9W} and \citet{Zurlo+20_2020MNRAS.496.5089Z} have also been excluded as they show circumbinary 1.33 mm emission (\citealt{Williams+19_2019ApJ...875L...9W}, see Fig.~1 in \citealt{Zurlo+20_2020MNRAS.496.5089Z}). As for \citet{Harris+12_2012ApJ...751..115H}, all close pairs for which the individual component emission was not resolved were excluded if no follow-up survey clearly solves the degeneracy. Moreover, we considered as circumbinary the emission coming from \citet{Harris+12_2012ApJ...751..115H} pairs closer than 0.15 arcsec, a half of the average angular resolution of SMA (\citealt{Ho+04_2004ApJ...616L...1H}, see also the sample selection criteria on binary separation in \citealt{Harris+12_2012ApJ...751..115H}).

As firstly recognised by \citet{Harris+12_2012ApJ...751..115H}, Fig.~\ref{fig.2.1a} suggests that a positive correlation between binary disc fluxes and their projected separation exists. Here we prove that this same relation holds for a larger sample of Taurus discs and if $\rho$~Ophiuchus and Lupus sources are considered as well. Moreover, $F_\text{pair}$ appears to increase continuously with $a_\text{p}$, rather than in discrete jumps as stated in \citet{Harris+12_2012ApJ...751..115H}. However, our cut in resolution does not allow to draw a general conclusion in the case of binaries closer than 30 au.

\begin{figure*}
    \centering
	\includegraphics[width=\textwidth]{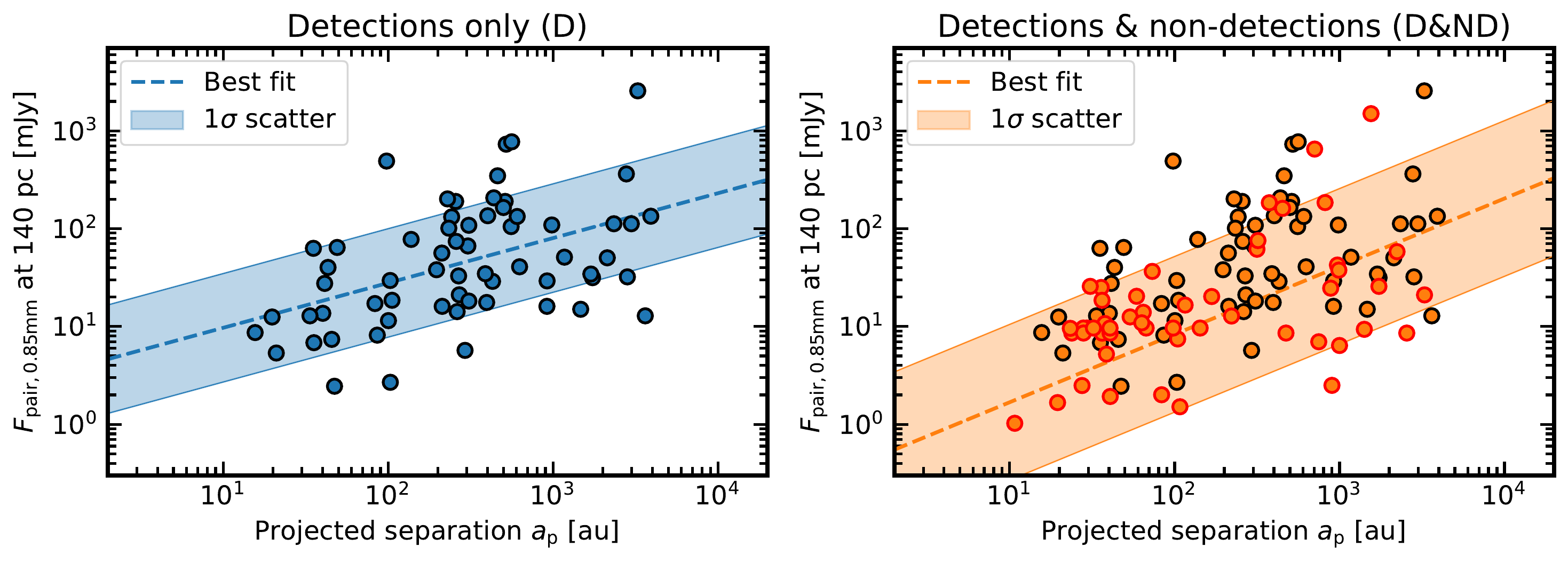}
    \caption{\textbf{Left-hand panel:} Linear regression in the case of detected pairs (both stars were detected: D). The dashed line identifies the best-fit correlation, while the shaded area refers to its Gaussian scatter standard deviation. \textbf{Right-hand panel:} Same as in the left-hand panel for both detected and non-detected pairs (only one star was detected: D$\&$ND). The marker edges have the same meaning as in Fig.~\ref{fig.2.1a}.}
    \label{fig.2.1b}
\end{figure*}

To quantitatively characterise the flux-separation correlation, we make the assumption that they are connected by a power-law relationships, which in log space reads: 
\begin{equation}\label{eq.1.1}
    \log\left[\dfrac{F_\mathrm{pair}}{\mathrm{mJy}}\left(\dfrac{d}{140\text{ pc}}\right)^2\right]=\alpha+\beta\log\left(\dfrac{a_\mathrm{p}}{\mathrm{au}}\right)+\epsilon,
\end{equation}
where $\epsilon$ is the Gaussian scatter standard deviation perpendicular to the linear scaling (this is to say that $\epsilon$ is Gaussian distributed with null mean and standard deviation $\sigma$). We use the \texttt{linmix} package\footnote{Code available at \texttt{github.com/jmeyers314/linmix}.} to perform a three-Gaussian hierarchical Bayesian linear regression of the data \citep{Kelly07_2007ApJ...665.1489K} in the log space. Our results are shown in Fig.~\ref{fig.2.1b}. In the left-hand panel the linear regression is performed only in the case of detected pairs (D), while in the right-hand panel also the non-detections (D$\&$ND) are considered. In the latter case, uncertainties on the non-detections have been assumed as a third of the quoted upper limit. The dashed lines identify the linear regression best-fits, while the shaded areas refer to their Gaussian scatter standard deviation (intrinsic scattering). We employed 10 chains and $2.5\times10^4$ steps. After convergence, the MCMC regression posteriors are single peaked and Gaussian-like. Our results for the coefficients $\alpha$, $\beta$ and $\sigma$, as well as for the correlation coefficient $\rho$ are reported in Tab.~\ref{tab:2.1}. 

We also performed a similar exercise with the pairs in \citet{Harris+12_2012ApJ...751..115H} only (using the same cut in resolution as above and assuming the same separation as in \citealt{Manara+19_2019A&A...628A..95M} for T~Tau~N and T~Tau~S). The best-fit parameters are reported in Tab.~\ref{tab:2.1} as for the complete sample under the label (H). The scattering and correlation coefficients are very similar between the two samples.

\begin{table}
    \centering
    \begin{tabular}{|c|c|c|c|c|}
        \hline
        & $\alpha$ & $\beta$ & $\sigma$ & $\rho$ \\
        \hline
        D    & $0.52\pm0.29$ & $0.46\pm0.11$ & $0.55_{-0.05}^{+0.06}$ & $0.47_{-0.11}^{+0.10}$ \\
        \hline
        D (H) & $0.80\pm0.29$ & $0.38\pm0.11$ & $0.54_{-0.05}^{+0.06}$ & $0.48_{-0.11}^{+0.10}$ \\
        \hline
        D\&ND    & $-0.45_{-0.33}^{+0.31}$ & $0.69_{-0.12}^{+0.13}$ & $0.80_{-0.07}^{+0.08}$ & $0.50_{-0.08}^{+0.07}$ \\
        \hline
        D\&ND (H)    & $0.06_{-0.36}^{+0.35}$ & $0.51\pm0.14$ & $0.78_{-0.08}^{+0.10}$ & $0.42_{-0.11}^{+0.09}$ \\
        \hline
    \end{tabular}
    \caption{Linear regression parameters for the lower panels of Fig.~\ref{fig.2.1a} and for the restricted sample (H) of \citet{Harris+12_2012ApJ...751..115H}. The median and the 16th and 84th percentiles of the intercept ($\alpha$), slope ($\beta$), scatter ($\sigma$), and correlation coefficient ($\rho$) posteriors are reported.}
    \label{tab:2.1}
\end{table}

Our results clearly show that in general we do expect a stellar companion to influence the disc (sub-)millimetre emission. However, following \citet{Harris+12_2012ApJ...751..115H}, who were motivated by the limited angular resolution of their survey, we considered a heterogeneous sample, made up of binary pairs in which circumstellar (primary and secondary discs), as well as circumbinary emission are mixed together. This could be the reason for the shallow correlation coefficients in Tab.~\ref{tab:2.1}. Furthermore, if we consider the results in Fig.s~\ref{fig.2.1a} and~\ref{fig.2.1b} as a tentative flux-radius correlation, it should be remarked that this analysis makes the assumption that there is a relationship between the disc sizes and the truncation radius, $R_\text{trunc}\propto a_\text{p}$. For this reason, we find that it is more useful to study the flux of individual sources as a function of the size of that given disc, rather than the binary separation. We will do this exercise in Section~\ref{sec.4}.

\section{Model description and numerical methods}\label{sec.3}
Let us now move on to the comparison between models and observations. Our models were obtained using the code firstly introduced in \citet{Booth+17_2017MNRAS.469.3994B}. We refer the reader to this paper for a detailed description of its architecture. In Paper~I (see Section~2 therein) it is discussed how this code was modified to take into account the effects of binarity on the secular evolution of the gas (following \citealt{Rosotti&Clarke18_2018MNRAS.473.5630R}) and the dust (as in \citealt{Rosotti+19a_2019MNRAS.486.4829R}) in a circumstellar disc. In this paper we take into account the same models described in Paper~I, spanning different values of the disc viscosity, $\alpha$, the initial disc scale radius, $R_0$, and the tidal truncation radius, $R_\text{trunc}$. For each set of the initial parameters we evolved the gas and the dust on secular time scales. The model results can be used to compute a synthetic surface brightness profile, $S_\text{b}$, at each time as \citep{Rosotti+19a_2019MNRAS.486.4829R}:
\begin{equation}\label{eq.3.1}
    S_\text{b}(R)=B_\nu(T)\bigl\{1-\exp{(-\kappa_\nu\Sigma_\text{d})}\bigr\},
\end{equation}
where $\Sigma_\text{d}$ is the dust surface density, while $B_\nu$ is the black body radiation spectrum at temperature $T$ and $\kappa_\nu$ is the dust opacity; both are computed at ALMA Band 7 wavelengths ($0.85\text{ mm}$). We have assumed face-on discs for the sake of simplicity.

\begin{figure}
    \centering
	\includegraphics[width=\columnwidth]{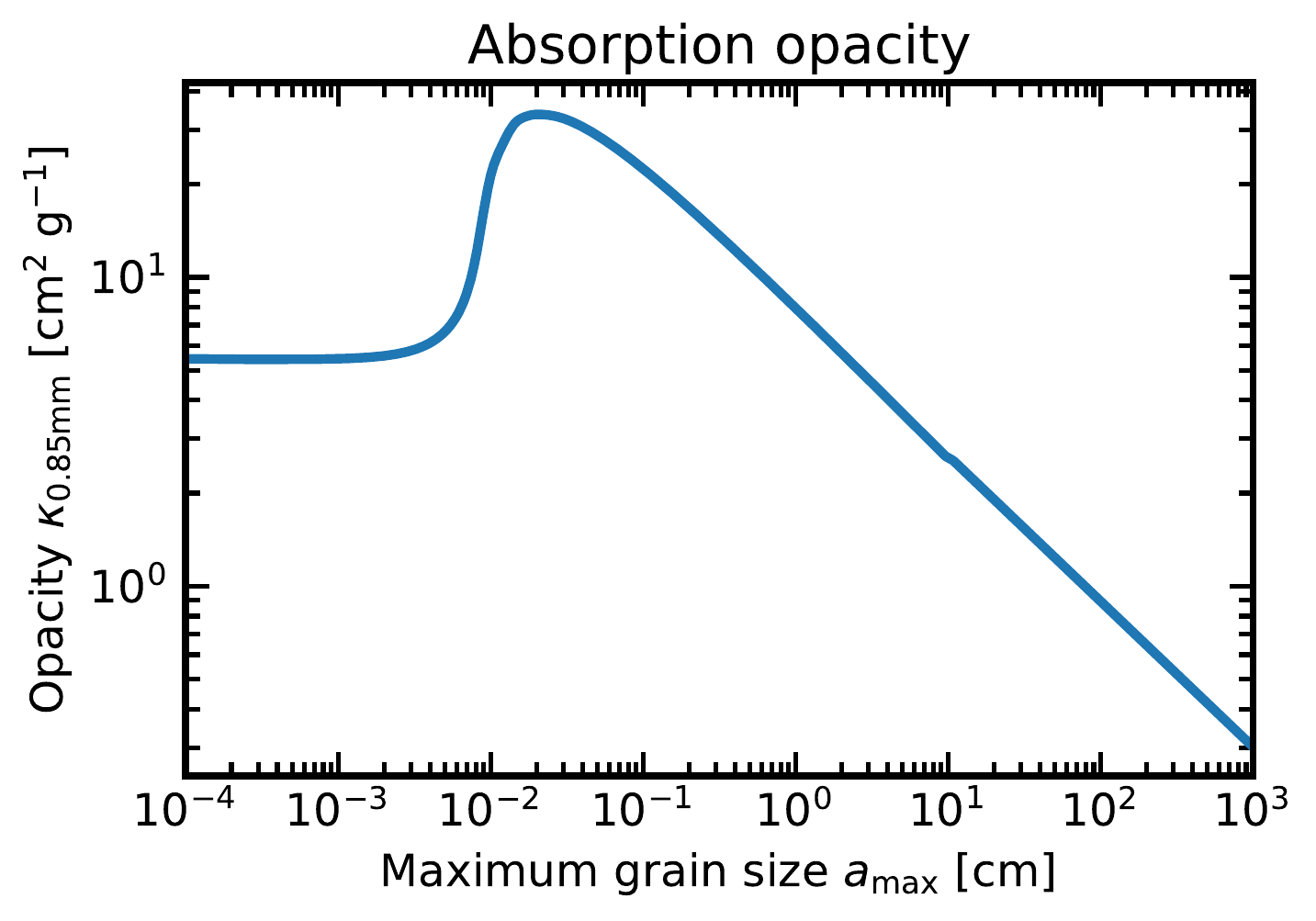}
    \caption{The $0.85$~mm model opacity, $\kappa_\mathrm{0.85mm}$, is plotted as a function of the maximum grain size, $a_\text{max}$. Around $a_\text{max}\sim10^{-2}\text{ cm}$ the opacity drops by a factor of $\sim10$ over a narrow range. Following \citet{Rosotti+19b_2019MNRAS.486L..63R,Rosotti+19a_2019MNRAS.486.4829R}, we will refer to this feature as to the \textit{opacity cliff}.}
    \label{fig.2.3}
\end{figure}

For the dust opacity we follow \citet{Tazzari+16_2016A&A...588A..53T}, employing the models of \citet{Natta&Testi04_2004ASPC..323..279N} and \citet{Natta+07_2007prpl.conf..767N}. We rely on Mie theory for compact spherical grains, assuming a composition of 10 per cent silicates, 30 per cent refractory organics and 60 per cent water ice \citep{Pollack+94_1994ApJ...421..615P}, and prescribe a power-law distribution of the grain size, $n(a)\propto a^{-q}$, with exponent $q=-3.5$ \citep{Mathis+77_1977ApJ...217..425M}. In Fig.~\ref{fig.2.3} the $0.85\text{ mm}$ opacity is plotted as a function of the maximum grain size. As it is clear from the figure, around $a_\text{max}\sim10^{-2}\text{ cm}$ the opacity plummets by an order of magnitude over a narrow range. Following \citet{Rosotti+19b_2019MNRAS.486L..63R,Rosotti+19a_2019MNRAS.486.4829R}, we will refer to this steep decrease of $\kappa_\nu$ as to the \textit{opacity cliff}.

Once the surface brightness profile has been determined, we compute the dust flux of a model disc as:
\begin{equation}\label{eq.3.2}
    F=\dfrac{1}{d^2}\int_{R_\text{in}}^{R_\text{trunc}}S_\text{b}(R)2\pi RdR,
\end{equation}
where $R_\text{in}=10^{-2}\text{ au}$ is the innermost grid cell radius and $d$ is the distance of the disc from the observer; we assume $d=140\text{ pc}$. As for the surface brightness, we only consider the case of face-on discs. In the case of optically thin discs, dust continuum emission is insensitive to the disc inclination. Instead, in the optically thick limit, we expect our results to be correct within a factor of $\langle\cos i\rangle=\pi/4\sim0.8$.

Following \citet{Rosotti+19b_2019MNRAS.486L..63R,Rosotti+19a_2019MNRAS.486.4829R}, we define the 68-per-cent-flux radius, $R_\mathrm{68,mod}$, and the 95-per-cent-flux radius, $R_\mathrm{95,mod}$, as the disc sizes enclosing at a given time 68 per cent and 95 per cent of the model dust flux, respectively. Similarly to the case of single-star discs, in which this choice was motivated by the possibility of comparing our theoretical models with the observational results of \citet{Tripathi+17_2017ApJ...845...44T} and \citet{Andrews+18a_2018ApJ...865..157A}, we retain the same flux fraction in the definition of the dust radius in order to directly test our models against the (circumstellar binary disc) observations of \citet{Manara+19_2019A&A...628A..95M} in Taurus and \citet{Cox+17_2017ApJ...851...83C} in $\rho$~Ophiuchus. In particular, \citet{Manara+19_2019A&A...628A..95M} call $R_\mathrm{eff}$ and $R_\mathrm{disc}$ the observational inferences for the 68-per-cent-flux and 95-per-cent-flux radii, respectively. Instead, we use the $R_\mathrm{68,obs}$ and $R_\mathrm{95,obs}$ symbols for the same observationally inferred sizes. 

In Appendix~\ref{app:1} the computation of the disc fluxes and sizes is also discussed following \citet{Rosotti+19a_2019MNRAS.486.4829R}. We (re-)analyse $\rho$~Ophiuchus observations in the visibility plane making use of the same functional form employed by \citet{Manara+19_2019A&A...628A..95M} in order to have a homogeneous sample of observational disc sizes in both regions. We refer to Appendix~\ref{app:2} for an insight into the analysis of \citet{Cox+17_2017ApJ...851...83C} discs in the visibility plane.

\section{Disc sizes from models and data}\label{sec.5}
Knowledge of the tidal truncation radius is central to study proto-planetary disc evolution in multiple systems. However, inferring $R_\text{trunc}$ from the data is prohibitive as it depends on several dynamical parameters of the systems, such as the binary separation, $a$, the mass ratio, $q$, and the orbital eccentricity, $e$, some of which are often unknown. Indeed, while it is almost always possible to provide reliable estimates of $q$\footnote{As for Taurus binaries, to determine stellar masses \citet{Long+19_2019ApJ...882...49L} and \citet{Manara+19_2019A&A...628A..95M} rely on spectroscopic optical/infrared measurements of the stellar effective temperature and luminosity \citep{Herczeg&Hillenbrand14_2014ApJ...786...97H}, coupled with pre main-sequence stellar evolution models \citep{Baraffe+15_2015A&A...577A..42B,Feiden16_2016A&A...593A..99F}. Orbital dynamics measurements are used for UZ~Tau~E \citep{Simon+00_2000ApJ...545.1034S} and HN~Tau~A \citep{Simon+17_2017ApJ...844..158S}, instead.}, binaries are generally too wide to infer $a$ and $e$ (e.g., \citealt{Harris+12_2012ApJ...751..115H}). As a consequence of our ignorance on $R_\text{trunc}$, it is often difficult to compare model predictions and observations consistently; in addition, some (limiting) assumption have to be made. In the quest for a canonical method to infer the tidal truncation radius, several routes have been attempted. Among those, the possibility of considering the disc-flux sizes as proxies for $R_\text{trunc}$ has been discussed in several papers. Hereafter we analyse the relationship between the tidal truncation radius and the 68- and 95-per-cent-flux radius from our models and the observations in Taurus and $\rho$~Ophiuchus.

\subsection{Inferring the disc truncation radius from the observations.} \citet{Harris+12_2012ApJ...751..115H} and \citet{Cox+17_2017ApJ...851...83C} compared the dust sizes, obtained with a 2D-Gaussian fit of the disc emission, and their tidal truncation radii, inferred following the model proposed by \citet{Pichardo+05_2005MNRAS.359..521P}. In particular, a Monte Carlo method is used to estimate the binary separation given $a_\text{p}$, assuming uniform eccentricities \citep[e.g.,][]{Raghavan+10_2010ApJS..190....1R,Duchene&Kraus13_2013ARA&A..51..269D} and a probability distribution of the orbital parameters. Then $R_\text{trunc}$ is computed estimating the Hill radius of each star. While, despite some exceptions, \citet{Harris+12_2012ApJ...751..115H} found on average dust radii similar or larger than $R_\text{trunc}$, \citet{Cox+17_2017ApJ...851...83C} obtained opposite results. This discrepancy can be attributed only in part to the $q=1$ assumption in \citet{Cox+17_2017ApJ...851...83C} and the different upper limits to the eccentricity distributions employed in the two works ($e_\text{max}=0.7$ in \citealt{Harris+12_2012ApJ...751..115H}, because of their biased sample selection, and $e_\mathrm{max}=1.0$ in \citealt{Cox+17_2017ApJ...851...83C}). Most likely it is due to the very different angular resolution of the two surveys ($\sim0.2$~arcsec in \citealt{Cox+17_2017ApJ...851...83C} and $\gtrsim0.4$~arcsec in \citealt{Harris+12_2012ApJ...751..115H}). For this reason, we will mainly rely on \citet{Cox+17_2017ApJ...851...83C} results. They suggest that the mismatch between the observed dust radii and $R_\text{trunc}$ can be motivated by the effects of radial drift, which determines a more compact dust emission with respect to the gas outermost radius. Alternatively, the agreement between the measured dust radii and $R_\text{trunc}$ could be improved assuming a distribution of $e$ more skewed towards larger values. However, neither \citet{Harris+12_2012ApJ...751..115H} nor \citet{Cox+17_2017ApJ...851...83C} were able to consider the effects of the disc viscous evolution in their works \citep[e.g.,][]{Artymowicz&Lubow94_1994ApJ...421..651A}.

More recently, \citet{Manara+19_2019A&A...628A..95M} used a complementary approach. For different viscosity models, they explored the possible values of the binary eccentricity, assuming that the disc truncation radius equals $R_\mathrm{68,obs}$ or $R_\mathrm{95,obs}$, %that is to say the \textit{measured} 68- and 95-per-cent-flux radius, 
respectively. \citet{Manara+19_2019A&A...628A..95M} found that only implausibly high values of $e$ were compatible with their assumption. This is in qualitative agreement with the results of \citet{Cox+17_2017ApJ...851...83C}. If instead a ratio of $\sim2$ between gas and dust sizes was considered (e.g., \citealt{Ansdell+18_2018ApJ...859...21A,Rodriguez+18_2018ApJ...859..150R}, in single-star and binary discs, respectively), they found that the inferred eccentricities substantially decreased, still falling in the higher tail of the distribution. All in all, assuming reasonable eccentricities, the measured disc dust sizes never trace the position of the truncation radius and a factor $\sim2$ correction is needed to obtain sensible results.

\begin{figure*}
    \centering
    \includegraphics[width=\textwidth]{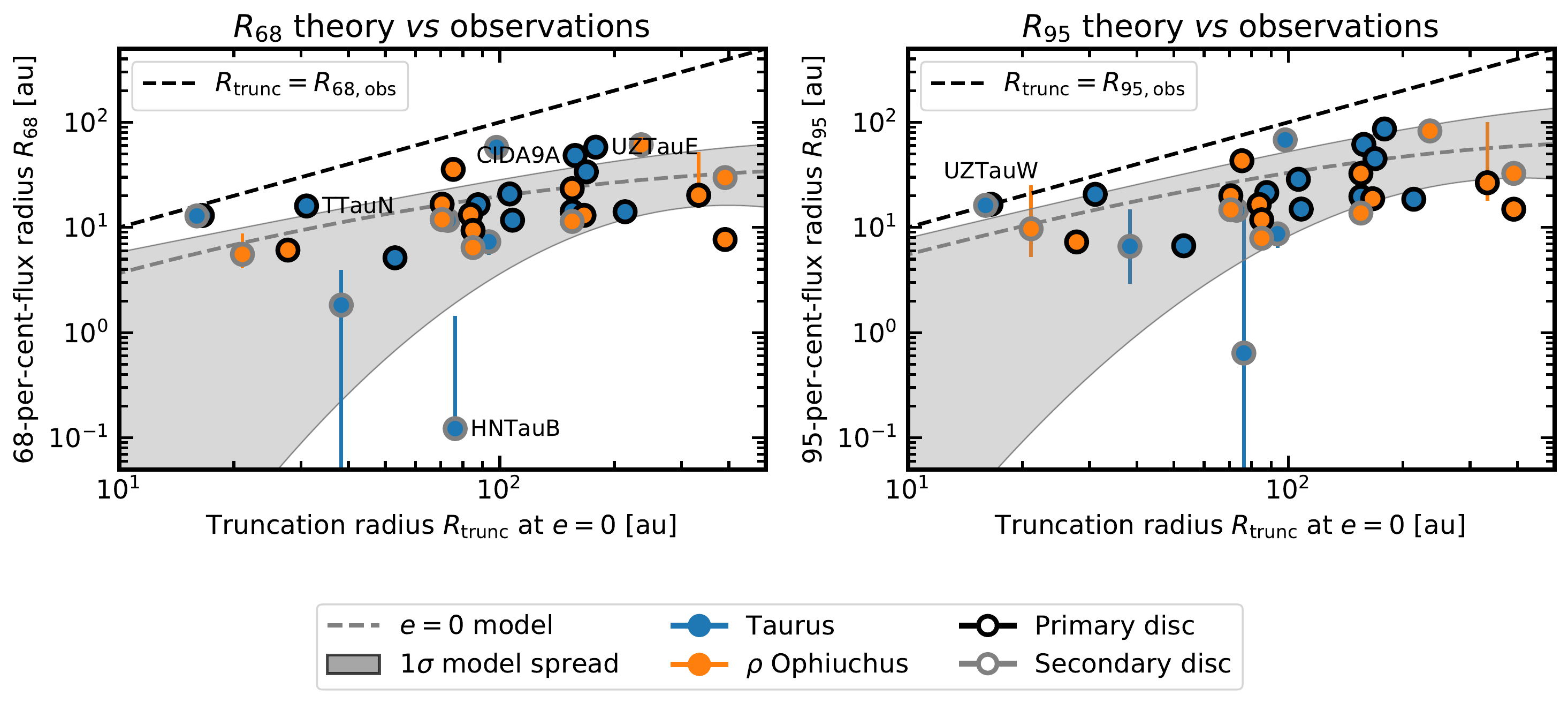}
    \caption{\textbf{Left-hand panel:} The 68-per-cent flux radius, $R_\mathrm{68,mod}$, as a function of the truncation radius, $R_\text{trunc}$, from our models. The dashed grey line and the shaded grey area identify the model best-fit and its $1\sigma$ spread, respectively. The region of the parameter space where the truncation radius equals the dust radius is shown by the black dashed line. The observed discs in Taurus and $\rho$~Ophiuchus are over-plotted as blue and orange dots, respectively. The black edges are used for the primaries, while the grey ones for the secondaries \textbf{Right-hand panel:} Same as in the left-hand panel for the 95-per-cent-flux radius, $R_\mathrm{95,mod}$.}
    \label{fig.2.6}
\end{figure*}

\subsection{Do models and observations agree?} It is clear from the previous considerations that the possible source of the unexpectedly high eccentricities \citep{Manara+19_2019A&A...628A..95M} or low disc radii \citep{Cox+17_2017ApJ...851...83C} is the (unfair) comparison between the location where the disc is tidally truncated in the \textit{gas} with the radius enclosing a given fraction of the \textit{dust} flux. Our aim is showing that, when radial drift is taken into account, the disc dust sizes predicted by our models are compatible with the observationally inferred ones, without the need of invoking very high eccentricities.

To do so, for every value of the disc viscosity, $\alpha$, the initial scale radius, $R_0$, and the tidal truncation radius, $R_\text{trunc}$, in our models in Paper I, we computed the 68- and 95-per-cent-flux radius after $t=1,\,2\text{ and }3\text{ Myr}$. In our calculations a surface brightness sensitivity cut was applied corresponding to the ALMA Band 7 sensitivity rms at $0.85\text{ mm}$ for observations with integration time of $\sim24\text{ s}$ and angular resolution of $\sim0.2\text{ arcsec}$ \citep{Cox+17_2017ApJ...851...83C}. For these values the ALMA sensitivity calculator provides a threshold\footnote{Employing \citet{Manara+19_2019A&A...628A..95M} values (45 antennas, 0.12 arcsec angular resolution and 8 to 10 min integration time) at the same wavelength gives a similar sensitivity threshold of $S_\text{b}=1.87\times10^8\text{ Jy sr}^{-1}=4.39\text{ mJy arcsec}^{-2}$. This small difference does not affect our final results.} of $S_\text{b}=3.33\times10^8\text{ Jy sr}^{-1}=7.85\text{ mJy arcsec}^{-2}$. The values of the disc viscosity and initial disc scale radius are observationally unconstrained. Also the age spread of Taurus and $\rho$~Ophiuchus discs is uncertain. For this reason, we considered the median of $R_\mathrm{68,mod}$ and $R_\mathrm{95,mod}$ over their possible ranges: $10^{-4}\leq\alpha\leq0.025$, $10\text{ au}\leq R_0\leq80\text{ au}$ and $1\text{ Myr}\leq t\leq3\text{ Myr}$ for every value of the tidal truncation radius. Then the distribution of the median 68- and 95-per-cent-flux radius with $R_\text{trunc}$ was fitted using a tapered power-law:
\begin{equation}\label{eq.5.1}
    R_{x,\mathrm{mod}}=aR_\text{trunc}^b\exp\{-cR_\text{trunc}^d+e\},
\end{equation}
where $R_{x,\mathrm{mod}}$ is the $x$-per-cent-flux radius. The $1\sigma$ model spread was obtained fitting the $16$th and $84$th percentiles of the $R_\mathrm{68,mod}$ and $R_\mathrm{95,mod}$ distributions at each truncation radius, again using eq.~\ref{eq.5.1}.

As for the observations, to estimate $R_\text{trunc}$ from the data we follow the approach of \citet{Pichardo+05_2005MNRAS.359..521P}, making use of the following relation (see Appendix C.1. in \citealt{Manara+19_2019A&A...628A..95M}):
\begin{equation}\label{eq.4.1}
    R_\text{trunc}(q,e,a)=R_{i,\text{Egg}}\cdot(be^c+h\mu^k),
\end{equation}
where $q$ is the binary mass ratio, $e$ is the binary orbital eccentricity, $a$ is the binary separation and $\mu=q/(1+q)$. We assume zero eccentricity, $e=0$, and $a\sim a_\text{p}$, with $a_\text{p}$ the observed projected separation of each system. $b,\,c,\,h$ and $k$ are free parameters, while $R_{i,\text{Egg}}$ is the Eggleton radius, which gives a rough estimate of the Roche lobe radius of the primaries and secondaries. It is defined as \citep{Eggleton83_1983ApJ...268..368E}:
\begin{equation}\label{eq.4.2}
    R_{i,\text{Egg}}=a\dfrac{0.49q_i^{2/3}}{0.6q_i^{2/3}+\ln\left(1+q_i^{1/3}\right)},
\end{equation}
where the subscript $i$ refers to the primary $(i=1)$ or secondary $(i=2)$ disc, $q_1=1/q$ and $q_2=q$. In the case of triple systems we consider each hierarchical pair. As for the free parameters $h$ and $k$, they can be determined by fitting the \citet{Papaloizou&Pringle77_1977MNRAS.181..441P} model results. We use $h=0.88$ and $k=0.01$ (see Appendix C.1. in \citealt{Manara+19_2019A&A...628A..95M}). Unfortunately, \citet{Cox+17_2017ApJ...851...83C} do not provide the binary mass ratio for their targets. Only in this case we rely on the classical estimate: $R_\text{trunc}\sim a_\text{p}/3$. This is valid in the case of circular binaries with equal mass stars \citep{Papaloizou&Pringle77_1977MNRAS.181..441P}. 

In Fig.~\ref{fig.2.6} the 68- and 95-per-cent-flux radius are plotted as a function of the truncation radius, $R_\text{trunc}$, in the left and right-hand panels, respectively. The dashed grey line and the shaded grey area identify the model best-fit from eq.~\ref{eq.5.1} and its $1\sigma$ spread, respectively. The dashed black line shows the condition where the dust radius is equal to the truncation radius. The observed discs in Taurus \citep[][with the exception of T~Tau~S showing noisy circumbinary emission]{Manara+19_2019A&A...628A..95M} and $\rho$~Ophiuchus \citep{Cox+17_2017ApJ...851...83C} are over-plotted as blue and orange dots, respectively. The black edges refer to the primaries, while the grey ones to the secondaries. We use $R_\mathrm{68,obs}$ and $R_\mathrm{95,obs}$ as proxies for $R_\mathrm{68,mod}$ and $R_\mathrm{95,mod}$, respectively\footnote{A subtlety that has to be mentioned is that $R_\mathrm{68,obs}$ in \citet{Manara+19_2019A&A...628A..95M} was determined at 1.33~mm, which is expected to be smaller than its 0.85~mm counterpart. In Lupus singles \citet{Tazzari20b_2020arXiv201002249T} provide a possible multi-band relationship that could be used to correct the radii under the assumption that it holds in Taurus, too. Nevertheless, as shown in the same paper, the $R_\mathrm{68,1.3mm}/R_\mathrm{68,0.9mm}$ ratio is almost always around unity.}.

\begin{figure*}
    \centering
    \includegraphics[width=\textwidth]{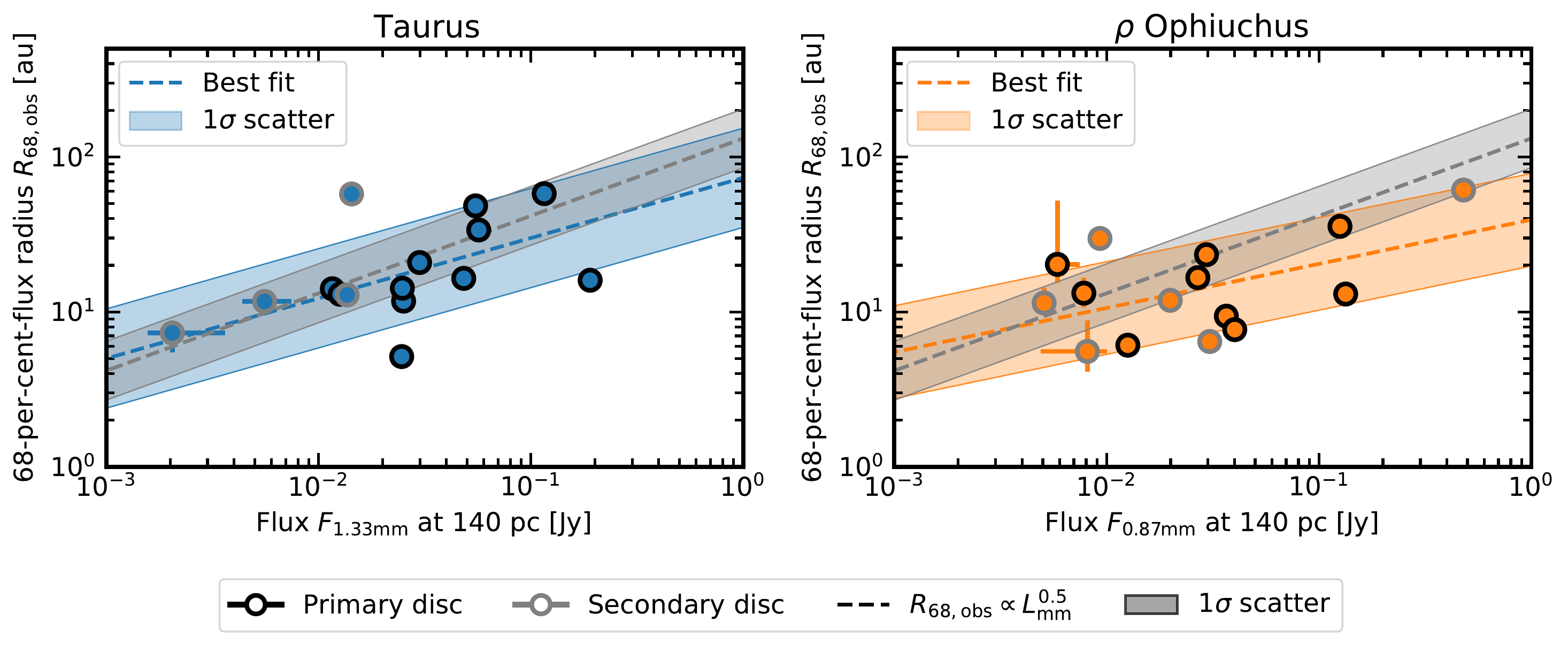}
    \caption{\textbf{Left-hand panel:} Flux-radius correlation for Taurus binary discs in \citet{Manara+19_2019A&A...628A..95M}. The black edges are used for the primaries, while the grey ones for the secondaries. The dashed grey line and the shaded grey area identify the $R_\mathrm{68,obs}\propto L_\mathrm{mm}^{0.5}$ relation in \citet{Tripathi+17_2017ApJ...845...44T} and its Gaussian scatter standard deviation, respectively. The dashed blue line refers to the linear regression best fit, while the shaded blue area to its Gaussian scatter standard deviation. \textbf{Right-hand panel:} Same as in the right-hand panel for $\rho$~Ophiuchus binaries in \citet{Cox+17_2017ApJ...851...83C} in orange.}
    \label{fig.2.4b}
\end{figure*}

Fig.~\ref{fig.2.6} shows that the \textit{measured} dust radii are compatible with our zero-eccentricity models within their $1\sigma$ spread. A notable exception is HN~Tau~B, in the bottom region of both the sub-plots. However, as remarked in \citet{Manara+19_2019A&A...628A..95M}, HN~Tau~B large errors suggest that its dust radii are not well constrained. In addition, some data points, particularly in the left panel, fall out of the $1\sigma$ model spread, above the grey area. Given the results of \citet{Manara+19_2019A&A...628A..95M}, observing larger disc radii than what is expected from tidal truncation theory is unexpected. It could be due to (unresolved) sub-structures in the outer part of the disc halting radial drift. Indeed, among the outliers we know that both UZ~Tau~E and CIDA~9~A show large inner cavities \citep[e.g.,][]{Long+18_2018ApJ...869...17L}. In addition, the large residuals in the T~Tau~N fit can also be interpreted as tentative evidence of the presence of gaps and rings \citep{Manara+19_2019A&A...628A..95M}. However, the larger the disc the easier it is to identify those sub-structures. UZ~Tau~Wa and UZ~Tau~Wb perfectly lie on the $R_\mathrm{95,obs}=R_\text{trunc}$ line, in agreement with the $e\sim0$ inference in \citet{Manara+19_2019A&A...628A..95M}.

For small values of the truncation radius ($R_\mathrm{trunc}\lesssim100$~au), $R_\mathrm{68,mod}$ and $R_\mathrm{95,mod}$ scale as a power law with $R_\text{trunc}$: they can be used as a proxy for the tidal truncation radius (see also the behaviour of $R_\mathrm{95,mod}$ in Fig.~\ref{fig:A1} in Appendix~\ref{app:1}). A simple check using \texttt{scipy.optimize.curve\_fit} and the median dust radii gives $R_\mathrm{68,mod}=0.65\times R_\mathrm{trunc}^{0.76}$ and $R_\mathrm{95,mod}=1.01\times R_\mathrm{trunc}^{0.77}$. We remark that those scaling relations were obtained for binary disc models around a solar mass star and are sensitive to our uncertainty on the initial disc radius and the disc viscosity. Indeed, in this same region the $1\sigma$ spread is very large: its lower limit significantly goes down because the smallest, most viscous discs are almost completely dispersed after $t\sim1-3\text{ Myr}$. As the tidal truncation radius increases, both $R_\mathrm{68,mod}$ and $R_\mathrm{95,mod}$ depart from the $R_\text{trunc}=R_\mathrm{95,obs}$ line. The saturation of the model disc sizes at large truncation radii can partly be due to our choice of the initial conditions, specifically of the initial disc scale radius, $R_0\leq80\text{ au}$, reflecting the absence of (many) discs larger than roughly $100\text{ au}$ in the dust \citep[e.g.,][]{Andrews20_2020arXiv200105007A}.

Rota et al. (subm.) recently analysed the \ce{^{12}CO} emission in a sub-sample of the Taurus binaries in \citet{Manara+19_2019A&A...628A..95M} with the aim of estimating \textit{gas} disc sizes and computing disc eccentricities. Their results confirm our previous finding that \textit{dust} disc sizes do not trace the truncation radius and are compatible with small values of $e$. We refer to Appendix~\ref{app:3} for further considerations on gas observations and the dust-to-gas size ratio from our models and data.

To summarise, dust radial drift naturally explains the low disc dust sizes in \citet{Manara+19_2019A&A...628A..95M} and \citet{Cox+17_2017ApJ...851...83C} without the necessity of invoking high orbital eccentricities. Moreover, dust disc sizes are always smaller than the disc truncation radius.

\section{Flux-radius correlation in models and data}\label{sec.4}
Let us now discuss if a correlation exists between (sub-)millimetre disc sizes and fluxes in binaries, whether this is the same $R_\mathrm{68,obs}\propto L_\mathrm{mm}^{0.5}$ relation followed by single-star discs \citep{Tripathi+17_2017ApJ...845...44T, Andrews+18a_2018ApJ...865..157A}, where $L_\mathrm{mm}$ is the disc luminosity (a flux re-scaled to a distance $d=140$~pc), as well as if and how this is influenced by the binary separation. First of all, we focus on the data, referring to the binary disc surveys in Taurus \citep{Manara+19_2019A&A...628A..95M} and $\rho$~Ophiuchus \citep{Cox+17_2017ApJ...851...83C}. As a subsequent step we take into account both models and observations together, assessing whether they agree and follow the flux-radius correlation in \citet{Tripathi+17_2017ApJ...845...44T}.

\begin{figure*}
    \centering
	\includegraphics[width=\textwidth]{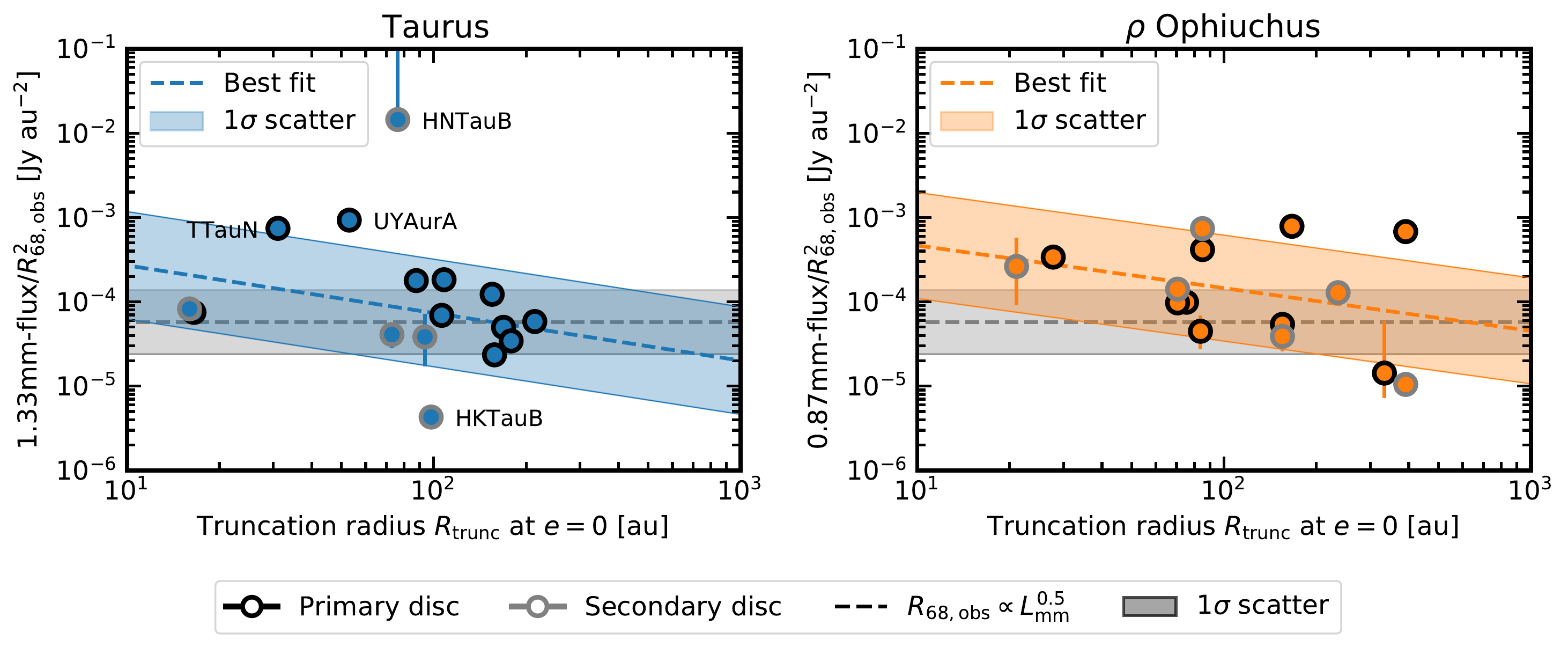}
    \caption{\textbf{Left-hand panel:} Flux-radius correlation for Taurus binary discs in \citet{Manara+19_2019A&A...628A..95M} as a function of the disc truncation radius inferred assuming zero eccentricity. The black edges are used for the primaries, while the grey ones for the secondaries. The dashed grey line and the grey shaded area identify the $R_\mathrm{68,obs}\propto L_\mathrm{mm}^{0.5}$ relation in \citet{Tripathi+17_2017ApJ...845...44T} and its Gaussian scatter standard deviation, respectively. The dashed blue line, instead refers to the linear regression best fit while the blue shaded area to its Gaussian scatter standard deviation. \textbf{Right-hand panel:} Same as in the right-hand panel for $\rho$~Ophiuchus binaries in \citet{Cox+17_2017ApJ...851...83C} in orange.
    }
    \label{fig.2.4a}
\end{figure*}

\subsection{Flux-radius correlation in binary disc observations}
In the left- and right-hand panel of Fig.~\ref{fig.2.4b} the \textit{measured} 68-per-cent-flux radius, $R_\mathrm{68,obs}$, is plotted as a function of the \textit{observed} disc flux, $F_\nu$, re-scaled to a distance $d=140\text{ pc}$, in Taurus and $\rho$~Ophiuchus, respectively. The black edges are used for the primaries, while the grey ones for the secondaries. The dashed grey line and the grey shaded area identify the \citet{Tripathi+17_2017ApJ...845...44T} flux-radius correlation and its Gaussian scatter standard deviation, respectively. 

\begin{table}
    \centering
    \begin{tabular}{|c|c|c|c|c|}
        \hline
        & $\alpha$ & $\beta$ & $\sigma$ & $\rho$ \\
        \hline
        Tau     & $1.86_{-0.29}^{+0.31}$ & $0.39_{-0.17}^{+0.18}$ & $0.32_{-0.06}^{+0.08}$ & $0.70_{-0.25}^{+0.16}$ \\
        \hline
        Oph     & $1.59\pm0.25$ & $0.29\pm0.15$ & $0.30_{-0.06}^{+0.08}$ & $0.54_{-0.28}^{+0.21}$ \\
        \hline
    \end{tabular}
    \caption{Linear regression parameters for Fig.~\ref{fig.2.4b}. The median and the 16th and 84th percentiles of the intercept ($\alpha$), slope ($\beta$), scatter ($\sigma$), and correlation coefficient ($\rho$) posteriors are reported.}
    \label{tab:5.1}
\end{table}

We fit the observed distributions in Taurus and $\rho$~Ophiuchus with a power-law relation, which in log space reads:
\begin{equation}\label{eq.4.3}
    \log\left(\dfrac{R_\mathrm{68,obs}}{\text{au}}\right)=\alpha+\beta\log\left[\dfrac{F_\nu}{\text{Jy}}\left(\dfrac{d}{140\text{ pc}}\right)^2\right]+\epsilon,
\end{equation}
where the symbols have the same meaning as in eq.~\ref{eq.1.1}. We perform a linear regression analysis of the data making use of the \texttt{linmix} package with the same set up as in Section~\ref{sec.2}. When the uncertainties on fluxes and radii are not symmetric, the highest between the two was chosen. T~Tau~S and UY~Aur~B were excluded from the sample; the first because of its noisy circumbinary emission, while the second as its $R_\mathrm{68,obs}$ is $1\sigma$ compatible with being negative. In Fig.~\ref{fig.2.4b} the dashed blue and orange lines identify the linear regression best fit, while the shaded areas of the same colours refer to their Gaussian scatter standard deviation. 

Our results are summarised in Tab.~\ref{tab:5.1} and tentatively suggest that a correlation between (sub-)millimetre binary disc sizes and fluxes exists. Indeed, the correlations coefficients are not very high, especially in $\rho$~Ophiuchus, where the low sensitivity and moderate resolution \citep{Cox+17_2017ApJ...851...83C} could have negatively affected our results. Moreover, the uncertainties on the linear regression parameters are large, probably because of the limited amount of available data. Assuming that a flux-radius correlation holds in binaries, this is not the same $R_\mathrm{68,obs}\propto L_\mathrm{mm}^{0.5}$ relation valid for single-star discs \citep{Tripathi+17_2017ApJ...845...44T}, neither in Taurus nor in $\rho$~Ophiuchus. Nevertheless, Taurus best-fit parameters (slope and intercept) in Tab.~\ref{tab:5.1} are compatible with the \citet{Tripathi+17_2017ApJ...845...44T} ones ($\alpha= 2.12\pm0.05$, $\beta=0.50\pm0.07$) within $1\sigma$ (as remarked in \citealt{Long+19_2019ApJ...882...49L}, even though their slope is larger and single-star discs are also included in their sample). However, this is not true for $\rho$~Ophiuchus intercept. Indeed, the two samples in this paper show slightly different correlation coefficients; this could be due to intrinsic properties of the two regions or more simply to the different observational set-up of the two surveys. However, the $\rho$ values are compatible within their $1\sigma$ uncertainty in Tab.~\ref{tab:5.1}. 

\begin{table}
    \centering
    \begin{tabular}{|c|c|c|c|c|}
        \hline
        & $\alpha$ & $\beta$ & $\sigma$ & $\rho$ \\
        \hline
        Tau     & $-3.00_{-0.93}^{+0.94}$ & $-0.57\pm0.48$ & $0.64_{-0.12}^{+0.17}$ & $-0.33_{-0.25}^{+0.27}$ \\
        \hline
        Oph     & $-2.83_{-1.00}^{+1.02}$ & $-0.50_{-0.50}^{+0.48}$ & $0.63_{-0.12}^{+0.17}$ & $-0.35_{-0.27}^{+0.33}$ \\
        \hline
    \end{tabular}
    \caption{Linear regression parameters for Fig.~\ref{fig.2.4a}. The median and the 16th and 84th percentiles of the intercept ($\alpha$), slope ($\beta$), scatter ($\sigma$), and correlation coefficient ($\rho$) posteriors are reported.}
    \label{tab:5.2}
\end{table}

Having assessed that discs in binaries tentatively follow a (potentially quadratic) flux-radius correlation, we now wish to determine if the properties of the correlation normalisation depend on the truncation radius. In the left- and right-hand panels of Fig.~\ref{fig.2.4a} we analyse the flux-radius correlations in \citet{Manara+19_2019A&A...628A..95M} and \citet{Cox+17_2017ApJ...851...83C} binary discs, respectively, as a function of their truncation radius, $R_\text{trunc}$, inferred as in Section~\ref{sec.5}, using eq.s~\ref{eq.4.1} and~\ref{eq.4.2}, and assuming zero binary orbital eccentricity ($e=0$). The black edges are used for the primaries, while the grey ones for the secondaries. Fluxes have been re-scaled to a distance $d=140\text{ pc}$. The dashed line and the shaded region identify the observational $R_\mathrm{68,obs}\propto L_\mathrm{mm}^{0.5}$ relation reported in \citet{Tripathi+17_2017ApJ...845...44T}, and the associated Gaussian scatter standard deviation, respectively, under the assumption that the correlation holds in multiple stellar discs regardless of $R_\mathrm{trunc}$. 

As it is clear from the figure, the binary discs in \cite{Manara+19_2019A&A...628A..95M} are compatible with the flux-radius correlation normalisation within the spread for $R_\mathrm{trunc}\gtrsim50$~au. However, as the inferred truncation radius decreases several discs depart from the correlation, even though some outliers are present: e.g., HN~Tau~B, the uppermost point with large error bars, shows a poor fit in the visibility plane \citep{Manara+19_2019A&A...628A..95M}, making its dust sizes difficult to estimate. Furthermore, primary discs apparently follow the correlation better than the secondary ones.
These results are in line with the $1\sigma$ compatibility of \citet{Tripathi+17_2017ApJ...845...44T} and Tab.~\ref{tab:5.1} correlation coefficients.  

Also \citet{Cox+17_2017ApJ...851...83C} binaries display no clear trend with $R_\text{trunc}$ and follow the correlation the most in the same range as the Taurus ones. However, at any truncation radius, several discs are far above the the $R_\mathrm{68,obs}\propto L_\mathrm{mm}^{0.5}$ relation normalisation, due to high dust fluxes being associated with small dust sizes. This tendency for some discs to fall above the correlation was already shown in Fig.~9 in \citet{Long+19_2019ApJ...882...49L} for Taurus discs (even though $R_\mathrm{95,obs}$, the \textit{measured} 95-per-cent-flux radius, is plotted instead of $R_\mathrm{68,obs}$ in their paper). No evidence for either primary or secondary components following the correlation more tightly can be seen in $\rho$~Ophiuchus.

To assess whether there is any tendency for binaries to depart from the correlation as $R_\text{trunc}$ varies we performed a Spearman test. The Spearman test estimates if the relation between two data-sets is monotonic: two monotonically increasing (decreasing) data-sets have Spearman rank coefficient $r_\mathrm{s}=+1(-1)$. The rank correlation coefficient is $r_\mathrm{s,Tau}=-0.25$ in Taurus and $r_\mathrm{s,Oph}=-0.20$ in $\rho$~Ophiuchus, suggesting a slightly monotonically decreasing correlation normalisation with $R_\text{trunc}$. However, the $p-$values for a null-hypothesis that the two sets are uncorrelated are high, 0.34 in Taurus and 0.48 in $\rho$~Ophiuchus, meaning that the flux-radius correlation normalisation has a high probability of being independent of $R_\text{trunc}$

Assuming that the relation inferred from the Spearman test is real, to determine its coefficients we performed a linear regression analysis similar to those in the previous paragraphs (again excluding UY~Aur~B from the sample), using the package \texttt{linmix}. The best fit parameters for the correlation:
\begin{equation}\label{eq.4.4}
    \log\left[\dfrac{F_\nu}{R_\mathrm{68,obs}^2}\dfrac{\text{au}^2}{\text{Jy}}\right]=\alpha+\beta\log\left(\dfrac{R_\text{trunc}}{\text{au}}\right)+\epsilon,
\end{equation}
are summarised in Tab.~\ref{tab:5.2}. The dashed blue and orange lines, as well as the shaded areas of the same colours in the left- and right-hand panels of Fig.~\ref{fig.2.4a} identify the linear regression best fit and its Gaussian scatter standard deviation, respectively. In general, the linear regression analysis shows that the observations are only marginally compatible with a flat distribution: apparently the flux-radius correlation in binary discs depends on the disc truncation radius. However the two quantities are very loosely (anti-)correlated. This is consistent with the results of the Spearman test previously described. 

We attempt a similar exercise employing the best-fit parameters in Tab.~\ref{tab:5.1} instead of the quadratic relation in \citet{Tripathi+17_2017ApJ...845...44T}. The match between the data and the correlation slightly improves due to the large intrinsic scatter of those relations. The Spearman test suggests a tighter negative trend of the correlation normalisation with rank correlation coefficients $r_\mathrm{s,Tau}=-0.28$ and $r_\mathrm{s,Oph}=-0.37$ for Taurus and $\rho$~Ophiuchus, respectively. The dependence on $R_\text{trunc}$ is confirmed by the lower $p-$values for the null hypothesis of no-correlation, 0.29 in Taurus and 0.17 in $\rho$~Ophiuchus as well as the (slightly) larger (anti-)correlation coefficients. 
%The results of a linear regression analysis in the log space agree with the behaviour indicated by the Spearman test. The best-fit parameter for eq.~\ref{eq.4.4} are summarised in Tab.~\ref{tab:5.2} and suggest a steeper trend than in Fig.~\ref{fig.2.4a}.}

%\textcolor{red}{Of similar advise is the Pearson test, which determines if two data-sets are linearly correlated. In the case of positively (negatively) correlated data-sets, the Pearson rank coefficient is $r=+1(-1)$. Assuming a $R_\mathrm{68,obs}\propto L_\mathrm{mm}^{0.5}$ relation, for Taurus and $\rho$~Ophiuchus discs $r_\mathrm{Tau}=0.06$, with $p$-value 0.82, and $r_\mathrm{Oph}=-0.37$, with $p$-value 0.17. Apparently, while \citet{Manara+19_2019A&A...628A..95M} data are uncorrelated with $\gtrsim80\%$ probability, \citet{Cox+17_2017ApJ...851...83C} ones are more likely to be correlated, probably because of a lower scatter in the flux-radius correlation normalisation. When the linear regression best-fit parameters in Tab.~\ref{tab:5.2} are used, the Pearson test gives $r_\mathrm{Tau}=0.03$, with $p$-value 0.90, and $r_\mathrm{Oph}=-0.49$, with $p$-value 0.06. suggesting a looser correlation in Taurus but a much stronger one in $\rho$~Ophiuchus.} 
The Pearson test, which determines if two data-sets are linearly correlated, gives similar results. In Taurus discs the $R_\mathrm{68,obs}\propto L_\mathrm{mm}^{0.5}$ relation normalisation has a high probability (80 to 90 per cent) of being uncorrelated with the truncation radius. Instead, in $\rho$ Ophiuchus an anti-correlation between the two is suggested. Similar results are obtained when the flux-radius correlation parameters in Tab.~\ref{tab:5.2} are considered.

To summarise, disc sizes and fluxes in binaries are tentatively correlated but a $R_\mathrm{68,obs}\propto L_\mathrm{mm}^{0.5}$ relation \citep{Tripathi+17_2017ApJ...845...44T} is broadly compatible only with the Taurus data. In general, there is evidence for a slight dependence of the correlation normalisation on $R_\text{trunc}$. However, it should be remarked that in Tab.s~\ref{tab:5.1} and~\ref{tab:5.2} the uncertainties on the parameters are large and the correlation coefficients small. Indeed, it is possible that our results are affected by the uncertainty in the determination of $R_\text{trunc}$ and the restricted sizes of the sample. Definitely larger data-sets are needed to draw more robust conclusions. For this reason, in the rest of the paper we will only compare our models with the \citet{Tripathi+17_2017ApJ...845...44T} results.

\subsection{Flux-radius correlation in binary disc models}
\begin{figure*}
	\includegraphics[width=\textwidth]{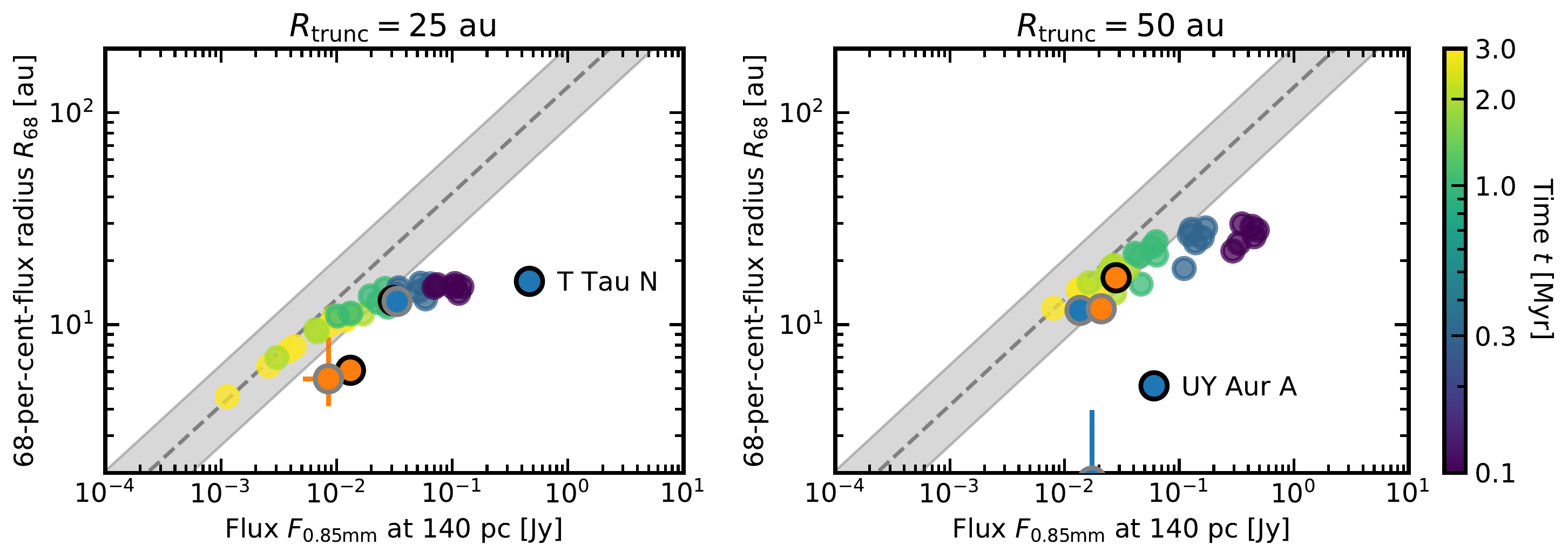}
	\includegraphics[width=\textwidth]{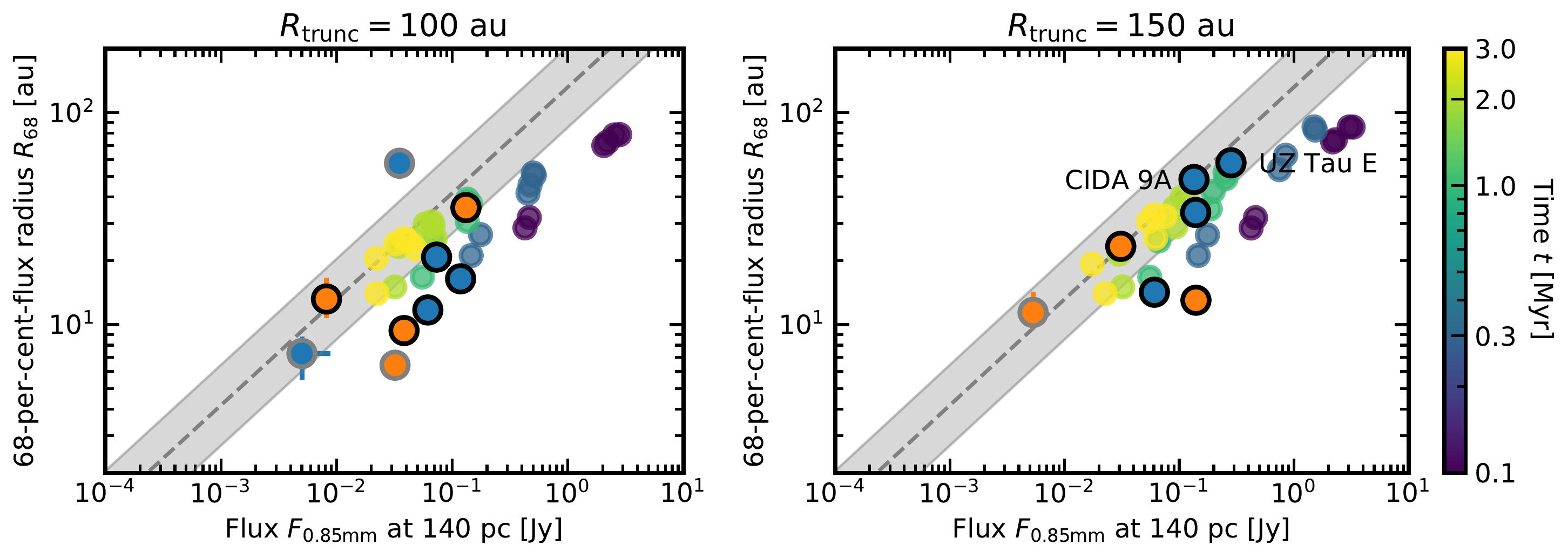}
	\includegraphics[width=\textwidth]{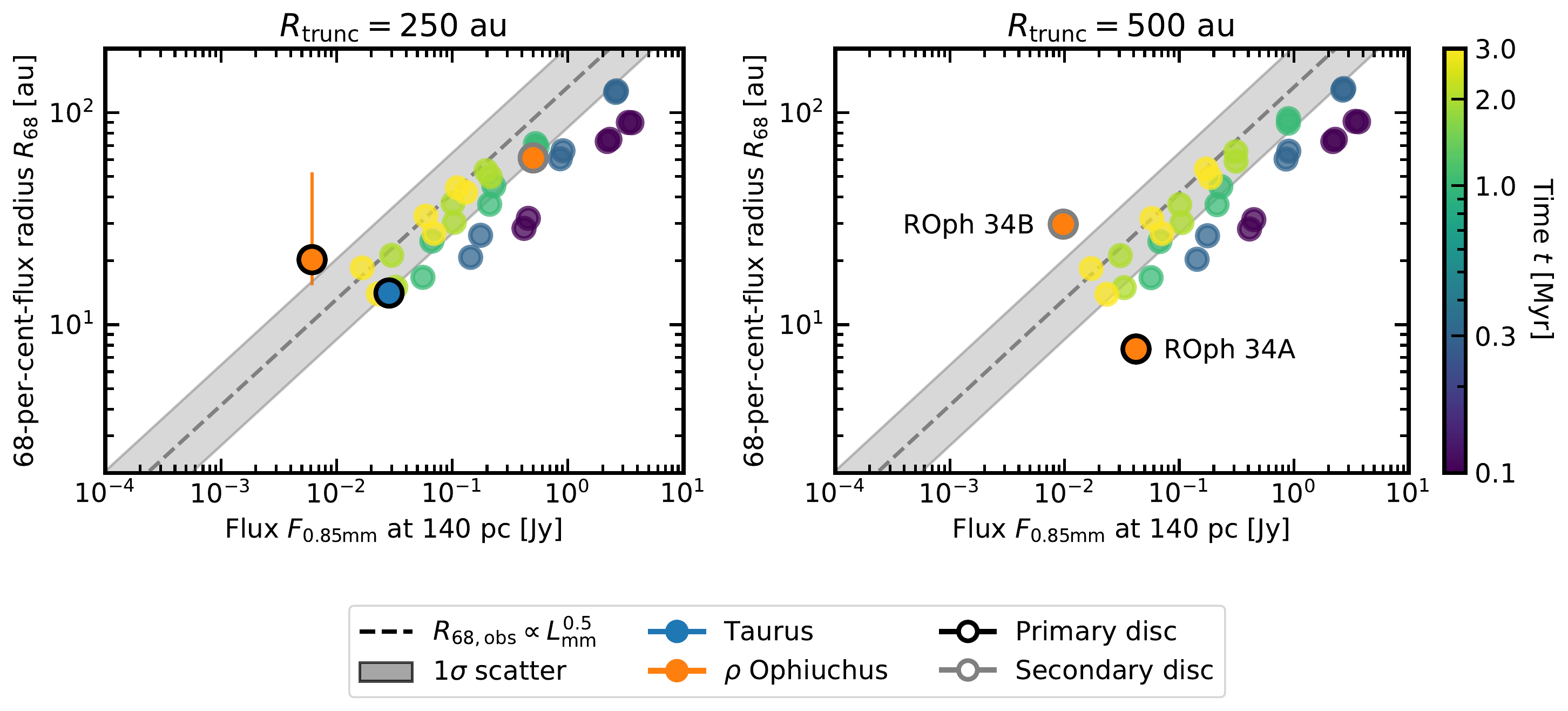}
    \caption{68-per-cent-flux radius, $R_\mathrm{68,mod}$, as a function of the $0.85\text{ mm}$ model flux, $F_\mathrm{0.85mm}$, re-scaled to a distance $d=140\text{ pc}$, for different values of the tidal truncation radius, $R_\text{trunc}$, and $t=0.1,\,0.3,\,1,\,2\text{ and }3\text{ Myr}$. Each series of dots corresponds to a disc with fixed viscosity, $\alpha=10^{-3}\text{ and }10^{-4}$, and initial scale radius, $R_0=10,\,30\text{ and }80\text{ au}$. The dashed grey line and the shaded grey area identify the $R_\mathrm{68,obs}\propto L_\mathrm{mm}^{0.5}$ relation and its Gaussian scatter standard deviation, respectively \citep{Tripathi+17_2017ApJ...845...44T}. The data points are over-plotted in each sub-plot whose model truncation radius is the closest to the observationally inferred one. Taurus and $\rho$~Ophiuchus observations are identified by large blue and orange dots, respectively. The black edges are used for the primaries, while the grey ones for the secondaries.}
    \label{fig.2.5}
\end{figure*}

Hereafter we discuss whether our models follow a flux-radius correlation and if this is the same as in \citet{Tripathi+17_2017ApJ...845...44T}. 
In Fig.~\ref{fig.2.5} we plot the 68-per-cent-flux radius, $R_\mathrm{68,mod}$,  against the $0.85\text{ mm}$ flux, $F_\mathrm{0.85 mm}$, for different values of the tidal truncation radius, $R_\mathrm{trunc}$. For each of those values, the dots highlight six binary disc models at different evolutionary stages: $t=0.1,\,0.3,\,1,\,2\text{ and }3\text{ Myr}$. Such discs share the same initial scale radius, $R_0=10,\,30\text{ and }80\text{ au}$, and viscosity, $\alpha=10^{-3}\text{ and }10^{-4}$, corresponding to radial drift being the main mechanism limiting grain growth. Models with higher disc viscosities have been excluded as they prove to be fragmentation-dominated and do not show a quadratic relation between fluxes and radii \citep{Rosotti+19b_2019MNRAS.486L..63R}\footnote{This choice is not based on our considerations in Paper~I on planet formation being viable in multiple stellar systems. Our aim is simply to test if \citet{Rosotti+19b_2019MNRAS.486L..63R} explanation for the flux-radius correlation holds in binaries.}. The dashed grey line and the shaded grey area identify the $R_\mathrm{68,obs}\propto L_\mathrm{mm}^{0.5}$ relation and its Gaussian scatter standard deviation, respectively \citep{Tripathi+17_2017ApJ...845...44T}.

As it is clear from Fig.~\ref{fig.2.5}, the larger the truncation radius, the brighter and larger the binary discs are. Such an evidence is in qualitative agreement with the observational results of \citet{Harris+12_2012ApJ...751..115H,Cox+17_2017ApJ...851...83C,Akeson+19_2019ApJ...872..158A,Zurlo+20_2020MNRAS.496.5089Z} and \citet{Zurlo+21_2021MNRAS.501.2305Z}, who found larger fluxes in wider binaries (recall Fig.~\ref{fig.2.1a}), as well as those of \citet{Manara+19_2019A&A...628A..95M} and \citet{Zurlo+20_2020MNRAS.496.5089Z,Zurlo+21_2021MNRAS.501.2305Z} who find smaller discs in multiple systems than around isolated stars. When $R_\text{trunc}\gtrsim100\text{ au}$ the model fluxes and radii are tightly correlated, behaving as in single-star discs\footnote{The discrepancies with Fig.~2 in \citet{Rosotti+19b_2019MNRAS.486L..63R} are due to the different temperature profile employed (cfr. Paper~I and \citealt{Rosotti+19b_2019MNRAS.486L..63R,Rosotti+19a_2019MNRAS.486.4829R}).} and following the quadratic relation in \citet{Tripathi+17_2017ApJ...845...44T} at almost every time. This last remark is consistent with our speculations in Paper~I (see e.g., Fig.~3 therein and the relative discussion). The smallest value of $R_\text{trunc}$ for which binary discs behave as singles depends on viscosity and generally increases with $\alpha$. Then, the flux-radius correlation can be considered as a further indication of low disc viscosities. On the contrary, as $R_\text{trunc}$ decreases the behaviour of our binary models and the singles in \citet{Rosotti+19b_2019MNRAS.486L..63R} start to differ. In particular, the binary discs with $R_\text{trunc}\lesssim50\text{ au}$ significantly depart from the \citet{Tripathi+17_2017ApJ...845...44T} correlation if $t\lesssim1\text{ Myr}$ and the models do not lie on the same power-law line. 

A comment is due on our models falling in the higher-flux region of the \citet{Tripathi+17_2017ApJ...845...44T} correlation (on the right of the dashed grey line in Fig.~\ref{fig.2.5}). As the model flux is mainly set by the (fixed) initial disc mass and temperature profile, we expect that fine-tuning those parameters could reproduce the observed normalisation better (as is discussed in the next sub-section in the case of individual sources). Nevertheless, disc population synthesis studies would be needed in order to properly compare models and observations.

It may be surprising that our models for $t\gtrsim1\text{ Myr}$ still have a significant flux (see Appendix~\ref{app:1} where we show that an even smaller disc with $R_\text{trunc}=10\text{ au}$ could be observed with ALMA using a set-up similar to the one employed in \citealt{Manara+19_2019A&A...628A..95M}) considering that they are substantially dust-depleted (with dust-to-gas ratio as small as $2\times10^{-6}$ after $1\text{ Myr}$; see e.g., Fig.~3 in Paper~I and the relative discussion). This can be motivated by the presence of grains with high absorption opacity in the inner disc that have not been accreted yet. Indeed, while in the models the \textit{global} dust-to-gas ratio is very low, the discs are not homogeneously fainter: they are smaller but almost as luminous as singles in the innermost disc regions.

We checked that the $R_\mathrm{68,obs}\propto L_\mathrm{mm}^{0.5}$ relation in our models is not due to optical depth effects by computing their optical depth fraction, $\mathcal{F}_\mathrm{0.85mm}$, according to the definition in \citet{Tazzari20b_2020arXiv201002249T}. The largest optical depth fractions are attained by the youngest, largest and most viscous discs, with $\mathcal{F}_\mathrm{0.85mm}\sim0.66$. By the time when the models match the correlation the most (after $t\sim1\text{ Myr}$), their optical depth fraction has substantially decreased, with $\mathcal{F}_\mathrm{0.85mm}\lesssim0.15$.

We also studied the flux-radius correlation in models with smaller truncation radii ($R_\text{trunc}=5\text{ au and }10\text{ au}$, not shown in Fig.~\ref{fig.2.5}). However, it should be considered that for such small binary separations, $a_\text{p}\lesssim15\text{ au}$, it is likely that circumbinary rather than circumstellar discs are formed. In this case a different modelling exercise is needed. Although their behaviour resembles the case of $R_\text{trunc}=25\text{ au}$, not only the youngest but also the most aged discs depart from the \citet{Tripathi+17_2017ApJ...845...44T} correlation, in particular in the smallest, most viscous cases. In those models the \textit{opacity cliff} lies inside $R\sim1\text{ au}$ when $t\gtrsim2\text{ Myr}$. This suggests that those discs have been almost dispersed and the largest contribution to their flux comes from the smallest grains beyond the cliff. As a consequence, disc dust fluxes are considerably reduced, making those models hard to be detected by ALMA. %Moreover, in the faintest discs the 68-per-cent-\textit{flux} radius does not trace the position of the \textit{opacity cliff}, yet it is closer to the 68-per-cent-\textit{mass} radius \citep{Rosotti+19a_2019MNRAS.486.4829R}. This happens because the opacity profile is almost flat beyond the \textit{opacity cliff} and the models are optically thin: then $S_\text{b}(R)\sim\Sigma_\text{d}T$. 
In these extreme cases a cut in sensitivity (see Appendix~\ref{app:1}) can reconcile our models with the flux-radius correlation for the older discs. Recently, \citet{Sanchis20_2020A&A...633A.114S} and \citet{Kurtovic+21_2021A&A...645A.139K} showed that brown dwarfs are compatible with the \citet{Andrews+18a_2018ApJ...865..157A} and \citet{Tripathi+17_2017ApJ...845...44T} flux-radius correlation. A proper comparison between our models and their results is potentially unfair as we considered only discs orbiting a Solar mass star (see Paper I). However, it is reassuring that also the models with the smallest truncation radii are compatible with the $R_\mathrm{68,obs}\propto L_\mathrm{mm}^{0.5}$ relation in \citet{Tripathi+17_2017ApJ...845...44T} in a region with both comparable and smaller fluxes and sizes than those explored in \citet{Sanchis20_2020A&A...633A.114S} and \citet{Kurtovic+21_2021A&A...645A.139K}.

%We performed a linear regression of the model distribution using the package \texttt{linmix}. With reference to eq.~\ref{eq.4.3}, using $R_\mathrm{68,mod}$ instead of $R_\mathrm{68,obs}$, our best fit parameters are $1.74\leq\alpha\leq1.83$, $0.29\leq\beta\leq0.37$ for $R_\text{trunc}\geq100\text{ au}$, while $\alpha\leq1.46$ and $\beta\leq0.23$ for $R_\text{trunc}<100\text{ au}$. 
Discs with small $R_\mathrm{trunc}$ falling above the \citet{Tripathi+17_2017ApJ...845...44T} correlation normalisation in Fig.~\ref{fig.2.4a} can be explained by a less steep flux-radius correlation ($\beta<0.5$), meaning brighter discs for a given disc radius. In fact, also the models in Fig.~\ref{fig.2.5} show a tendency for brighter discs in closer binaries, in particular if younger. However, as the observed discs in Fig.~\ref{fig.2.4b} are expected to be much older than those young models, the previous comparison could be unfair.

To summarise, our models follow the flux-radius correlation in \citet{Tripathi+17_2017ApJ...845...44T} in wide binaries ($R_\text{trunc}\gtrsim100\text{ au}$) at almost all times. However, for a smaller $R_\text{trunc}$ this is true only after $t\sim1\text{ Myr}$. In general, the tentative trend of a higher correlation normalisation in closer binaries seen in the data is reproduced by the models, even though more observations are needed to better understand and constrain the dependence of the flux-radius correlation on $R_\text{trunc}$.

\subsection{Do models and observations agree?}
Having looked at the general trends and correlations, we now investigate whether we can reproduce individual sources. To test our theoretical predictions, in each panel of Fig.~\ref{fig.2.5} we over-plot the observational data of \citet{Manara+19_2019A&A...628A..95M}, and \citet{Cox+17_2017ApJ...851...83C} as large blue and orange dots, respectively. The black edges are used for the primaries, while the grey ones for the secondaries. We include each observed binary disc in the sub-plot whose reference tidal truncation radius is the closest to the one estimated from the observations using eq.s~\ref{eq.4.1} and~\ref{eq.4.2}, and assuming zero eccentricity. Fluxes were re-scaled to ALMA Band 7 frequencies using the same $F_\nu\propto\nu^2$ relationship introduced in Sec.~\ref{sec.2}, and a distance $d=140\text{ pc}$; $R_\mathrm{68,obs}$ was used as a proxy for the 68-per-cent-flux radius\footnote{The same considerations in Section~\ref{sec.5} for the dust sizes of \citet{Manara+19_2019A&A...628A..95M} binary discs apply.}.

From a quick look at Fig.~\ref{fig.2.5} one can see that the Taurus and $\rho$~Ophiuchus discs roughly match our models. However, some discs do not lie close to the model distribution, consistently with our findings in Fig.~\ref{fig.2.4b}. \citet{Cox+17_2017ApJ...851...83C} data display a worse agreement with our discs: the $\rho$~Ophiuchus binaries well above the $R_\mathrm{68,obs}\propto L_\mathrm{mm}^{0.5}$ relation in Fig.~\ref{fig.2.4a}, appear to be too bright or compact with respect to our models (see for example ROph~34~A~or~B in the $R_\text{trunc}=500\text{ au}$ sub-plot). This is consistent with the results of the linear regression analysis in Fig.~\ref{fig.2.4b} where $\rho$~Ophiuchus discs show a worse agreement with the \citet{Tripathi+17_2017ApJ...845...44T} relation than the Taurus ones.  % For example, neither ROph~34~A~nor~B in the $R_\text{trunc}=500\text{ au}$ sub-plot agree both with the theoretical distribution and the quadratic relation in \citet{Tripathi+17_2017ApJ...845...44T}. %If taken at face value this evidence could lead to the puzzling results that discs in wider binaries are less similar to those around singles, in contrast with previous results (e.g., \citealt{Harris+12_2012ApJ...751..115H} on fluxes). Nonetheless, in this case the observations are too few to draw robust conclusions.

The differences between models and data can be partly motivated by our choice of a fixed initial disc mass and temperature profile (see Paper~I for the details). For example, in the case of $R_\text{trunc}=100\text{ au}$, models with a lower initial disc mass ($M_0=0.01M_\odot$) or temperature ($T_0=44.10\text{ K}$) agree better with the lower flux data (not shown in Fig.~\ref{fig.2.5}). However, in order for models to reproduce the higher flux data, implausibly high temperatures ($T_0=176.39\text{ K}$) are required, suggesting that those discs are optically thick (but see also e.g., \citealt{Nelson00_2000ApJ...537L..65N} and \citealt{Picogna&Marzari13_2013A&A...556A.148P} for temperature increase in the inner binary discs due to tidal interactions). The differences between models and observations can also be influenced by the uncertainties in $R_\text{trunc}$ and our assumption of zero eccentricity. 

Another possibility to partly explain some of the data with highest fluxes and radii would be the presence of disc substructures. However, there are at least two binaries in the sample of \citet{Manara+19_2019A&A...628A..95M} that show evidence of substructures but agree with our models and the $R_\mathrm{68,obs}\propto L_\mathrm{mm}^{0.5}$ relation in \citet{Tripathi+17_2017ApJ...845...44T}: UZ~Tau~E and CIDA~9~A \citep[][for a discussion]{Long+18_2018ApJ...869...17L}. This last evidence is consistent with the results in \citet{Rosotti+19b_2019MNRAS.486L..63R} of substructures not influencing the flux-radius correlation. However, some specific work is needed to address this issue in detail.

\textit{Outliers:} Some data points in Fig.~\ref{fig.2.5} fall very far from the models. Those are the same binaries showing poor agreement with the flux-radius correlation in Fig.s~\ref{fig.2.4b} and~\ref{fig.2.4a}. In particular, this is the case of HN~Tau~B, whose flux and 68-per-cent-flux radius are so small that the disc lies out on the left of the $R_\text{trunc}=150\text{ au}$ panel, as well as UY~Aur~A and B at the bottom of the $R_\text{trunc}=50\text{ au}$ sub-plot, with UY~Aur~B partly out of the figure. Also T~Tau~N, at the right of the $R_\text{trunc}=25\text{ au}$ panel, shows an anomalously high flux. However, this could be due to eccentricity effects: \citet{Harris+12_2012ApJ...751..115H} and \citet{Kohler+08_2008A&A...482..929K} suggest that $a\sim1500\text{ au}\gg a_\text{p}\sim100\text{ au}$ in \citet{Manara+19_2019A&A...628A..95M}. In this case, T~Tau~N should be compared with the $R_\text{trunc}=500\text{ au}$ models, with whom it shows a tighter agreement. It is also possible to explain the strange behaviour of T~Tau~N as due to the presence of substructures. This would be consistent with the large residuals in the fits (see Appendix A in \citealt{Manara+19_2019A&A...628A..95M}).

To summarise, models of dusty binary discs with large truncation radii show high single-disc-like fluxes that decrease as $R_\text{trunc}$ does. In general, models and observations in Taurus and $\rho$~Ophiuchus agree, with some notable exceptions. Too few points are available to draw robust conclusions.

\section{Conclusions}\label{sec.6}
Following up on our previous study that focused on the theoretical and numerical modelling of dust evolution in circumstellar binary discs, in this paper we took into account the same topic from the observational point of view with the aim of discussing if our models and the data in Taurus \citep{Manara+19_2019A&A...628A..95M} and $\rho$~Ophiuchus \citep{Cox+17_2017ApJ...851...83C} agree. To deal with the observations consistently, we analysed \citet{Cox+17_2017ApJ...851...83C} data in the visibility plane in order to compute disc dust sizes as \citet{Manara+19_2019A&A...628A..95M} did in Taurus (see Appendix~\ref{app:2}). This allowed for a study of the disc sizes and the flux-radius correlation in binary discs.

\begin{itemize}
    \item Under the assumption the the measured disc dust sizes trace the position of the truncation radius, \citet{Cox+17_2017ApJ...851...83C} and \citet{Manara+19_2019A&A...628A..95M} showed that implausibly high orbital eccentricities are required to explain their data. In this paper we suggest that this is due to a potentially unfair comparison between dust and gas quantities. In particular, the measured disc dust sizes are always lower than the truncation radius and never trace $R_\text{trunc}$. What is more, when radial drift is taken into account, our zero eccentricity model results are compatible within $1\sigma$ with the measured disc sizes;
    %\item We show that the ratio between the \textit{measured} gas and dust disc sizes is expected to decrease as $R_\text{trunc}$ does and, consequently, to be smaller in binaries than in single-star discs. However, this inference needs to be confirmed by comparison with gas observations as well as modelling of the gas emission;
    \item As for the flux-radius correlation, we found that both in Taurus and $\rho$-Ophiuchus, binary (sub-)millimetre sizes and fluxes are tentatively correlated and, in the former region, also marginally compatible with the \citet{Tripathi+17_2017ApJ...845...44T} relation. Moreover, the correlation normalisation shows a slightly decreasing trend with $R_\text{trunc}$. However, larger data-sets are needed to draw more robust conclusions;
    \item We compared our model prediction for the flux-radius correlation with Taurus \citep{Manara+19_2019A&A...628A..95M} and $\rho$~Ophiuchus \citep{Cox+17_2017ApJ...851...83C} data. We found that our models roughly reproduce the tentative trend in the observations and follow the \citet{Tripathi+17_2017ApJ...845...44T} relation after $t\sim1\text{ Myr}$ regardless of $R_\text{trunc}$;
    \item The binary surveys in the literature allowed us to confirm the \citet{Harris+12_2012ApJ...751..115H} correlation between fluxes in binary pairs and their projected separation in a larger sample of Taurus, $\rho$~Ophiuchus and Lupus discs. This is in qualitative agreement with our modes, whose fluxes and sizes are larger in wider binaries.
\end{itemize}

\section*{Acknowledgements}
We thank the anonymous referee for their helpful comments. This paper makes use of the following ALMA data: 

ADS/JAO.ALMA\#2013.1.00157.S 

ADS/JAO.ALMA\#2016.1.01164.S.

\noindent ALMA is a partnership of ESO (representing its member states), NSF (USA) and NINS (Japan), together with NRC (Canada), MOST and ASIAA (Taiwan), and KASI (Republic of Korea), in cooperation with the Republic of Chile. The Joint ALMA Observatory is operated by ESO, AUI/NRAO and NAOJ.
F.Z. is grateful to Cathie Clarke and the IoA group for insightful discussions. He acknowledges support from the Erasmus+ Traineeship program and IUSS for his MSc thesis internship in Leiden as well as a Science and Technology Facilities Council (STFC) studentship and the Cambridge European Scholarship. G.R. acknowledges support from the Netherlands Organisation for Scientific Research (NWO, program number 016.Veni.192.233) and from an STFC Ernest Rutherford Fellowship (grant number ST/T003855/1). This project has received funding from the European Union’s Horizon 2020 research and innovation programme under the Marie Sklodowska-Curie grant agreement No 823823 (Dustbusters RISE project). Software: \texttt{numpy} \citep{numpy20_2020Natur.585..357H}, \texttt{matplotlib} \citep{matplotlib_Hunter:2007}, \texttt{scipy} \citep{scipy_2020SciPy-NMeth}, \texttt{JupyterNotebook} \citep{Jupyter}, \texttt{uvplot} \citep{Tazzari_uvplot}.

%%%%%%%%%%%%%%%%%%%%%%%%%%%%%%%%%%%%%%%%%%%%%%%%%%
\section*{Data Availability}
The code used in this paper is publicly available on GitHub at \texttt{github.com/rbooth200/DiscEvolution}. The data underlying this paper are available in the ALMA archive:

ADS/JAO.ALMA\#2013.1.00157.S and 

ADS/JAO.ALMA\#2016.1.01164.S.



\bibliographystyle{mnras}
\bibliography{references} 

\begin{thebibliography}{}
\makeatletter
\relax
\def\mn@urlcharsother{\let\do\@makeother \do\$\do\&\do\#\do\^\do\_\do\%\do\~}
\def\mn@doi{\begingroup\mn@urlcharsother \@ifnextchar [ {\mn@doi@}
  {\mn@doi@[]}}
\def\mn@doi@[#1]#2{\def\@tempa{#1}\ifx\@tempa\@empty \href
  {http://dx.doi.org/#2} {doi:#2}\else \href {http://dx.doi.org/#2} {#1}\fi
  \endgroup}
\def\mn@eprint#1#2{\mn@eprint@#1:#2::\@nil}
\def\mn@eprint@arXiv#1{\href {http://arxiv.org/abs/#1} {{\tt arXiv:#1}}}
\def\mn@eprint@dblp#1{\href {http://dblp.uni-trier.de/rec/bibtex/#1.xml}
  {dblp:#1}}
\def\mn@eprint@#1:#2:#3:#4\@nil{\def\@tempa {#1}\def\@tempb {#2}\def\@tempc
  {#3}\ifx \@tempc \@empty \let \@tempc \@tempb \let \@tempb \@tempa \fi \ifx
  \@tempb \@empty \def\@tempb {arXiv}\fi \@ifundefined
  {mn@eprint@\@tempb}{\@tempb:\@tempc}{\expandafter \expandafter \csname
  mn@eprint@\@tempb\endcsname \expandafter{\@tempc}}}

\bibitem[\protect\citeauthoryear{{Akeson} \& {Jensen}}{{Akeson} \&
  {Jensen}}{2014}]{Akeson&Jensen14_2014ApJ...784...62A}
{Akeson} R.~L.,  {Jensen} E.~L.~N.,  2014, \mn@doi [\apj]
  {10.1088/0004-637X/784/1/62}, \href
  {https://ui.adsabs.harvard.edu/abs/2014ApJ...784...62A} {784, 62}

\bibitem[\protect\citeauthoryear{{Akeson}, {Jensen}, {Carpenter}, {Ricci},
  {Laos}, {Nogueira}  \& {Suen-Lewis}}{{Akeson}
  et~al.}{2019}]{Akeson+19_2019ApJ...872..158A}
{Akeson} R.~L.,  {Jensen} E. L.~N.,  {Carpenter} J.,  {Ricci} L.,  {Laos} S.,
  {Nogueira} N.~F.,   {Suen-Lewis} E.~M.,  2019, \mn@doi [\apj]
  {10.3847/1538-4357/aaff6a}, \href
  {https://ui.adsabs.harvard.edu/abs/2019ApJ...872..158A} {872, 158}

\bibitem[\protect\citeauthoryear{{Andrews}}{{Andrews}}{2020}]{Andrews20_2020arXiv200105007A}
{Andrews} S.~M.,  2020, arXiv e-prints, \href
  {https://ui.adsabs.harvard.edu/abs/2020arXiv200105007A} {p. arXiv:2001.05007}

\bibitem[\protect\citeauthoryear{{Andrews} \& {Williams}}{{Andrews} \&
  {Williams}}{2005}]{Andrews&Williams05_2005ApJ...631.1134A}
{Andrews} S.~M.,  {Williams} J.~P.,  2005, \mn@doi [\apj] {10.1086/432712},
  \href {https://ui.adsabs.harvard.edu/abs/2005ApJ...631.1134A} {631, 1134}

\bibitem[\protect\citeauthoryear{{Andrews}, {Terrell}, {Tripathi}, {Ansdell},
  {Williams}  \& {Wilner}}{{Andrews}
  et~al.}{2018}]{Andrews+18a_2018ApJ...865..157A}
{Andrews} S.~M.,  {Terrell} M.,  {Tripathi} A.,  {Ansdell} M.,  {Williams}
  J.~P.,   {Wilner} D.~J.,  2018, \mn@doi [\apj] {10.3847/1538-4357/aadd9f},
  \href {https://ui.adsabs.harvard.edu/abs/2018ApJ...865..157A} {865, 157}

\bibitem[\protect\citeauthoryear{{Ansdell} et~al.,}{{Ansdell}
  et~al.}{2016}]{Ansdell+16_2016ApJ...828...46A}
{Ansdell} M.,  et~al., 2016, \mn@doi [\apj] {10.3847/0004-637X/828/1/46}, \href
  {https://ui.adsabs.harvard.edu/abs/2016ApJ...828...46A} {828, 46}

\bibitem[\protect\citeauthoryear{{Ansdell} et~al.,}{{Ansdell}
  et~al.}{2018}]{Ansdell+18_2018ApJ...859...21A}
{Ansdell} M.,  et~al., 2018, \mn@doi [\apj] {10.3847/1538-4357/aab890}, \href
  {https://ui.adsabs.harvard.edu/abs/2018ApJ...859...21A} {859, 21}

\bibitem[\protect\citeauthoryear{{Artymowicz} \& {Lubow}}{{Artymowicz} \&
  {Lubow}}{1994}]{Artymowicz&Lubow94_1994ApJ...421..651A}
{Artymowicz} P.,  {Lubow} S.~H.,  1994, \mn@doi [\apj] {10.1086/173679}, \href
  {https://ui.adsabs.harvard.edu/abs/1994ApJ...421..651A} {421, 651}

\bibitem[\protect\citeauthoryear{{Baraffe}, {Homeier}, {Allard}  \&
  {Chabrier}}{{Baraffe} et~al.}{2015}]{Baraffe+15_2015A&A...577A..42B}
{Baraffe} I.,  {Homeier} D.,  {Allard} F.,   {Chabrier} G.,  2015, \mn@doi
  [\aap] {10.1051/0004-6361/201425481}, \href
  {https://ui.adsabs.harvard.edu/abs/2015A&A...577A..42B} {577, A42}

\bibitem[\protect\citeauthoryear{{Barenfeld}, {Carpenter}, {Sargent}, {Isella}
  \& {Ricci}}{{Barenfeld} et~al.}{2017}]{Barenfeld+17_2017ApJ...851...85B}
{Barenfeld} S.~A.,  {Carpenter} J.~M.,  {Sargent} A.~I.,  {Isella} A.,
  {Ricci} L.,  2017, \mn@doi [\apj] {10.3847/1538-4357/aa989d}, \href
  {https://ui.adsabs.harvard.edu/abs/2017ApJ...851...85B} {851, 85}

\bibitem[\protect\citeauthoryear{{Barenfeld} et~al.,}{{Barenfeld}
  et~al.}{2019}]{Barenfeld+19_2019ApJ...878...45B}
{Barenfeld} S.~A.,  et~al., 2019, \mn@doi [\apj] {10.3847/1538-4357/ab1e50},
  \href {https://ui.adsabs.harvard.edu/abs/2019ApJ...878...45B} {878, 45}

\bibitem[\protect\citeauthoryear{{Booth}, {Clarke}, {Madhusudhan}  \&
  {Ilee}}{{Booth} et~al.}{2017}]{Booth+17_2017MNRAS.469.3994B}
{Booth} R.~A.,  {Clarke} C.~J.,  {Madhusudhan} N.,   {Ilee} J.~D.,  2017,
  \mn@doi [\mnras] {10.1093/mnras/stx1103}, \href
  {https://ui.adsabs.harvard.edu/abs/2017MNRAS.469.3994B} {469, 3994}

\bibitem[\protect\citeauthoryear{{Bruderer}}{{Bruderer}}{2013}]{Bruderer13_2013A&A...559A..46B}
{Bruderer} S.,  2013, \mn@doi [\aap] {10.1051/0004-6361/201321171}, \href
  {https://ui.adsabs.harvard.edu/abs/2013A&A...559A..46B} {559, A46}

\bibitem[\protect\citeauthoryear{{Bruderer}, {van Dishoeck}, {Doty}  \&
  {Herczeg}}{{Bruderer} et~al.}{2012}]{Bruderer12_2012A&A...541A..91B}
{Bruderer} S.,  {van Dishoeck} E.~F.,  {Doty} S.~D.,   {Herczeg} G.~J.,  2012,
  \mn@doi [\aap] {10.1051/0004-6361/201118218}, \href
  {https://ui.adsabs.harvard.edu/abs/2012A&A...541A..91B} {541, A91}

\bibitem[\protect\citeauthoryear{{Bruderer}, {van der Marel}, {van Dishoeck}
  \& {van Kempen}}{{Bruderer} et~al.}{2014}]{Bruderer14_2014A&A...562A..26B}
{Bruderer} S.,  {van der Marel} N.,  {van Dishoeck} E.~F.,   {van Kempen}
  T.~A.,  2014, \mn@doi [\aap] {10.1051/0004-6361/201322857}, \href
  {https://ui.adsabs.harvard.edu/abs/2014A&A...562A..26B} {562, A26}

\bibitem[\protect\citeauthoryear{{Chen} et~al.,}{{Chen}
  et~al.}{2013}]{Chen+13_2013ApJ...768..110C}
{Chen} X.,  et~al., 2013, \mn@doi [\apj] {10.1088/0004-637X/768/2/110}, \href
  {https://ui.adsabs.harvard.edu/abs/2013ApJ...768..110C} {768, 110}

\bibitem[\protect\citeauthoryear{{Cieza} et~al.,}{{Cieza}
  et~al.}{2019}]{Cieza+19_2019MNRAS.482..698C}
{Cieza} L.~A.,  et~al., 2019, \mn@doi [\mnras] {10.1093/mnras/sty2653}, \href
  {https://ui.adsabs.harvard.edu/abs/2019MNRAS.482..698C} {482, 698}

\bibitem[\protect\citeauthoryear{{Cox} et~al.,}{{Cox}
  et~al.}{2017}]{Cox+17_2017ApJ...851...83C}
{Cox} E.~G.,  et~al., 2017, \mn@doi [\apj] {10.3847/1538-4357/aa97e2}, \href
  {https://ui.adsabs.harvard.edu/abs/2017ApJ...851...83C} {851, 83}

\bibitem[\protect\citeauthoryear{{Duch{\^e}ne} \& {Kraus}}{{Duch{\^e}ne} \&
  {Kraus}}{2013}]{Duchene&Kraus13_2013ARA&A..51..269D}
{Duch{\^e}ne} G.,  {Kraus} A.,  2013, \mn@doi [\araa]
  {10.1146/annurev-astro-081710-102602}, \href
  {https://ui.adsabs.harvard.edu/abs/2013ARA&A..51..269D} {51, 269}

\bibitem[\protect\citeauthoryear{{Eggleton}}{{Eggleton}}{1983}]{Eggleton83_1983ApJ...268..368E}
{Eggleton} P.~P.,  1983, \mn@doi [\apj] {10.1086/160960}, \href
  {https://ui.adsabs.harvard.edu/abs/1983ApJ...268..368E} {268, 368}

\bibitem[\protect\citeauthoryear{{Facchini}, {Birnstiel}, {Bruderer}  \& {van
  Dishoeck}}{{Facchini} et~al.}{2017}]{Facchini+17_2017A&A...605A..16F}
{Facchini} S.,  {Birnstiel} T.,  {Bruderer} S.,   {van Dishoeck} E.~F.,  2017,
  \mn@doi [\aap] {10.1051/0004-6361/201630329}, \href
  {https://ui.adsabs.harvard.edu/abs/2017A&A...605A..16F} {605, A16}

\bibitem[\protect\citeauthoryear{{Feiden}}{{Feiden}}{2016}]{Feiden16_2016A&A...593A..99F}
{Feiden} G.~A.,  2016, \mn@doi [\aap] {10.1051/0004-6361/201527613}, \href
  {https://ui.adsabs.harvard.edu/abs/2016A&A...593A..99F} {593, A99}

\bibitem[\protect\citeauthoryear{{Foreman-Mackey}, {Hogg}, {Lang}  \&
  {Goodman}}{{Foreman-Mackey}
  et~al.}{2013}]{Foreman-Mackey+13_2013PASP..125..306F}
{Foreman-Mackey} D.,  {Hogg} D.~W.,  {Lang} D.,   {Goodman} J.,  2013, \mn@doi
  [\pasp] {10.1086/670067}, \href
  {https://ui.adsabs.harvard.edu/abs/2013PASP..125..306F} {125, 306}

\bibitem[\protect\citeauthoryear{{Foreman-Mackey} et~al.,}{{Foreman-Mackey}
  et~al.}{2019}]{Foreman-Mackey+19_2019JOSS....4.1864F}
{Foreman-Mackey} D.,  et~al., 2019, \mn@doi [The Journal of Open Source
  Software] {10.21105/joss.01864}, \href
  {https://ui.adsabs.harvard.edu/abs/2019JOSS....4.1864F} {4, 1864}

\bibitem[\protect\citeauthoryear{{Goldreich} \& {Tremaine}}{{Goldreich} \&
  {Tremaine}}{1979}]{Goldreich&Tremaine79_1979ApJ...233..857G}
{Goldreich} P.,  {Tremaine} S.,  1979, \mn@doi [\apj] {10.1086/157448}, \href
  {https://ui.adsabs.harvard.edu/abs/1979ApJ...233..857G} {233, 857}

\bibitem[\protect\citeauthoryear{{Goldreich} \& {Tremaine}}{{Goldreich} \&
  {Tremaine}}{1980}]{Goldreich&Tremaine80_1980ApJ...241..425G}
{Goldreich} P.,  {Tremaine} S.,  1980, \mn@doi [\apj] {10.1086/158356}, \href
  {https://ui.adsabs.harvard.edu/abs/1980ApJ...241..425G} {241, 425}

\bibitem[\protect\citeauthoryear{{Harris}, {Andrews}, {Wilner}  \&
  {Kraus}}{{Harris} et~al.}{2012}]{Harris+12_2012ApJ...751..115H}
{Harris} R.~J.,  {Andrews} S.~M.,  {Wilner} D.~J.,   {Kraus} A.~L.,  2012,
  \mn@doi [\apj] {10.1088/0004-637X/751/2/115}, \href
  {https://ui.adsabs.harvard.edu/abs/2012ApJ...751..115H} {751, 115}

\bibitem[\protect\citeauthoryear{{Harris} et~al.,}{{Harris}
  et~al.}{2020}]{numpy20_2020Natur.585..357H}
{Harris} C.~R.,  et~al., 2020, \mn@doi [\nat] {10.1038/s41586-020-2649-2},
  \href {https://ui.adsabs.harvard.edu/abs/2020Natur.585..357H} {585, 357}

\bibitem[\protect\citeauthoryear{{Hatzes}}{{Hatzes}}{2016}]{Hatzes16_2016SSRv..205..267H}
{Hatzes} A.~P.,  2016, \mn@doi [\ssr] {10.1007/s11214-016-0246-3}, \href
  {https://ui.adsabs.harvard.edu/abs/2016SSRv..205..267H} {205, 267}

\bibitem[\protect\citeauthoryear{{Hendler}, {Pascucci}, {Pinilla}, {Tazzari},
  {Carpenter}, {Malhotra}  \& {Testi}}{{Hendler}
  et~al.}{2020}]{Hendler20_2020ApJ...895..126H}
{Hendler} N.,  {Pascucci} I.,  {Pinilla} P.,  {Tazzari} M.,  {Carpenter} J.,
  {Malhotra} R.,   {Testi} L.,  2020, \mn@doi [\apj]
  {10.3847/1538-4357/ab70ba}, \href
  {https://ui.adsabs.harvard.edu/abs/2020ApJ...895..126H} {895, 126}

\bibitem[\protect\citeauthoryear{{Herczeg} \& {Hillenbrand}}{{Herczeg} \&
  {Hillenbrand}}{2014}]{Herczeg&Hillenbrand14_2014ApJ...786...97H}
{Herczeg} G.~J.,  {Hillenbrand} L.~A.,  2014, \mn@doi [\apj]
  {10.1088/0004-637X/786/2/97}, \href
  {https://ui.adsabs.harvard.edu/abs/2014ApJ...786...97H} {786, 97}

\bibitem[\protect\citeauthoryear{{Ho}, {Moran}  \& {Lo}}{{Ho}
  et~al.}{2004}]{Ho+04_2004ApJ...616L...1H}
{Ho} P. T.~P.,  {Moran} J.~M.,   {Lo} K.~Y.,  2004, \mn@doi [\apjl]
  {10.1086/423245}, \href
  {https://ui.adsabs.harvard.edu/abs/2004ApJ...616L...1H} {616, L1}

\bibitem[\protect\citeauthoryear{Hunter}{Hunter}{2007}]{matplotlib_Hunter:2007}
Hunter J.~D.,  2007, \mn@doi [Computing in Science \& Engineering]
  {10.1109/MCSE.2007.55}, 9, 90

\bibitem[\protect\citeauthoryear{{Kelly}}{{Kelly}}{2007}]{Kelly07_2007ApJ...665.1489K}
{Kelly} B.~C.,  2007, \mn@doi [\apj] {10.1086/519947}, \href
  {https://ui.adsabs.harvard.edu/abs/2007ApJ...665.1489K} {665, 1489}

\bibitem[\protect\citeauthoryear{Kluyver et~al.,}{Kluyver
  et~al.}{2016}]{Jupyter}
Kluyver T.,  et~al., 2016, in Loizides F.,  Scmidt B.,  eds, Positioning and
  Power in Academic Publishing: Players, Agents and Agendas. IOS Press, pp
  87--90, \url {https://eprints.soton.ac.uk/403913/}

\bibitem[\protect\citeauthoryear{{K{\"o}hler}, num{Ratzka}, {Herbst}  \&
  {Kasper}}{{K{\"o}hler} et~al.}{2008}]{Kohler+08_2008A&A...482..929K}
{K{\"o}hler} R.,  num{Ratzka} T.,  {Herbst} T.~M.,   {Kasper} M.,  2008,
  \mn@doi [\aap] {10.1051/0004-6361:20079269}, \href
  {https://ui.adsabs.harvard.edu/abs/2008A&A...482..929K} {482, 929}

\bibitem[\protect\citeauthoryear{{Kraus}, {Ireland}, {Hillenbrand}  \&
  {Martinache}}{{Kraus} et~al.}{2012}]{Kraus+12_2012ApJ...745...19K}
{Kraus} A.~L.,  {Ireland} M.~J.,  {Hillenbrand} L.~A.,   {Martinache} F.,
  2012, \mn@doi [\apj] {10.1088/0004-637X/745/1/19}, \href
  {https://ui.adsabs.harvard.edu/abs/2012ApJ...745...19K} {745, 19}

\bibitem[\protect\citeauthoryear{{Kraus}, {Ireland}, {Huber}, {Mann}  \&
  {Dupuy}}{{Kraus} et~al.}{2016}]{Kraus+16_2016AJ....152....8K}
{Kraus} A.~L.,  {Ireland} M.~J.,  {Huber} D.,  {Mann} A.~W.,   {Dupuy} T.~J.,
  2016, \mn@doi [\aj] {10.3847/0004-6256/152/1/8}, \href
  {https://ui.adsabs.harvard.edu/abs/2016AJ....152....8K} {152, 8}

\bibitem[\protect\citeauthoryear{{Kurtovic} et~al.,}{{Kurtovic}
  et~al.}{2021}]{Kurtovic+21_2021A&A...645A.139K}
{Kurtovic} N.~T.,  et~al., 2021, \mn@doi [\aap] {10.1051/0004-6361/202038983},
  \href {https://ui.adsabs.harvard.edu/abs/2021A&A...645A.139K} {645, A139}

\bibitem[\protect\citeauthoryear{{Lin} \& {Papaloizou}}{{Lin} \&
  {Papaloizou}}{1986}]{Lin&Papaloizou86_1986ApJ...309..846L}
{Lin} D.~N.~C.,  {Papaloizou} J.,  1986, \mn@doi [\apj] {10.1086/164653}, \href
  {https://ui.adsabs.harvard.edu/abs/1986ApJ...309..846L} {309, 846}

\bibitem[\protect\citeauthoryear{{Long} et~al.,}{{Long}
  et~al.}{2018}]{Long+18_2018ApJ...869...17L}
{Long} F.,  et~al., 2018, \mn@doi [\apj] {10.3847/1538-4357/aae8e1}, \href
  {https://ui.adsabs.harvard.edu/abs/2018ApJ...869...17L} {869, 17}

\bibitem[\protect\citeauthoryear{{Long} et~al.,}{{Long}
  et~al.}{2019}]{Long+19_2019ApJ...882...49L}
{Long} F.,  et~al., 2019, \mn@doi [\apj] {10.3847/1538-4357/ab2d2d}, \href
  {https://ui.adsabs.harvard.edu/abs/2019ApJ...882...49L} {882, 49}

\bibitem[\protect\citeauthoryear{{Lubow}, {Martin}  \& {Nixon}}{{Lubow}
  et~al.}{2015}]{Lubow+15_2015ApJ...800...96L}
{Lubow} S.~H.,  {Martin} R.~G.,   {Nixon} C.,  2015, \mn@doi [\apj]
  {10.1088/0004-637X/800/2/96}, \href
  {https://ui.adsabs.harvard.edu/abs/2015ApJ...800...96L} {800, 96}

\bibitem[\protect\citeauthoryear{{Manara} et~al.,}{{Manara}
  et~al.}{2019}]{Manara+19_2019A&A...628A..95M}
{Manara} C.~F.,  et~al., 2019, \mn@doi [\aap] {10.1051/0004-6361/201935964},
  \href {https://ui.adsabs.harvard.edu/abs/2019A&A...628A..95M} {628, A95}

\bibitem[\protect\citeauthoryear{{Martin}}{{Martin}}{2018}]{Martin18_2018haex.bookE.156M}
{Martin} D.~V.,  2018, {Populations of Planets in Multiple Star Systems}.
p.~156, \mn@doi{10.1007/978-3-319-55333-7_156}

\bibitem[\protect\citeauthoryear{{Marzari} \& {Thebault}}{{Marzari} \&
  {Thebault}}{2019}]{Marzari&Thebault20_2019Galax...7...84M}
{Marzari} F.,  {Thebault} P.,  2019, \mn@doi [Galaxies]
  {10.3390/galaxies7040084}, \href
  {https://ui.adsabs.harvard.edu/abs/2019Galax...7...84M} {7, 84}

\bibitem[\protect\citeauthoryear{{Mathis}, {Rumpl}  \& {Nordsieck}}{{Mathis}
  et~al.}{1977}]{Mathis+77_1977ApJ...217..425M}
{Mathis} J.~S.,  {Rumpl} W.,   {Nordsieck} K.~H.,  1977, \mn@doi [\apj]
  {10.1086/155591}, \href
  {https://ui.adsabs.harvard.edu/abs/1977ApJ...217..425M} {217, 425}

\bibitem[\protect\citeauthoryear{{Moe} \& {Di Stefano}}{{Moe} \& {Di
  Stefano}}{2017}]{Moe&DiStefano17_2017ApJS..230...15M}
{Moe} M.,  {Di Stefano} R.,  2017, \mn@doi [\apjs] {10.3847/1538-4365/aa6fb6},
  \href {https://ui.adsabs.harvard.edu/abs/2017ApJS..230...15M} {230, 15}

\bibitem[\protect\citeauthoryear{{Monin}, {Whelan}, {Lefloch}, {Dougados}  \&
  {Alves de Oliveira}}{{Monin} et~al.}{2013}]{Monin+13_2013A&A...551L...1M}
{Monin} J.~L.,  {Whelan} E.~T.,  {Lefloch} B.,  {Dougados} C.,   {Alves de
  Oliveira} C.,  2013, \mn@doi [\aap] {10.1051/0004-6361/201220973}, \href
  {https://ui.adsabs.harvard.edu/abs/2013A&A...551L...1M} {551, L1}

\bibitem[\protect\citeauthoryear{{Natta} \& {Testi}}{{Natta} \&
  {Testi}}{2004}]{Natta&Testi04_2004ASPC..323..279N}
{Natta} A.,  {Testi} L.,  2004, {Grain Growth in Circumstellar Disks}.
p.~279

\bibitem[\protect\citeauthoryear{{Natta}, {Testi}, {Calvet}, {Henning},
  {Waters}  \& {Wilner}}{{Natta} et~al.}{2007}]{Natta+07_2007prpl.conf..767N}
{Natta} A.,  {Testi} L.,  {Calvet} N.,  {Henning} T.,  {Waters} R.,   {Wilner}
  D.,  2007, in {Reipurth} B.,  {Jewitt} D.,   {Keil} K.,  eds, Protostars and
  Planets V. p.~767 (\mn@eprint {arXiv} {astro-ph/0602041})

\bibitem[\protect\citeauthoryear{{Nelson}}{{Nelson}}{2000}]{Nelson00_2000ApJ...537L..65N}
{Nelson} A.~F.,  2000, \mn@doi [\apjl] {10.1086/312752}, \href
  {https://ui.adsabs.harvard.edu/abs/2000ApJ...537L..65N} {537, L65}

\bibitem[\protect\citeauthoryear{{Papaloizou} \& {Pringle}}{{Papaloizou} \&
  {Pringle}}{1977}]{Papaloizou&Pringle77_1977MNRAS.181..441P}
{Papaloizou} J.,  {Pringle} J.~E.,  1977, \mn@doi [\mnras]
  {10.1093/mnras/181.3.441}, \href
  {https://ui.adsabs.harvard.edu/abs/1977MNRAS.181..441P} {181, 441}

\bibitem[\protect\citeauthoryear{{Pichardo}, {Sparke}  \& {Aguilar}}{{Pichardo}
  et~al.}{2005}]{Pichardo+05_2005MNRAS.359..521P}
{Pichardo} B.,  {Sparke} L.~S.,   {Aguilar} L.~A.,  2005, \mn@doi [\mnras]
  {10.1111/j.1365-2966.2005.08905.x}, \href
  {https://ui.adsabs.harvard.edu/abs/2005MNRAS.359..521P} {359, 521}

\bibitem[\protect\citeauthoryear{{Picogna} \& {Marzari}}{{Picogna} \&
  {Marzari}}{2013}]{Picogna&Marzari13_2013A&A...556A.148P}
{Picogna} G.,  {Marzari} F.,  2013, \mn@doi [\aap]
  {10.1051/0004-6361/201321860}, \href
  {https://ui.adsabs.harvard.edu/abs/2013A&A...556A.148P} {556, A148}

\bibitem[\protect\citeauthoryear{{Pollack}, {Hollenbach}, {Beckwith},
  {Simonelli}, {Roush}  \& {Fong}}{{Pollack}
  et~al.}{1994}]{Pollack+94_1994ApJ...421..615P}
{Pollack} J.~B.,  {Hollenbach} D.,  {Beckwith} S.,  {Simonelli} D.~P.,  {Roush}
  T.,   {Fong} W.,  1994, \mn@doi [\apj] {10.1086/173677}, \href
  {https://ui.adsabs.harvard.edu/abs/1994ApJ...421..615P} {421, 615}

\bibitem[\protect\citeauthoryear{{Pringle}}{{Pringle}}{1981}]{Pringle81_1981ARA&A..19..137P}
{Pringle} J.~E.,  1981, \mn@doi [\araa] {10.1146/annurev.aa.19.090181.001033},
  \href {https://ui.adsabs.harvard.edu/abs/1981ARA&A..19..137P} {19, 137}

\bibitem[\protect\citeauthoryear{{Raghavan} et~al.,}{{Raghavan}
  et~al.}{2010}]{Raghavan+10_2010ApJS..190....1R}
{Raghavan} D.,  et~al., 2010, \mn@doi [\apjs] {10.1088/0067-0049/190/1/1},
  \href {https://ui.adsabs.harvard.edu/abs/2010ApJS..190....1R} {190, 1}

\bibitem[\protect\citeauthoryear{{Rodriguez} et~al.,}{{Rodriguez}
  et~al.}{2018}]{Rodriguez+18_2018ApJ...859..150R}
{Rodriguez} J.~E.,  et~al., 2018, \mn@doi [\apj] {10.3847/1538-4357/aac08f},
  \href {https://ui.adsabs.harvard.edu/abs/2018ApJ...859..150R} {859, 150}

\bibitem[\protect\citeauthoryear{{Rosotti} \& {Clarke}}{{Rosotti} \&
  {Clarke}}{2018}]{Rosotti&Clarke18_2018MNRAS.473.5630R}
{Rosotti} G.~P.,  {Clarke} C.~J.,  2018, \mn@doi [\mnras]
  {10.1093/mnras/stx2769}, \href
  {https://ui.adsabs.harvard.edu/abs/2018MNRAS.473.5630R} {473, 5630}

\bibitem[\protect\citeauthoryear{{Rosotti}, {Booth}, {Tazzari}, {Clarke},
  {Lodato}  \& {Testi}}{{Rosotti}
  et~al.}{2019a}]{Rosotti+19b_2019MNRAS.486L..63R}
{Rosotti} G.~P.,  {Booth} R.~A.,  {Tazzari} M.,  {Clarke} C.,  {Lodato} G.,
  {Testi} L.,  2019a, \mn@doi [\mnras] {10.1093/mnrasl/slz064}, \href
  {https://ui.adsabs.harvard.edu/abs/2019MNRAS.486L..63R} {486, L63}

\bibitem[\protect\citeauthoryear{{Rosotti}, {Tazzari}, {Booth}, {Testi},
  {Lodato}  \& {Clarke}}{{Rosotti}
  et~al.}{2019b}]{Rosotti+19a_2019MNRAS.486.4829R}
{Rosotti} G.~P.,  {Tazzari} M.,  {Booth} R.~A.,  {Testi} L.,  {Lodato} G.,
  {Clarke} C.,  2019b, \mn@doi [\mnras] {10.1093/mnras/stz1190}, \href
  {https://ui.adsabs.harvard.edu/abs/2019MNRAS.486.4829R} {486, 4829}

\bibitem[\protect\citeauthoryear{{Sanchis} et~al.,}{{Sanchis}
  et~al.}{2020}]{Sanchis20_2020A&A...633A.114S}
{Sanchis} E.,  et~al., 2020, \mn@doi [\aap] {10.1051/0004-6361/201936913},
  \href {https://ui.adsabs.harvard.edu/abs/2020A&A...633A.114S} {633, A114}

\bibitem[\protect\citeauthoryear{{Sanchis} et~al.,}{{Sanchis}
  et~al.}{2021}]{Sanchis+21_2021arXiv210111307S}
{Sanchis} E.,  et~al., 2021, arXiv e-prints, \href
  {https://ui.adsabs.harvard.edu/abs/2021arXiv210111307S} {p. arXiv:2101.11307}

\bibitem[\protect\citeauthoryear{{Simon}, {Dutrey}  \& {Guilloteau}}{{Simon}
  et~al.}{2000}]{Simon+00_2000ApJ...545.1034S}
{Simon} M.,  {Dutrey} A.,   {Guilloteau} S.,  2000, \mn@doi [\apj]
  {10.1086/317838}, \href
  {https://ui.adsabs.harvard.edu/abs/2000ApJ...545.1034S} {545, 1034}

\bibitem[\protect\citeauthoryear{{Simon} et~al.,}{{Simon}
  et~al.}{2017}]{Simon+17_2017ApJ...844..158S}
{Simon} M.,  et~al., 2017, \mn@doi [\apj] {10.3847/1538-4357/aa78f1}, \href
  {https://ui.adsabs.harvard.edu/abs/2017ApJ...844..158S} {844, 158}

\bibitem[\protect\citeauthoryear{{Tazzari}}{{Tazzari}}{2017}]{Tazzari_uvplot}
{Tazzari} M.,  2017, {Mtazzari/Uvplot: V0.1.1}, \mn@doi{10.5281/zenodo.1003113}

\bibitem[\protect\citeauthoryear{{Tazzari} et~al.,}{{Tazzari}
  et~al.}{2016}]{Tazzari+16_2016A&A...588A..53T}
{Tazzari} M.,  et~al., 2016, \mn@doi [\aap] {10.1051/0004-6361/201527423},
  \href {https://ui.adsabs.harvard.edu/abs/2016A&A...588A..53T} {588, A53}

\bibitem[\protect\citeauthoryear{{Tazzari} et~al.,}{{Tazzari}
  et~al.}{2017}]{Tazzari+17_2017A&A...606A..88T}
{Tazzari} M.,  et~al., 2017, \mn@doi [\aap] {10.1051/0004-6361/201730890},
  \href {https://ui.adsabs.harvard.edu/abs/2017A&A...606A..88T} {606, A88}

\bibitem[\protect\citeauthoryear{{Tazzari}, {Beaujean}  \& {Testi}}{{Tazzari}
  et~al.}{2018}]{Tazzari+18_2018MNRAS.476.4527T}
{Tazzari} M.,  {Beaujean} F.,   {Testi} L.,  2018, \mn@doi [\mnras]
  {10.1093/mnras/sty409}, \href
  {https://ui.adsabs.harvard.edu/abs/2018MNRAS.476.4527T} {476, 4527}

\bibitem[\protect\citeauthoryear{{Tazzari} et~al.,}{{Tazzari}
  et~al.}{2020a}]{Tazzari20a_2020arXiv201002248T}
{Tazzari} M.,  et~al., 2020a, arXiv e-prints, \href
  {https://ui.adsabs.harvard.edu/abs/2020arXiv201002248T} {p. arXiv:2010.02248}

\bibitem[\protect\citeauthoryear{{Tazzari}, {Clarke}, {Testi}, {Williams},
  {Facchini}, {Manara}, {Natta}  \& {Rosotti}}{{Tazzari}
  et~al.}{2020b}]{Tazzari20b_2020arXiv201002249T}
{Tazzari} M.,  {Clarke} C.~J.,  {Testi} L.,  {Williams} J.~P.,  {Facchini} S.,
  {Manara} C.~F.,  {Natta} A.,   {Rosotti} G.,  2020b, arXiv e-prints, \href
  {https://ui.adsabs.harvard.edu/abs/2020arXiv201002249T} {p. arXiv:2010.02249}

\bibitem[\protect\citeauthoryear{{Thebault} \& {Haghighipour}}{{Thebault} \&
  {Haghighipour}}{2015}]{Thebault&Haghighipour15_2015pes..book..309T}
{Thebault} P.,  {Haghighipour} N.,  2015, {Planet Formation in Binaries}.
pp 309--340, \mn@doi{10.1007/978-3-662-45052-9_13}

\bibitem[\protect\citeauthoryear{{Trapman}, {Facchini}, {Hogerheijde}, {van
  Dishoeck}  \& {Bruderer}}{{Trapman}
  et~al.}{2019}]{Trapman+19_2019A&A...629A..79T}
{Trapman} L.,  {Facchini} S.,  {Hogerheijde} M.~R.,  {van Dishoeck} E.~F.,
  {Bruderer} S.,  2019, \mn@doi [\aap] {10.1051/0004-6361/201834723}, \href
  {https://ui.adsabs.harvard.edu/abs/2019A&A...629A..79T} {629, A79}

\bibitem[\protect\citeauthoryear{{Tripathi}, {Andrews}, {Birnstiel}  \&
  {Wilner}}{{Tripathi} et~al.}{2017}]{Tripathi+17_2017ApJ...845...44T}
{Tripathi} A.,  {Andrews} S.~M.,  {Birnstiel} T.,   {Wilner} D.~J.,  2017,
  \mn@doi [\apj] {10.3847/1538-4357/aa7c62}, \href
  {https://ui.adsabs.harvard.edu/abs/2017ApJ...845...44T} {845, 44}

\bibitem[\protect\citeauthoryear{{Virtanen} et~al.,}{{Virtanen}
  et~al.}{2020}]{scipy_2020SciPy-NMeth}
{Virtanen} P.,  et~al., 2020, \mn@doi [Nature Methods]
  {https://doi.org/10.1038/s41592-019-0686-2}, \href {https://rdcu.be/b08Wh}
  {17, 261}

\bibitem[\protect\citeauthoryear{{Williams}, {Cieza}, {Hales}, {Ansdell},
  {Ruiz-Rodriguez}, {Casassus}, {Perez}  \& {Zurlo}}{{Williams}
  et~al.}{2019}]{Williams+19_2019ApJ...875L...9W}
{Williams} J.~P.,  {Cieza} L.,  {Hales} A.,  {Ansdell} M.,  {Ruiz-Rodriguez}
  D.,  {Casassus} S.,  {Perez} S.,   {Zurlo} A.,  2019, \mn@doi [\apjl]
  {10.3847/2041-8213/ab1338}, \href
  {https://ui.adsabs.harvard.edu/abs/2019ApJ...875L...9W} {875, L9}

\bibitem[\protect\citeauthoryear{{Winn} \& {Fabrycky}}{{Winn} \&
  {Fabrycky}}{2015}]{Winn&Fabrycky15_2015ARA&A..53..409W}
{Winn} J.~N.,  {Fabrycky} D.~C.,  2015, \mn@doi [\araa]
  {10.1146/annurev-astro-082214-122246}, \href
  {https://ui.adsabs.harvard.edu/abs/2015ARA&A..53..409W} {53, 409}

\bibitem[\protect\citeauthoryear{{Zagaria}, {Rosotti}  \& {Lodato}}{{Zagaria}
  et~al.}{2021}]{Zagaria+21_2021MNRAS.504.2235Z}
{Zagaria} F.,  {Rosotti} G.~P.,   {Lodato} G.,  2021, \mn@doi [\mnras]
  {10.1093/mnras/stab985}, \href
  {https://ui.adsabs.harvard.edu/abs/2021MNRAS.504.2235Z} {504, 2235}

\bibitem[\protect\citeauthoryear{{Zhu} et~al.,}{{Zhu}
  et~al.}{2019}]{Zhu+19_2019ApJ...877L..18Z}
{Zhu} Z.,  et~al., 2019, \mn@doi [\apjl] {10.3847/2041-8213/ab1f8c}, \href
  {https://ui.adsabs.harvard.edu/abs/2019ApJ...877L..18Z} {877, L18}

\bibitem[\protect\citeauthoryear{{Zsom}, {S{\'a}ndor}  \& {Dullemond}}{{Zsom}
  et~al.}{2011}]{Zsom+11_2011A&A...527A..10Z}
{Zsom} A.,  {S{\'a}ndor} Z.,   {Dullemond} C.~P.,  2011, \mn@doi [\aap]
  {10.1051/0004-6361/201015434}, \href
  {https://ui.adsabs.harvard.edu/abs/2011A&A...527A..10Z} {527, A10}

\bibitem[\protect\citeauthoryear{{Zurlo} et~al.,}{{Zurlo}
  et~al.}{2020}]{Zurlo+20_2020MNRAS.496.5089Z}
{Zurlo} A.,  et~al., 2020, \mn@doi [\mnras] {10.1093/mnras/staa1886}, \href
  {https://ui.adsabs.harvard.edu/abs/2020MNRAS.496.5089Z} {496, 5089}

\bibitem[\protect\citeauthoryear{{Zurlo} et~al.,}{{Zurlo}
  et~al.}{2021}]{Zurlo+21_2021MNRAS.501.2305Z}
{Zurlo} A.,  et~al., 2021, \mn@doi [\mnras] {10.1093/mnras/staa3674}, \href
  {https://ui.adsabs.harvard.edu/abs/2021MNRAS.501.2305Z} {501, 2305}

\makeatother
\end{thebibliography}

%%%%%%%%%%%%%%%%%%%%%%%%%%%%%%%%%%%%%%%%%%%%%%%%%%

%%%%%%%%%%%%%%%%% APPENDICES %%%%%%%%%%%%%%%%%%%%%

\appendix
\section{Disc-integrated spectral indices in binaries}\label{app:0}
In this Section we motivate our choice of a quadratic multi-band scaling relation for binary disc fluxes. Disc-integrated spectral indices are considered: we are interested in a general trend, rather than a detailed analysis (e.g., both in Lupus and $\rho$~Ophiuchus a similar study as in \citet{Tazzari20a_2020arXiv201002248T,Tazzari20b_2020arXiv201002249T} can be performed) which is deferred to a subsequent paper.

\begin{figure*}
    \centering
    \includegraphics[width=\textwidth]{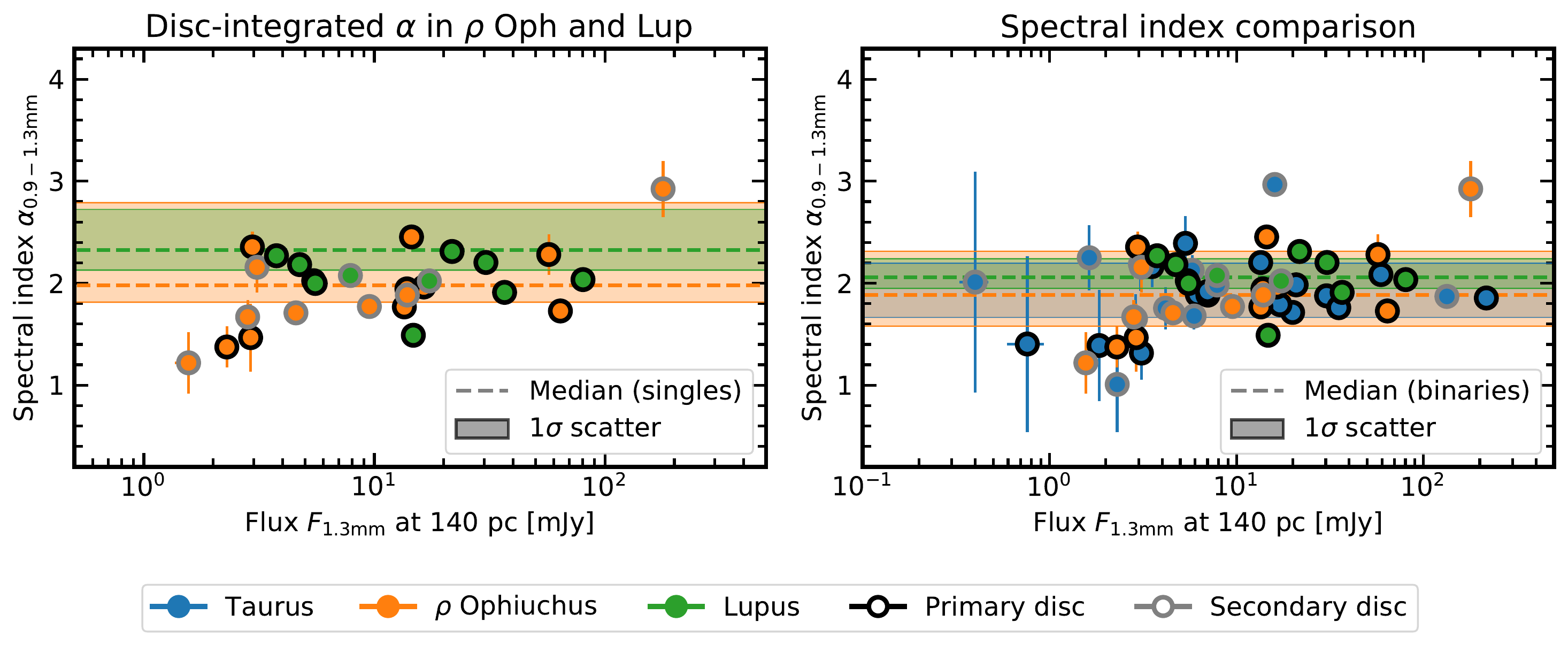}
    \caption{\textbf{Left-hand panel:} The disc-integrated spectral index in binaries, $\alpha_\text{0.9-1.3mm}$, as a function of the 1.3 mm flux, $F_\text{1.3mm}$, for $\rho$~Ophiuchus and Lupus binary discs in orange and green, respectively. The black edges are used for the primaries, while the grey ones for the secondaries. The dashed lines identify the median of the \textit{single-star disc} population in each region (same colours) while the shaded areas its $1\sigma$ scatter. \textbf{Right-hand panel:} Same as in the left-hand panel for Taurus, $\rho$~Ophiuchus and Lupus binaries, as blue, orange and green dots, respectively. The dashed lines indicate the median of the \textit{binary disc} $\alpha_\text{0.9-1.3mm}$ distribution in each region (same colours) while the shaded region identifies its 1$\sigma$ scatter.}
    \label{fig.2.2}
\end{figure*}

Disc-integrated spectral indices in binaries were studied for the first time by \citet{Akeson&Jensen14_2014ApJ...784...62A} in Taurus. They found that, on average, $\alpha_\text{0.9-1.3mm}\sim2$, a value compatible with the results of \citet{Andrews&Williams05_2005ApJ...631.1134A} in the case of single-star discs in the same region. Hereafter, we carry out a similar analysis in $\rho$~Ophiuchus and Lupus. In the former region we make use of ALMA Band 7 observations in \citet{Cox+17_2017ApJ...851...83C}, as well as ALMA Band 6 observations in \citet{Cieza+19_2019MNRAS.482..698C} and \citet{Williams+19_2019ApJ...875L...9W}. In the latter one, instead, we rely on \citet{Ansdell+16_2016ApJ...828...46A,Ansdell+18_2018ApJ...859...21A} data. The disc-integrated spectral indices are computed according to:
\begin{equation}\label{eq.2.1}
    \alpha_\text{0.9-1.3mm}=\dfrac{\log F_\text{0.9mm} - \log F_\text{1.3mm}}{\log 1.3 - \log 0.9},
\end{equation}
where $F_\text{1.3mm}$ and $F_\text{0.9mm}$ are the ALMA Band 6 and 7 dust fluxes.

For $\rho$~Ophiuchus and Lupus sources, in the left-hand panel of Fig.~\ref{fig.2.2} the disc-integrated spectral index, $\alpha_\text{0.9-1.3mm}$, is plotted as a function of the 1.3 mm flux, $F_\text{1.3mm}$, re-scaled to a distance $d=140\text{ pc}$. The dots are used for multiple disc components (excluding the circumbinary ones), while the dashed lines and shaded areas identify the median and $1\sigma$ scatter of the single-star disc population in the same regions, respectively. The black edges are used for the primaries, while the grey ones for the secondaries. The flux calibration uncertainty was not included. 

As can be seen from the figure, on average single-star and binary disc spectral indices are very similar. In Tab.~\ref{tab:2.2} the median and the 16th and 84th percentiles of the observed distribution, in $\rho$~Ophiuchus and Lupus are reported. In both regions the median disc-integrated spectral index in binaries, $\alpha_\text{bin}$, is generally lower than   its single-star discs analogue, $\alpha_\text{sing}$. Nevertheless, they are compatible within $1\sigma$. In Lupus $\alpha_\text{sing}$ is different from the median in \citet{Ansdell+18_2018ApJ...859...21A}, $\alpha=2.25$, as transition discs were excluded from their sample but not known multiples. No big differences can be witnessed between discs around primaries or higher order components. Instead, all circumbinary discs (not plotted in Fig.~\ref{fig.2.2}) show spectral indices larger than the median of the circumstellar binary disc ones. We do not attempt a more quantitative analysis (e.g., a Kolmogorov-Smirnov test) due to the small number of sources for each stellar component, particularly in the case of Lupus, where only two secondaries have been detected at 0.89 mm \citep{Ansdell+16_2016ApJ...828...46A}.

In the right-hand panel of Fig.~\ref{fig.2.2} the disc-integrated spectral indices for multiple stellar sources in Taurus (blue dots, data from \citealt{Akeson&Jensen14_2014ApJ...784...62A}), $\rho$~Ophiuchus (orange dots) and Lupus (green dots) are compared. Even though \citet{Akeson&Jensen14_2014ApJ...784...62A} consider only binary sources, in the case of $\rho$~Ophiuchus we also retain triple components to avoid restricting the sample too much. The dashed lines indicate the median of the $\alpha_\text{0.9-1.3mm}$ distribution in each region while the shaded areas identify their $1\sigma$ error. As it is clear from the plots the three distributions almost perfectly overlap. As reported in Tab.~\ref{tab:2.2}, the median $\alpha_\text{0.9-1.3mm}$ in Taurus is the same as in $\rho$~Ophiuchus and compatible within $1\sigma$ with the Lupus one. %As final remark, while for low disc fluxes in Taurus and $\rho$~Ophiuchus $\alpha_\mathrm{0.9-1.3mm}$ decreases, it increases in Lupus (in agreement with \citealt{Ansdell+18_2018ApJ...859...21A}, who show that single star discs follow the same trend).

The highest values of $\alpha_\text{0.9-1.3mm}$ can be explained assuming disc emission to be optically thin and disc grains to be mm- to cm-sized. %, with absorption spectral index $\beta=\alpha-2\sim0.5-1.0$. This is compatible with grains larger than a few mm (using DSHARP opacities for compact spherical grains with size distribution exponent $q\leq3.0$, \citealt{Birnstiel+18_2018ApJ...869L..45B}).Values of 
$\alpha_\text{0.9-1.3mm}\sim2$ can be interpreted assuming that discs are optically thick and in the Rayleigh-Jeans limit. The smallest values of $\alpha_\text{0.9-1.3mm}<2$ could be still interpreted as due to optically thick emission if the Rayleigh-Jeans approximation does not apply e.g., because of a low disc temperature. However, assuming that the temperature profile is the same in multiple- and single-star discs, we would expect this feature to be more important in the latter, being those discs more extended. %Apparently, this is not the case as the median disc-integrated spectral index is higher in single-star discs than in the binary ones. 
Another possible interpretation relies on continuum emission being dominated by dust self-scattering and (sub-)mm-sized grains \citep[e.g.,][]{Zhu+19_2019ApJ...877L..18Z}. %In particular, an optically thick disc model with high scattering contribution and albedo increasing with the wavelength \citep[][]{Zhu+19_2019ApJ...877L..18Z} can explain those \textquotedblleft anomalously\textquotedblright~low \citep{Liu19_2019ApJ...877L..22L} spectral indices. In this case, the expected maximum grain size would be as big as $0.04\text{ mm}\lesssim a_\text{max}\lesssim0.3\text{ mm}$ (using DSHARP opacities for compact spherical grains with size distribution exponent $q=3.5$ \citep{Birnstiel+18_2018ApJ...869L..45B}; $a_\text{max}$ decreases for more top-heavy grain size distributions). This estimate is also consistent with the results of \citet{Liu19_2019ApJ...877L..22L}. Nevertheless, in the optically thick case we are probing dust sizes of grains emitting at the optically thick layer (the surface in the disc where the optical depth is nearly one): dust grains in the mid-plane could be much larger.

%Both \citet{Cox+17_2017ApJ...851...83C} as well as \citet{Cieza+19_2019MNRAS.482..698C} and \citet{Williams+19_2019ApJ...875L...9W} data have enough resolution for disc dust sizes to be computed. This makes it possible to carry out an analysis similar to the one in \citet{Tazzari20a_2020arXiv201002248T,Tazzari20b_2020arXiv201002249T}. Here the disc optical depth fraction, $\mathcal{F}_\mathrm{1.3mm}$, was computed so that models and data could be compared in the $\mathcal{F}$~\textit{vs}~$\alpha$, plane. This way we should be able to provide better constraints on the maximum grain sizes in discs in multiple systems. This however goes beyond the aims of the paper.

To sum up, disc-integrated spectral indices are consistent with the $F_\nu\propto\nu^2$ scaling ration assumed in this paper in all the star-forming regions taken into account.

\section{Surface brightness and size determination - a model case}\label{app:1}
To compare theoretical predictions and observations we need to compute dust fluxes from the binary disc models in Paper~I. Here we show how this is done following \citet{Rosotti+19a_2019MNRAS.486.4829R}. We focus on the single-star and the binary disc models with initial parameters $\alpha=10^{-3}$ and $R_0=R_\text{trunc}=10\text{ au}$ introduced in Section 3 in Paper~I.

\begin{table}
    \centering
    \begin{tabular}{|c|c|c|}
    \hline
                         & $\alpha_\text{sing}$ & $\alpha_\text{bin}$ \\
    \hline
    Taurus               & - & $1.88~(+0.31,\,-0.22)$ \\
    $\rho$~Ophiuchus     &  $1.98~(+0.81,\,-0.17)$ & $1.88~(+0.43,\,-0.30)$ \\
    Lupus                &  $2.32~(+0.40,\,-0.20)$ & $2.06~(+0.18,\,-0.11)$ \\
    \hline
    \end{tabular}
    \caption{Disc-integrated spectral indices in singles, $\alpha_\text{sing}$, and binaries, $\alpha_\text{bin}$, as plotted in Fig.~\ref{fig.2.2}. The median, as well as the 16th and 84th percentiles of the observed distributions are reported.}
    \label{tab:2.2}
\end{table}

\begin{figure*}
	\centering
    \includegraphics[width=0.995\textwidth]{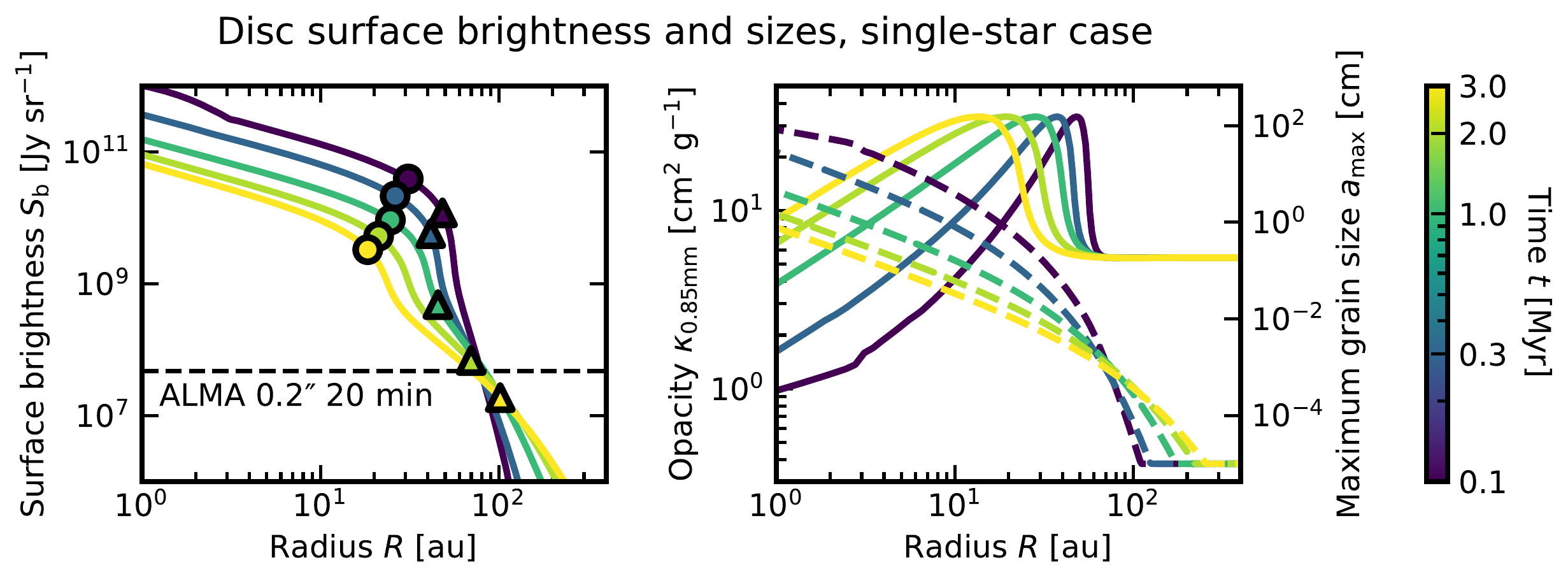}
    \includegraphics[width=0.995\textwidth]{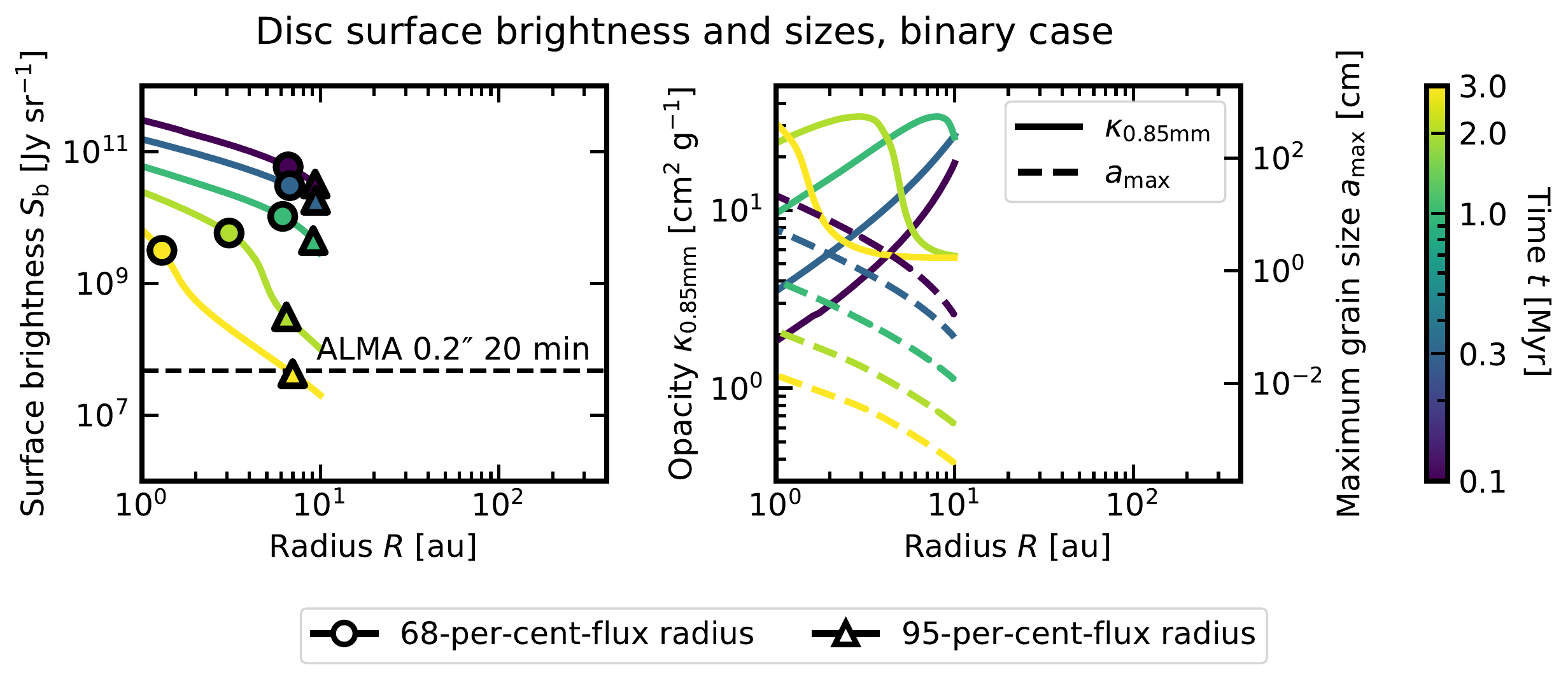}
    \caption{\textbf{Upper panels:} In the left-hand panel the dust surface brightness, $S_\text{b}$, is displayed as a function of the disc radius, for our single-star model. The profiles are evaluated at $t=0.1,\,0.3,\,1,\,2\text{ and }3\text{ Myr}$. The dots and the triangles identify $R_\mathrm{68,mod}$ and $R_\mathrm{95,mod}$, respectively. In addition, we plot the $0.85\text{ mm}$ ALMA sensitivity threshold with integration time of $20\text{ min}$ and angular resolution of $0.2\text{ arcsec}$. In the right-hand panel the single-star model dust opacity, $\kappa_\mathrm{0.85 mm}$ - solid line - and maximum grain size, $a_\text{max}$ - dashed line - are plotted as a function of the disc radius at the same times. \textbf{Lower panels:} Same as in the upper panels but for the binary disc model.}
    \label{fig:A1}
\end{figure*}

In Fig.~\ref{fig:A1} on the left-hand side the $0.85$~mm surface brightness radial profile, $S_\text{b}$, is plotted after $t=0.1,\,0.3,\,1,\,2\text{ and }3\text{ Myr}$. The dots and the triangles identify $R_\mathrm{68,mod}$ and $R_\mathrm{95,mod}$, respectively. Similarly, on the right-hand side the dust opacity, $\kappa_\mathrm{0.85mm}$, and the maximum grain size, $a_\text{max}$, are displayed as a function of the disc radius at the same times. Solid and dashed lines are employed, respectively. The top panels refer to the single-star disc case, while the bottom ones to the binary disc models. A sensitivity threshold corresponding to an angular resolution of $0.2\text{ arcsec}$ and an integration time of $20\text{ min}$ is over-plotted\footnote{Given these initial parameters, if 43 antennas are employed, at $0.85\text{ mm}$ the ALMA sensitivity calculator reports a sensitivity rms of $\sim50.34~\mu\text{Jy~beam}^{-1}$ (\texttt{almascience.eso.org/proposing/sensitivity-calculator}), corresponding to $\sim4.73\times10^7\text{~Jy~sr}^{-1}=1.11\text{~mJy~arcsec}^{-2}$.} on the surface brightness profiles in Fig.~\ref{fig:A1}. 

As it is clear from Fig.~\ref{fig:A1}, in the single-star model the surface brightness profile is characterised at any time by two smoothly varying regions connected by a short interval in which it undergoes an abrupt change. This can be explained in terms of the opacity profile which also experiences a rapid variation at the same radii. \citet{Rosotti+19b_2019MNRAS.486L..63R,Rosotti+19a_2019MNRAS.486.4829R} call this feature the \textit{opacity cliff}. 

In the earliest stages of the binary disc evolution, the surface brightness behaves as its single-star counterpart does in the innermost regions of the disc. Indeed, as it is evident from the lower-right panel in Fig~\ref{fig:A1}, in our binary model initially all grains are large enough to overcome the \textit{opacity cliff} and no abrupt reduction of the surface brightness can be seen due to the absence of small grains. However, as time goes on, the surface brightness in the binary model resembles that attained by the single-star model in outer and outer regions of the disc. In fact, as we showed in Paper~I, dust depletion takes place more rapidly in binary discs rather than in single-star ones. This is why, at later times only small dust grains are retained. Such small grains are not able to overcome the \textit{opacity cliff}. 

Let us now focus on the behaviour of the 68- and 95-per-cent-flux radius. In the single-star model $R_\mathrm{68,mod}$ decreases with time and always traces the position of the peak in the opacity profile. On the contrary, $R_\mathrm{95,mod}$ increases as time goes on. In the binary model $R_\mathrm{68,mod}$ decreases more sharply than in the single-star disc case, yet it still traces the position of the opacity cliff, as a look at the bottom panels in Fig.~\ref{fig:A1} suggests. In other words, the 68-per-cent-flux radius gives a measure of how fast large grains are depleted. From this point of view, its faster decrease is consistent with radial drift being more efficient in binaries, as shown in Paper~I. On the other hand, while $R_\mathrm{95,mod}$ increases with time in the single-star model, it remains roughly constant with time in the binary case. This feature can be explained in terms of the closed-outer-boundary condition that we imposed on gas and dust dynamics in binary discs. Indeed, as the zero-flux condition prevents disc spreading, the 95-per-cent-flux radius cannot increase with time: it traces the position of the tidal truncation radius. Finally, it is worth noticing that both in the single-star and binary disc model even surveys with long integration time will have difficulties to detect the tails of the disc emission. In particular, at late times the \textit{measured} 95-per-cent-flux radius will underestimated $R_\mathrm{95,mod}$. As a consequence, we expect that in binaries the measured disc sizes will decrease with time and will not trace the $R_\mathrm{trunc}$. 

To sum up, the faster time scale of dust depletion in binaries determines a sharper dependence of the surface brightness with time than in single-star discs. Consequently, both $R_\mathrm{68,mod}$ and $R_\mathrm{95,mod}$ always decreases with time if a cut in sensitivity is introduced.

\section{Analysis of $\rho$~Ophiuchus binary discs in the visibility plane}\label{app:2}
To compare our binary disc models with observations, in Section~\ref{sec.4} we relied on the observed discs in \citet{Manara+19_2019A&A...628A..95M} for Taurus and in \citet{Cox+17_2017ApJ...851...83C} for $\rho$~Ophiuchus. \citet{Manara+19_2019A&A...628A..95M} obtained $1.33\,\text{mm}$ fluxes and 68-per-cent-flux radii from fits in visibility plane. On the other hand, \citet{Cox+17_2017ApJ...851...83C} employed Gaussian fits in the image plane to compute disc sizes at 0.87~mm. 

In order to deal with observations consistently we proceeded to a fit of the binary discs in \citet{Cox+17_2017ApJ...851...83C} in the visibility plane to compute the 68-per-cent-flux radius, $R_\mathrm{68,obs}$, and the 95-per-cent-flux radius, $R_\mathrm{95,obs}$, according to \citep[e.g.,][]{Tripathi+17_2017ApJ...845...44T,Manara+19_2019A&A...628A..95M}:
\begin{equation}\label{eq:B1}
    x\cdot F_\text{tot}=\int_0^{R_x}I2\pi RdR,
\end{equation}
where $I$ is the disc intensity profile and $F_\text{tot}$ is the total inferred disc flux, with $x=0.68$ for $R_\mathrm{68,obs}\text{ and }x=0.95$ for $R_\mathrm{95,obs}$, respectively. Our sample is made up of all the binary and triple discs in \citet{Cox+17_2017ApJ...851...83C} not showing evidence for circumbinary emission, coherently with our exclusion of T~Tau~S from the \citet{Manara+19_2019A&A...628A..95M} sample. Two other sources, namely ROph~3 and ROph~4, have been subsequently added as \citet{Zurlo+20_2020MNRAS.496.5089Z} in their disc survey in multiple stellar systems in $\rho$~Ophiuchus showed that those discs are part of binary systems.

First of all, we averaged the continuum disc visibilities in each spectral window and re-scaled the $uv$-distances in units of the observation wavelength. To perform the fit, we assumed the following model \citep[e.g.,][]{Long+18_2018ApJ...869...17L,Long+19_2019ApJ...882...49L,Manara+19_2019A&A...628A..95M,Tazzari20b_2020arXiv201002249T} for the intensity profile of a single-star disc:
\begin{equation}\label{eq:B2}
    I=F_\text{tot}\dfrac{\bigl(R/R_\text{C}\bigr)^{-\gamma_1}\exp\bigl\{-\bigl(R/R_\text{C}\bigr)^{\gamma_2}\bigr\}}{\int_0^\infty \bigl(R/R_\text{C}\bigr)^{-\gamma_1}\exp\bigl\{-\bigl(R/R_\text{C}\bigr)^{\gamma_2}\bigr\}\,2\pi RdR}.
\end{equation}
Here $R_\text{C}$ is a characteristic scale radius and the exponents $\gamma_1$ and $\gamma_2$ describe the decay of the intensity of the dust emission in the inner and outer disc, respectively. Then, we employed \texttt{galario} \citep{Tazzari+18_2018MNRAS.476.4527T} to compute the model visibilities, first of a single-star disc, $\mathcal{V}_{\text{model},i}$, and then of any multiple stellar systems made up of $n$ single components, as:
\begin{equation}\label{eq:B3}
    \mathcal{V}_\text{model} = \sum_{i=1}^{n}\mathcal{V}_{\text{model},i}.
\end{equation}
$\mathcal{V}_{\text{model},i}$ is a function of the intensity profile, the disc offset from the phase centre of the observations given by $\Delta\alpha$ and $\Delta\delta$, as well as the disc inclination $i$ and position angle PA. Consequently, a total of 8 parameters for each single-disc needed to be determined.

We explored the $8\times n$-dimensional parameter space making use of the latest version of \texttt{emcee}  \citep{Foreman-Mackey+13_2013PASP..125..306F,Foreman-Mackey+19_2019JOSS....4.1864F}, a Markov-chain Monte Carlo sampler, adopting uniform priors and a Gaussian likelihood function to compute the posteriors for our model parameters. To achieve convergence we employed between $1\times10^2$ and $3\times10^2$ walkers and from $5\times10^3$ to $2\times10^4$ steps. The values of the reduced $\chi^2$ are around $\sim0.6-0.7$, suggesting that over-fitting occurred due to the fact that some of the discs are not resolved or are only marginally resolved. Over-fitting can also be explained as a consequence of error overestimation in the data. The best-fit values for the model parameters were chosen as the median of the last $1\times10^3$ steps for each chain (well beyond the burn-in phase) and the difference from the median and the $16$th and $84$th percentiles were used to determine the $1\sigma$ uncertainties, consistently with \citet{Manara+19_2019A&A...628A..95M}. The best fit values of the model parameters with their uncertainties are reported for each fitted disc in Tab.~\ref{tab:1}. 

% In Fig.s~\ref{fig:B1} to \ref{fig:B11} the corner plot and visibility $uv$-plot are shown for the fitted discs in Tab.~\ref{tab:1}. In multiple systems, when both discs were resolved and fitted together, the de-projection in the $uv$-plots was performed using the inclination and position angle of the brightest component.

$R_\mathrm{68,obs}$ and $R_\mathrm{95,obs}$ were computed as in eq.~\ref{eq:B1}: the best-fit values and the uncertainties were determined as for the model parameters. In Tab.~\ref{tab:2} the disc flux, $R_\mathrm{68,obs}$ and $R_\mathrm{95,obs}$ are reported. A direct comparison between the inferred fluxes in \citet{Cox+17_2017ApJ...851...83C}, $F_\text{Cox}$, and those from our fit $F=F_\text{tot}\cos i$ from Tab.~\ref{tab:2} proves a general agreement within $2\sigma$.

\begin{landscape}
\begin{table}
    \centering
    \begin{tabular}{|c|c|c|c|c|c|c|c|c|}
    \hline
    \hline
    \multirow{2}{*}{Source} & $\log F_\text{tot}$ & \multirow{2}{*}{$\gamma_1$} & \multirow{2}{*}{$\gamma_2$} & $R_\text{C}$ & $i$ & PA & $\Delta\alpha$ & $\Delta\delta$ \\
    & [mJy] & & & [arcsec] & [deg] & [deg] & [arcsec] & [arcsec] \\
    \hline
    \hline
    ROph 3 & 
    $2.1590^{+0.0068}_{-0.0070}$ &
    $-2.1829^{+0.5645}_{-0.9167}$ &
    $6.1060^{+5.3831}_{-2.1263}$ &
    $0.2339^{+0.0175}_{-0.0331}$ &
    $48.5507^{+0.7041}_{-0.7385}$ &
    $81.6223^{+1.0043}_{-1.0710}$ &
    $0.2536^{+0.0015}_{-0.0016}$ &
    $-0.3662^{+0.0012}_{-0.0011}$ \\
    \hline
    ROph 4 &
    $1.0392^{+0.2084}_{-0.1588}$ &
    $-0.2231^{+1.3169}_{-3.0188}$ &
    $26.1506^{+15.3586}_{-16.8515}$ &
    $0.1110^{+0.0571}_{-0.0294}$ &
    $50.4819^{+16.2839}_{-30.2850}$ &
    $67.9550^{+22.3491}_{-17.6096}$ &
    $0.1797^{+0.0084}_{-0.0083}$ &
    $-0.2223^{+0.0066}_{-0.0067}$ \\
    \hline
    \hline
    \multicolumn{9}{|c|}{Primary discs} \\       
    \hline
    \hline
    ROph 5 A & 
    $1.4327^{+0.0255}_{-0.0134}$ &
    $-0.5573^{+0.3131}_{-0.6425}$ &
    $17.5067^{+21.5132}_{-12.6048}$ &
    $0.1384^{+0.0076}_{-0.0135}$ &
    $15.3882^{+9.1259}_{-9.6077}$ &
    $134.2374^{+21.3717}_{-53.9183}$ &
    $0.2708^{+0.0026}_{-0.0026}$ &
    $-0.7758^{+0.0023}_{-0.0022}$ \\
    \hline
    ROph 7 A & 
    $1.9475^{+ 0.0350}_{-0.0391}$ &
    $0.3608^{+ 0.2539}_{-0.5233}$ &
    $25.4785^{+ 17.6340}_{-15.8818}$ &
    $0.0964^{+ 0.0090}_{-0.0075}$ &
    $59.3786^{+ 2.5161}_{-4.1039}$ &
    $164.2007^{+ 2.3730}_{-2.5552}$ &
    $-0.5148^{+ 0.0009}_{-0.0010}$ &
    $1.0143^{+ 0.0010}_{-0.0008}$ \\
    \hline
    ROph 21 A & 
    $1.2046^{+0.0697}_{-0.0620}$ &
    $-0.8128^{+1.7708}_{-2.2975}$ &
    $27.3682^{+15.0205}_{-16.7344}$ &
    $0.0557^{+0.0203}_{-0.0096}$ &
    $21.5719^{+17.1898}_{-15.0340}$ &
    $84.3149^{+65.3861}_{-54.7677}$ &
    $-0.3431^{+0.0024}_{-0.0022}$ &
    $-0.5648^{+0.0023}_{-0.0024}$ \\
    \hline
    ROph 27 A & 
    $2.0091^{+0.0340}_{-0.0312}$ &
    $0.6843^{+0.0888}_{-0.2134}$ &
    $10.6839^{+24.4570}_{-7.8231}$ &
    $0.2471^{+0.0135}_{-0.0285}$ &
    $70.5950^{+1.4462}_{-1.5106}$ &
    $36.6648^{+1.3323}_{-1.3798}$ &
    $-0.3626^{+0.0022}_{-0.0021}$ &
    $-0.2696^{+0.0024}_{-0.0023}$ \\
    \hline
    ROph 31 A & 
    $1.9799^{+0.0087}_{-0.0045}$ &
    $0.9219^{+0.0718}_{-0.4513}$ &
    $16.1519^{+23.3272}_{-13.5832}$ &
    $0.1161^{+0.0042}_{-0.0268}$ &
    $7.9016^{+6.4005}_{-5.2729}$ &
    $103.8223^{+38.3643}_{-51.3438}$ &
    $0.1512^{+0.0006}_{-0.0006}$ &
    $-0.0331^{+0.0006}_{-0.0006}$ \\
    \hline
    ROph 34 A & 
    $1.5893^{+0.0483}_{-0.0435}$ &
    $1.4074^{+0.1440}_{-0.3599}$ &
    $23.1388^{+17.7057}_{-16.5946}$ &
    $0.1059^{+0.0270}_{-0.0285}$ &
    $32.6436^{+11.2399}_{-18.5356}$ &
    $141.2977^{+15.5242}_{-30.1650}$ &
    $0.6558^{+0.0011}_{-0.0011}$ &
    $-0.2948^{+0.0011}_{-0.0011}$ \\
    \hline
    ROph 45 A & 
    $0.9273^{+0.1972}_{-0.1380}$ &
    $0.5408^{+0.8152}_{-2.5992}$ &
    $22.8281^{+17.8106}_{-17.1487}$ &
    $0.1874^{+0.5381}_{-0.0639}$ &
    $39.7740^{+22.6999}_{-26.3198}$ &
    $124.1943^{+28.5930}_{-50.4143}$ &
    $-0.0848^{+0.0179}_{-0.0185}$ &
    $-1.0186^{+0.0150}_{-0.0157}$ \\
    \hline
    \hline
    \multicolumn{9}{|c|}{Secondary discs} \\ 
    \hline
    \hline
    ROph 5 B & 
    $1.3476^{+0.0371}_{-0.0397}$ &
    $0.0800^{+0.4684}_{-1.7187}$ &
    $22.4107^{+17.6811}_{-14.2915}$ &
    $0.1056^{+0.0135}_{-0.0189}$ &
    $30.1983^{+7.2209}_{-12.8258}$ &
    $64.5025^{+14.2418}_{-12.6032}$ &
    $-1.2027^{+0.0023}_{-0.0022}$ &
    $-0.9843^{+0.0021}_{-0.0020}$ \\
    \hline
    ROph 7 B &
    $1.7127^{+0.0533}_{-0.0523}$ &
    $-0.4379^{+1.0128}_{-2.2095}$ &
    $20.0676^{+21.4032}_{-12.9557}$ &
    $0.0623^{+0.0119}_{-0.0136}$ &
    $41.8173^{+7.0737}_{-7.4926}$ &
    $148.6950^{+7.3016}_{-6.1686}$ &
    $0.4454^{+0.0010}_{-0.0010}$ &
    $-0.7347^{+0.0010}_{-0.0009}$ \\
    \hline
    ROph 27 B & 
    $0.9104^{+0.1699}_{-0.1226}$ &
    $-0.9298^{+1.8152}_{-2.5680}$ &
    $24.4957^{+17.8595}_{-15.9691}$ &
    $0.1023^{+0.0478}_{-0.0215}$ &
    $43.7275^{+17.9191}_{-29.7463}$ &
    $46.4491^{+43.2915}_{-20.8190}$ &
    $1.7371^{+0.0094}_{-0.0085}$ &
    $2.5064^{+0.0088}_{-0.0090}$ \\
    \hline
    ROph 34 B & 
    $1.4384^{+0.1842}_{-0.1327}$ &
    $-3.4663^{+1.3529}_{-1.0492}$ &
    $30.6522^{+13.5708}_{-15.6028}$ &
    $0.2083^{+0.0228}_{-0.0183}$ &
    $73.6989^{+5.2271}_{-5.4155}$ &
    $155.8769^{+4.2742}_{-4.2667}$ &
    $-2.539^{+0.0095}_{-0.0119}$ &
    $6.5106^{+0.0105}_{-0.0117}$ \\
    \hline
    \hline
    \multicolumn{9}{|c|}{Ternary discs} \\
    \hline
    \hline
    ROph 11 B &
    $1.4250^{+0.2772}_{-0.1320}$ &
    $1.1432^{+0.5486}_{-2.3592}$ &
    $24.4413^{+17.1997}_{-16.2745}$ &
    $0.0937^{+0.1900}_{-0.0467}$ &
    $61.7057^{+17.5783}_{-26.8859}$ &
    $112.7737^{+14.5817}_{-16.4206}$ &
    $0.3762^{+0.0023}_{-0.0025}$ &
    $-0.4397^{+0.0019}_{-0.0019}$ \\
    \hline
    ROph 13 B & 
    $2.8629^{+0.0017}_{-0.0017}$ &
    $-2.2015^{+0.1123}_{-0.1166}$ &
    $2.5373^{+0.1076}_{-0.1022}$ &
    $0.3381^{+0.0127}_{-0.0131}$ &
    $48.4986^{+0.1800}_{-0.1783}$ &
    $26.4426^{+0.2517}_{-0.2511}$ &
    $0.3031^{+0.0007}_{-0.0007}$ &
    $-1.5424^{+0.0007}_{-0.0007}$ \\
    \hline
    \hline
    \end{tabular}
    \caption{Parameters from the fits of the continuum visibilities of discs in multiple stellar systems in $\rho$~Ophiuchus from \citet{Cox+17_2017ApJ...851...83C}.}
    \label{tab:1}
\end{table}
\end{landscape}

\begin{landscape}
\begin{table}
    \centering
    \begin{tabular}{|c|c|c|c|c|c|c|c|c|c|}
    \hline
    \hline
    \multirow{2}{*}{ALMA name} &  Source Name & \multirow{2}{*}{$\alpha$} & \multirow{2}{*}{$\delta$} & Separation & $F_\text{Cox}$ & $F$ & $R_\mathrm{68,obs}$ & $R_\mathrm{95,obs}$ & $d$ \\
    & SSTc2d & & & [arsec] & [mJy] & [mJy] & [arcsec] & [arcsec] & [pc] \\
    \hline
    \hline
    \multicolumn{9}{|c|}{Binary disc, fit of both components} \\
    \hline
    \hline
    ROph 5 A & J162502.1-245932a & 16:25:02.119 & -24:59:32.798 & 1.490 & 27.74$\pm$0.77 & $26.0986^{+0.3973}_{-0.4073}$ & $0.1171^{+0.0037}_{-0.0040}$ & $0.1408^{+0.0100}_{-0.0070}$ & 142.04 \\
    \hline
    ROph 5 B & J162502.1-245932b & 16:25:02.011 & -24:59:33.004 & 1.490 & 19.60$\pm$0.70 & $19.3541^{+0.3523}_{-0.4492}$ & $0.0836^{+0.0062}_{-0.0070}$ & $0.1036^{+0.0117}_{-0.0144}$ & 142.04 \\
    \hline
    ROph 7 A & J162623.4-242101a & 16:26:23.362 & -24:20:59.997 & 2.030 & 46.75$\pm$0.40 & $45.3971^{+1.2018}_{-1.0932}$ & $0.0746^{+0.0035}_{-0.0034}$ & $0.0941^{+0.0068}_{-0.0066}$ & 125.69 \\
    \hline
    ROph 7 B & J162623.4-242101b & 16:26:23.432 & -24:21:01.749 & 2.030 & 38.36$\pm$0.38 & $37.9256^{+1.9133}_{-1.4758}$ & $0.0512^{+0.0041}_{-0.0056}$ & $0.0628^{+0.0089}_{-0.0111}$ & 125.69 \\
    \hline
    ROph 27 A & J163130.9-242440a & 16:31:30.873 & -24:24:40.288 & 3.560 & 36.29$\pm$0.78 & $33.8995^{+0.7475}_{-0.6761}$ & $0.1791^{+0.0066}_{-0.0065}$ & $0.2493^{+0.0303}_{-0.0152}$ & 130.61 \\
    \hline
    ROph 27 B & J163130.9-242440b & 16:31:31.025 & -24:24:37.484 & 3.560 & 6.58$\pm$0.32 & $5.8274^{+0.5074}_{-0.5356}$ & $0.0874^{+0.0219}_{-0.0140}$ & $0.1049^{+0.0437}_{-0.0209}$ & 130.61 \\
    \hline
    ROph 34 A & J163221.0-243036a & 16:32:21.047 & -24:30:36.309 & 7.560 & 33.43$\pm$0.96 & $32.6494^{+1.1996}_{-1.4906}$ & $0.0495^{+0.0053}_{-0.0053}$ & $0.0961^{+0.0195}_{-0.0193}$ & 155.13\\
    \hline
    ROph 34 B & J163221.0-243036b & 16:32:20.811 & -24:30:29.487 & 7.560 & 6.32$\pm$0.25 & $7.5725^{+0.7182}_{-0.6441}$ & $0.1920^{+0.0187}_{-0.0157}$ & $0.2105^{+0.0232}_{-0.0191}$ & 155.13 \\
    \hline
    \hline
    \multicolumn{9}{|c|}{Binary discs, fit of one component} \\
    \hline
    \hline
    ROph 3 & J162309.2-241705 & 16:23:09.219 & -24:17:05.364 & 1.650\tablefootnote{2MASS J16230923-2417047 in \text{\citet{Zurlo+20_2020MNRAS.496.5089Z}}} & $114.3\pm9.7$ & $95.4603^{+0.6081}_{-0.5906}$ & $0.2220^{+0.0021}_{-0.0021}$ & $0.2683^{+0.0080}_{-0.0097}$ & 160.52 \\
    \hline
    ROph 4 & J162336.1-240221 &  16:23:36.113 & -24:02:21.227 & 1.832\tablefootnote{2MASS J16233609-2402209 in \text{\citet{Zurlo+20_2020MNRAS.496.5089Z}}} & $7.12\pm0.31$ & $6.9157^{+0.5712}_{-0.6309}$ & $0.0888^{+0.0232}_{-0.0168}$ & $0.1111^{+0.0491}_{-0.0272}$ & 148.71 \\
    \hline
    ROph 21 A & J162740.3-242204 & 16:27:40.275 & -24:22:04.568 & 0.638 & 14.83$\pm$0.48 & $14.4805^{+1.1061}_{-1.5993}$ & $0.0467^{+0.0052}_{-0.0057}$ & $0.0558^{+0.0172}_{-0.0084}$ & 130.46 \\
    \hline
    ROph 45 A & J162751.8-243145 & 16:27:51.796 & -24:31:46.048 & 7.170 & 5.46$\pm$0.72 & $5.9081^{+1.6318}_{-0.6604}$ & $0.1453^{+0.2296}_{-0.0351}$ & $0.1908^{+0.5257}_{-0.0618}$ & 139.40 \\
    \hline
    \hline
    \multicolumn{9}{|c|}{Ternary discs, fit of one component} \\
    \hline
    \hline
    ROph 11 B & J162646.4-241160 & 16:26:46.427 & -24:12:00.443 & 0.577 & 15.41$\pm$0.38 & $13.2425^{+3.1133}_{-5.2698}$ & $0.0507^{+0.0297}_{-0.0131}$ & $0.0887^{+0.1399}_{-0.0411}$ & 109.54 \\
    \hline
    ROph 13 B & J162658.4-244532 & 16:26:58.504 & -24:45:37.220 & 5.065 & 624$\pm$73 & $483.2569^{+0.8476}_{-0.8513}$ & $0.4384^{+0.0010}_{-0.0010}$ & $0.5932^{+0.0027}_{-0.0026}$ & 139.40 \\
    \hline
    ROph 31 A & J163152.1-245616 & 16:31:52.111 & -24:56:16.030 & 3.000 & 94.4$\pm$2.7 & $94.4169^{+0.5072}_{-0.5525}$ & $0.0783^{+0.0018}_{-0.0041}$ & $0.1126^{+0.0083}_{-0.0037}$ & 166.46\\
    \hline
    \hline
    \end{tabular}
    \caption{Fitting parameters employed in this paper for $\rho$~Ophiuchus binaries in \citet{Cox+17_2017ApJ...851...83C}. We refer back to the original paper for the relevant literature for the binary position and separation. We employ the same Gaia distances of the corresponding sources in \citet{Williams+19_2019ApJ...875L...9W}.}
    \label{tab:2}
\end{table}
\end{landscape}

\section{Tentative relation between dust-to-gas size ratio and truncation radius}\label{app:3}
As outlined in the main text a recently submitted paper (Rota et al. subm.) studied the \ce{^{12}CO} emission in a sub-sample of the Taurus binaries in \citet{Manara+19_2019A&A...628A..95M} with the aim of confronting the radial extent of the gas and dust emission. Their main result is that, choosing the 68-per-cent radius as a metric, on average $R_\mathrm{gas}/R_\mathrm{dust}=2.8\pm1.2$. This is in line with the results of \citet{Sanchis+21_2021arXiv210111307S} in Lupus, who took into account single star discs and binaries with separation larger than 2~arcsec (roughly 317~au at the Lupus median distance), inferring a median of $R_\mathrm{gas}/R_\mathrm{dust}=2.5\pm1.5$. However, when the 95-per-cent radius is considered, Rota et al. (subm.) on average estimate $R_\mathrm{gas}/R_\mathrm{dust}=3.7\pm1.5$, which is in the upper end of the \citet{Sanchis+21_2021arXiv210111307S} distribution (whose median\footnote{For simplicity we are using the quoted size ratio employing the 90-per-cent-flux radius.} is the same as for the 68-per-cent radius).

\begin{figure*}
    \centering
    \includegraphics[width=\textwidth]{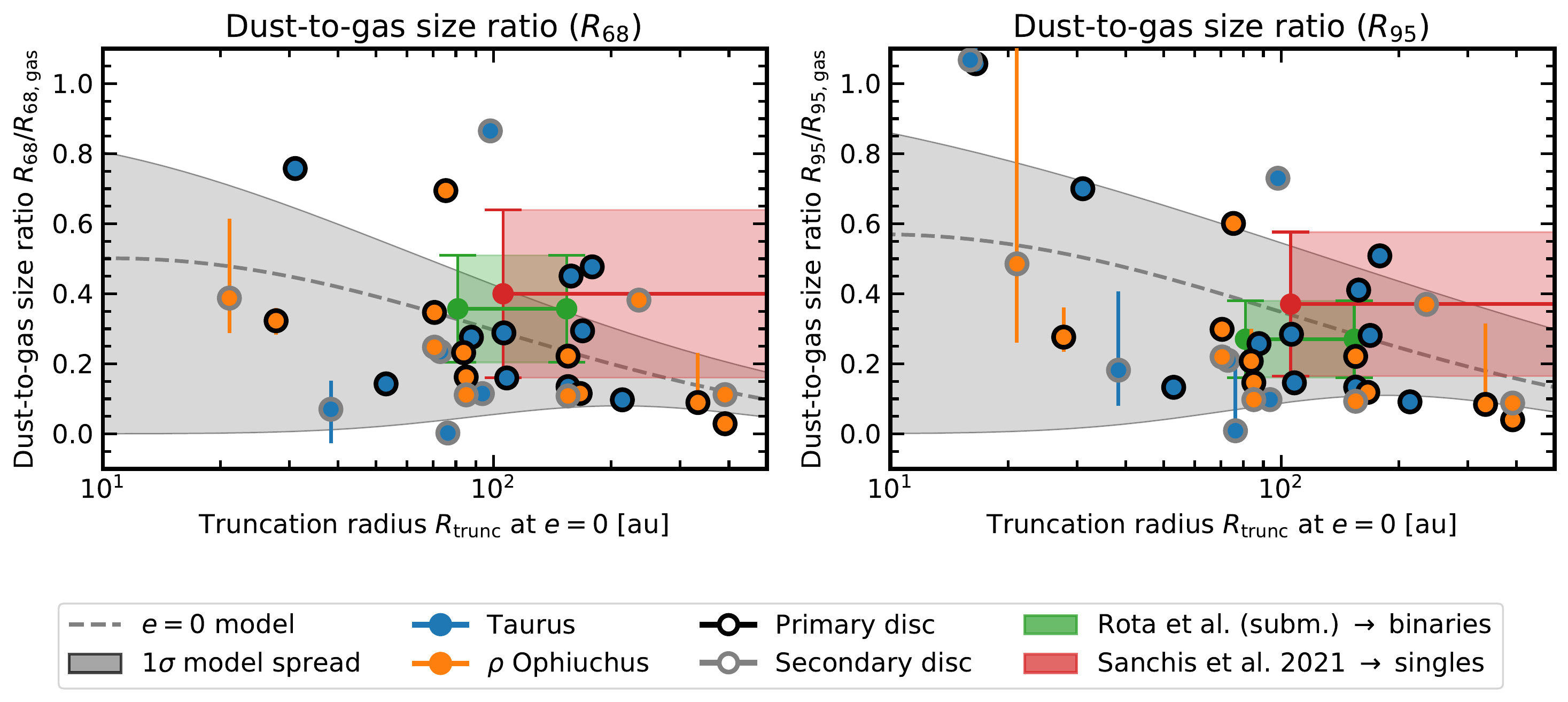}
    \caption{Same as in Fig.~\ref{fig.2.6} but showing the dust-to-gas disc size ratio (see text for definitions). The green and red shaded areas represent the median and mean estimate for $R_\mathrm{dust}/R_\mathrm{gas}$  with the relative uncertainty in Rota et al. (subm.) and \citet{Sanchis+21_2021arXiv210111307S}, respectively. The truncation radii those estimates apply for are roughly estimated as a third of the projected separation.}
    \label{fig.2.6_ratio}
\end{figure*}

Here we plot our estimate for the median dust-to-gas disc size ratio as a function of the truncation radius so as to see if a general trend can be observed. To infer $R_\mathrm{dust}/R_\mathrm{gas}$ we use the same dust radii computed in the main body of the paper (see Sec.~\ref{sec.5}) and determine the 68- and 95-per-cent \textit{gas} radii simply prescribing $R_\mathrm{68,gas}=0.68\times R_\mathrm{trunc}$ and $R_\mathrm{95,gas}=0.95\times R_\mathrm{trunc}$, where $R_\mathrm{trunc}$ is either the theoretical or zero-eccentricity observationally estimated truncation radius. It is our primary concern to underline that this method is definitely not rigorous: it neglects molecular dynamics, diffusion, any chemical network reactions, freeze-out, photo-dissociation and several other processes that do affect \ce{CO} secular evolution. Therefore our results should be considered as highly tentative. They are shown in Fig.~\ref{fig.2.6_ratio}: here the dashed grey line and the shaded grey area identify the model best-fit and its 1$\sigma$ spread, respectively. The Taurus and $\rho$~Ophiuchus discs observed in the continuum are over-plotted as blue and orange dots, respectively. The black edges are used for the primaries, while the grey ones for the secondaries. The red and green shaded areas represent the \textit{median} \citep{Sanchis+21_2021arXiv210111307S} and \textit{average} (Rota et al. subm.) inferred size ratio with their uncertainty in single-star and binary discs, respectively. They span the same truncation radius interval ($R_\mathrm{trunc}=a_\mathrm{p}/3$) between the closest and furthest systems taken into account in those papers (excluding the UZ~Tau~Wab sub-system in Rota et al. subm.).

If taken at face value, our results suggest that the dust-to-gas size ratio decreases with $R_\mathrm{trunc}$. This does not mean that we are predicting larger discs in binaries as clearly our gas radii scale with the truncation radius, yet simply that $R_\mathrm{gas}$ and $R_\mathrm{dust}$ are closer in binary than in single-star discs. However this trend could be indicative of our limiting assumptions on gas evolution. \ce{CO} photo-dissociation \citep[e.g.,][]{Facchini+17_2017A&A...605A..16F,Trapman+19_2019A&A...629A..79T} in the outer disc regions could reduce the \ce{CO} emission leading to $R_{x,\mathrm{gas}}$ being much smaller than $xR_\text{trunc}$, with $x=0.68,\,0.95$. This effect is expected to be more important in single-star discs, as they are more radially extended (Toci et al. in prep.) and could partially mitigate our downward trend of the dust-to-gas size ratio with $R_\mathrm{trunc}$. Detailed modelling of the gas evolution must be carried on e.g., studying the \ce{CO} emission in binary systems with full radiative transfer thermo-chemical codes such as \texttt{DALI} \citep{Bruderer12_2012A&A...541A..91B,Bruderer13_2013A&A...559A..46B,Bruderer14_2014A&A...562A..26B}. Furthermore, disc population synthesis, possibly taking into account observational biases, would be necessary for a proper comparison with the data.

This is why we would avoid a direct comparison with the observations. Tentatively, if only the two population-averaged values are considered (the green and red shaded areas in Fig.~\ref{fig.2.6_ratio}), they show an increasing dust-to-gas size ratio between binary (Rota et al. subm.) and single-star \citep{Sanchis+21_2021arXiv210111307S} discs, which is in contrast with our rough trend.
However, both the observationally inferred values are broadly compatible with our median dust-to-gas size ratio within their (large) uncertainty, suggesting that larger data samples (in addition to a dedicated modelling effort) are needed for more thorough comparisons.

In addition to the large uncertainties, the targeted samples are generally not complete (both gas and dust sizes are easily estimated in larger and brighter discs). To reduce the possible biases due to non-completeness, \citet{Sanchis+21_2021arXiv210111307S} considered a sub-sample of discs with stellar masses closer to the solar one (as in our models). However, only 8 discs in this group have both measured gas and dust sizes, with similar inferences for the median dust-to-gas size ratios and larger uncertainties ($R_\mathrm{gas}/R_\mathrm{dust}=2.5\pm2.0$, when the 68-per-cent flux radius is used as a metric, and $R_\mathrm{gas}/R_\mathrm{dust}=2.6\pm2.2$, when the 90-per-cent-flux radius is employed, have a better agreement with our estimates and trend). Finally, in the case of Rota et al. (subm.) results we underline that the \textit{median} is a more stable operator than the \textit{mean value} and should be used with consistency with \citet{Sanchis+21_2021arXiv210111307S}.

%\textit{Caveats and expectations from gas modelling} One of our most limiting assumptions is that a given fraction $x$ of $R_\mathrm{trunc}$ traces $R_{x,\mathrm{CO}}$, the disc radius enclosing the same fraction of the \ce{^{12}CO} flux. Indeed, several processes, such as \ce{CO} photo-dissociation in the outer disc regions, could reduce the \ce{CO} emission leading to $R_{x,\mathrm{CO}}$ being much smaller than $xR_\text{trunc}$. This effect is expected to be more important in single-star discs, as they are more radially extended (Toci et al. in prep.). This would partially mitigate our downward trend of the dust-to-gas size ratio with $R_\mathrm{trunc}$.
%Detailed modelling of the gas evolution must be carried on e.g., studying the \ce{CO} emission in binary systems with full radiative transfer thermo-chemical codes such as \texttt{DALI} \citep{Bruderer12_2012A&A...541A..91B,Bruderer13_2013A&A...559A..46B,Bruderer14_2014A&A...562A..26B}. Furthermore, disc population synthesis, possibly taking into account observational biases, would be necessary for a proper comparison with the data.

%%%%%%%%%%%%%%%%%%%%%%%%%%%%%%%%%%%%%%%%%%%%%%%%%%

% Don't change these lines
\bsp	

% typesetting comment
\label{lastpage}

% End of mnras_template.tex

\end{document}